\documentclass[11pt,english]{article}

\usepackage{times}
\usepackage{amsfonts}
\usepackage{amsmath,amssymb}
\usepackage{amsthm}
\usepackage{natbib}
\usepackage{bm}
\usepackage{graphicx}
\usepackage{subcaption}
\usepackage{appendix}
\usepackage{rotating}
\usepackage{float}
\usepackage{authblk}
\usepackage{color}
\usepackage[colorlinks,linkcolor=blue,citecolor=blue,bookmarks=false,pagebackref]{hyperref}
\usepackage[top=1.0in, bottom=1.3in, left=1.2in, right=1.2in]{geometry}
\usepackage{setspace}
\usepackage{titling}
\allowdisplaybreaks

\theoremstyle{plain}
\newtheorem{theorem}{Theorem}[section]
\theoremstyle{plain}
\newtheorem{lemma}{Lemma}[section]
\theoremstyle{plain}
\newtheorem{assumption}{Assumption}[section]
\theoremstyle{plain}
\newtheorem{proposition}{Proposition}[section]
\theoremstyle{plain}

\theoremstyle{remark}

\theoremstyle{definition}
\newtheorem{definition}{Definition}[section]

\newcommand{\argmin}{\operatornamewithlimits{argmin\,}}

\newcommand{\E}{\mathbb{E}}

\newcommand{\was}{\mathcal{W}}
\newcommand{\pdist}{\mathcal{D}}
\mathchardef\mhyphen="2D

\floatstyle{ruled}
\newfloat{algorithm}{tbp}{loa}
\providecommand{\algorithmname}{Algorithm}
\floatname{algorithm}{\protect\algorithmname}

\title{On parameter estimation with the Wasserstein distance}
\author{Espen Bernton\thanks{Department of Statistics, Harvard University, USA.
Address correspondence to ebernton@g.harvard.edu.} , Pierre E. $\text{Jacob}^*$, Mathieu
Gerber\thanks{School of Mathematics, University of Bristol, UK.} , Christian P.
Robert\thanks{CEREMADE, Universit\'e Paris-Dauphine and Paris Sciences \& Lettres - PSL Research University, France, and
Department of Statistics, University of Warwick, UK.}}
\date{}

\begin{document}

\maketitle

\begin{abstract}
\noindent     {Statistical inference can be performed by
    minimizing, over the parameter space, the Wasserstein distance between
    model distributions and the empirical distribution of the data. 
    We study
    asymptotic properties of such minimum Wasserstein distance estimators,
    complementing results derived by Bassetti, Bodini and Regazzini in 2006.
    In particular, our results cover the misspecified setting, in which
    the data-generating process is not assumed to be part of the family
    of distributions described by the model. 
    Our results are motivated by recent 
    applications of minimum Wasserstein estimators to complex generative models.
    We discuss some difficulties arising in the approximation of these estimators and illustrate their behavior in several numerical experiments. 
    Two of our examples are taken from the literature on approximate Bayesian computation 
    and have likelihood functions that are not analytically tractable. 
    Two other examples involve misspecified models.}
\end{abstract}


\section{Introduction \label{ch1:sec:introduction}}

We consider a statistical estimation approach for parametric models that is
based on minimizing the Wasserstein distance between the empirical distribution
of the data and the model distributions
\citep{belili1999estimate,bassetti2006minimum}. We study two different point
estimators, where the first, called the minimum Wasserstein estimator (MWE),
arises as the most important special case of the estimator introduced by
\citet{bassetti2006minimum}. The second, which we term the minimum expected
Wasserstein estimator (MEWE), is better suited to numerical approximations. 

We derive theoretical properties of the estimators, such as existence, measurability, and consistency, in
the misspecified setting. That is, we do not assume that the observations are generated 
from the working model. For one-dimensional data, we also study the convergence rate and asymptotic distribution of the minimum Wasserstein
estimator of order 1, extending the work of \citet{bassetti_regazzini2006} on
location-scale models. Our proofs are based on epi-convergence
\citep{rockafellar2009variational} and general results on minimum distance estimation \citep{pollard1980}, and are as such different from those presented by
Bassetti and coauthors.

There are two main motivations for developing these results.
Firstly, recent advances in computational optimal transport have led to the application of minimum Wasserstein distance estimators in increasingly complicated settings, where the models are likely to be misspecified. For instance, \citet{genevay2018learning} apply the MEWE in the tuning of
image generation models, and \citet{genevay2017gan} show that a version of the
MEWE also appears in the popular Wasserstein GAN method
\citep{arjovsky2017wasserstein}. This development has been driven by the advent of efficient numerical algorithms to approximate the Wasserstein distance \citep[see
e.g.][]{peyre2018computational,cuturi2013sinkhorn,benamou2014iterative,aude2016stochastic,li2018parallel,altschuler2018massively}.

Secondly, minimum Wasserstein distance
estimators, which are particular instances of minimum distance estimators 
\citep{basu2011statistical}, appear to be practical and robust alternatives to
likelihood-based estimation in the setting of generative models. In these
models, synthetic observations can be generated given a parameter, but the
likelihood function and associated maximum likelihood estimators might be intractable
\citep{gourieroux1993,marin2012approximate,bernton2019approximate}. 
Some comments on the comparison between the Wasserstein distance and other distances commonly
used in minimum distance estimation are provided.

The rest of this article is organized as follows: we review the definitions of minimum distance
estimation, of the Wasserstein distance, and of the 
estimators of interest in the rest of this section. Theoretical results, whose proofs can be found in the Appendix, and some open questions are stated in Section \ref{ch1:sec:theory}. We briefly review computational strategies to compute the
Wasserstein distance and the estimators in Section \ref{ch1:sec:computation},
before illustrating their behavior on various examples in Section
\ref{ch1:sec:illustrations}. We conclude in Section \ref{ch1:sec:discussion}. Code to reproduce the numerical results can be found at \href{https://github.com/pierrejacob/winference}{github.com/pierrejacob/winference}.

\subsection{Notation}\label{ch1:sec:notation}

Throughout this article we consider a probability space $(\Omega,\mathcal{F},\mathbb{P})$, with
associated expectation operator $\mathbb{E}$, on which all the random variables are defined. The set of probability measures on a space $\mathcal{X}$ is denoted by $\mathcal{P}(\mathcal{X})$. 
 The data take values in
$\mathcal{Y}$, a subset of $\mathbb{R}^{d}$ for some $d\in\mathbb{N}$, and is endowed with the Borel $\sigma$-algebra. We
observe $n\in\mathbb{N}$ data points, $y_{1:n} = y_1,\ldots,y_n$,
that are distributed according to
$\mu_\star^{(n)}\in\mathcal{P}(\mathcal{Y}^{n})$. Let $\hat{\mu}_n = n^{-1}\sum_{i=1}^n \delta_{y_i}$, where $\delta_y$ is the Dirac distribution with mass on $y\in\mathcal{Y}$. We refer to $\hat{\mu}_n$  as the empirical distribution  of $y_{1:n}$, even in settings where the observations are not i.i.d.

A model refers to a collection of distributions on 
$\mathcal{Y}^n$,
denoted by $$\mathcal{M}^{(n)}=\{\mu^{(n)}_\theta : \theta \in \mathcal{H}\} \subset\mathcal{P}(\mathcal{Y}^n),$$
where $\mathcal{H}\subset \mathbb{R}^{d_\theta}$ is the parameter space, endowed
with a distance $\rho_\mathcal{H}$ and of dimension $d_\theta\in \mathbb{N}$.  However, we will often assume that the sequence of models $(\mathcal{M}^{(n)})_{n\geq 1}$ is such that, for every $\theta\in\mathcal{H}$, the sequence $(\hat{\mu}_{\theta,n})_{n\geq 1}$ of random  probability measures on $\mathcal{Y}$ converges (in some sense) to a distribution $\mu_\theta\in\mathcal{P}(\mathcal{Y})$,  where $\hat{\mu}_{\theta,n}=n^{-1}\sum_{i=1}^n \delta_{z_i}$ with $z_{1:n}\sim \mu_\theta^{(n)}$. Similarly, we will often assume  that    $\hat{\mu}_n$ converges   to some distribution $\mu_\star\in  \mathcal{P}(\mathcal{Y})$ as $n\rightarrow \infty$. Whenever the notation $\mu_\star$ and $\mu_\theta$ is used, it is implicitly assumed that these objects exist. In such cases, we instead refer to $\mathcal{M} = \{\mu_\theta : \theta \in \mathcal{H}\} \subset \mathcal{P}(\mathcal{Y})$ as the model. 
We say that it is well-specified if there exists $\theta_\star\in\mathcal{H}$ such that $\mu_\star = \mu_{\theta_\star}$; otherwise it is misspecified. Parameters are identifiable if $\theta =\theta^\prime$ is implied by $\mu_\theta = \mu_{\theta^\prime}$.
The weak convergence of a sequence of measures $\mu_n$ to $\mu$ is denoted by $\mu_n \Rightarrow \mu$. The Kullback-Leibler (KL) divergence
between $\mu$ and $\nu$ is defined as 
$$\text{KL}(\mu |\nu) = \int \log \frac{d\mu}{d\nu} d\mu$$
if $\mu$ is absolutely continuous with respect to $\nu$, and $+\infty$ otherwise.

\subsection{Minimum distance estimation \label{ch1:sec:mde}}

Minimum distance estimation refers to the minimization, over the
parameter $\theta\in\mathcal{H}$, of a distance between the empirical distribution
$\hat{\mu}_n$ and the model distribution $\mu_\theta$ \citep{wolfowitz1957,basu2011statistical}. 
More formally, denoting by $\pdist$ a distance or divergence on $\mathcal{P}(\mathcal{Y})$, the associated minimum distance
estimator (MDE)  can be defined as
\begin{equation}\label{ch1:eq:mke}
    \hat{\theta}_n = \argmin_{\theta \in \mathcal{H}} \pdist(\hat{\mu}_n, \mu_\theta).
\end{equation}

In broad terms, the minimum distance estimation principle captures the idea of many statistical paradigms. For instance, the generalized method of moments \citep{hansen1982large} consists in minimizing a discrepancy $\pdist$ defined as the weighted Euclidean distance between moments of $\hat{\mu}_n$ and $\mu_\theta$. In the empirical likelihood method \citep{owen2001empirical},  $\pdist$ is taken to be the KL divergence, and the model is supported strictly on the set of observed data and subject to moment conditions. 
The maximum likelihood estimator minimizes the KL divergence
between $\mu_\star$ and $\mu_\theta$ in the limit of the number of observations going to
infinity.

However, it is worth noting that the definition in \eqref{ch1:eq:mke} precludes the
naive application of some discrepancy measures. For instance, one could not
directly choose $\pdist$ to be the KL divergence or the total variation
distance, since for any model distribution $\mu_\theta$ not supported solely on
the observed data, they would evaluate to $+\infty$ and $1$ respectively. To
apply  discrepancies of this kind, one would first need to build sample-based
estimators of the underlying population quantity $\pdist(\mu_\star,
\mu_\theta)$, assuming it is well-defined. Many such approaches have been
studied in detail by \citet{basu2011statistical}.

The computation of the minimum distance estimator might be intractable, especially in settings where it is assumed that one can simulate data from the model distribution but not evaluate its density. For such generative models, the following minimum expected distance estimator might be more computationally convenient:
\begin{equation}\label{ch1:eq:altmke}
    \hat{\theta}_{n,m} = \argmin_{\theta \in \mathcal{H}} \mathbb{E}_m
    \pdist(\hat{\mu}_n, \hat{\mu}_{\theta,m}),
\end{equation}
where the expectation $\mathbb{E}_m$ is taken over the distribution of the sample $z_{1:m}
\sim \mu_\theta^{(m)}$ giving rise to $\hat{\mu}_{\theta,m}=m^{-1}\sum_{i=1}^m
\delta_{z_i}$.  When $n$ is fixed and $m$ is large, or when $n=m$  and $n$ is
large, one might hope that the expectation is close to $\pdist(\hat{\mu}_n,
\mu_\theta)$, and that the estimators $\hat{\theta}_n$ and $\hat{\theta}_{n,m}$ have similar properties. 
Inference techniques such as the method of simulated moments \citep{mcfadden1989} and indirect inference \citep{gourieroux1993} often (implicitly) use estimators
of this form, in which $\pdist$ defined as the weighted Euclidean distance between sample moments or summary statistics of $y_{1:n}$ and $z_{1:m}$, and the expectation
in \eqref{ch1:eq:altmke} is replaced with a Monte Carlo approximation.

\subsection{Minimum Wasserstein estimation \label{ch1:subsec:mwe}}
In this work, we focus on minimum distance estimation with the Wasserstein distance. Let $\rho$ be a distance on the
observation space $\mathcal{Y}$, and let
$\mathcal{P}_p(\mathcal{Y})$ with $p\geq1$ (e.g. $p=1$ or $2$) be the set of
distributions $\mu\in\mathcal{P}(\mathcal{Y})$ with finite $p$-th moment, i.e. there
exists $y_0\in\mathcal{Y}$ such that $\int_\mathcal{Y} \rho(y,y_0)^p d\mu(y) < \infty$.  
The $p$-Wasserstein distance, also called the Monge-Kantorovich, Mallows, or Gini distance, is a finite metric on
$\mathcal{P}_p(\mathcal{Y})$, defined by the optimal transport problem
\begin{equation} \label{ch1:eq:wass_def} 
\was_p(\mu,\nu)^p = \inf_{\gamma \in \Gamma(\mu,\nu)} \int_{\mathcal{Y}\times \mathcal{Y}} \rho(x,y)^p d\gamma(x,y),
\end{equation}
where $\Gamma(\mu,\nu)$ is the set of probability measures on $\mathcal{Y}\times \mathcal{Y}$ with marginals $\mu$ and $\nu$ respectively;
see Chapter 6 of \citet{villani2008} for a brief history of this distance and its central role in optimal transport.

A useful property of the Wasserstein distance is that it is well-defined for distributions with non-overlapping supports.
This allows us to define the minimum Wasserstein estimator (MWE)  of order $p$, denoted $\hat{\theta}_n$, by simply plugging $\was_p$ into \eqref{ch1:eq:mke} in place of $\pdist$. 
Some properties of the MWE have been studied in \citet{bassetti2006minimum}, for well-specified models and i.i.d.~data; we
derive new results in Section \ref{ch1:sec:theory:mwe} under weaker assumptions. We also propose the minimum expected Wasserstein estimator (MEWE), obtained by replacing $\pdist$ with $\was_p$ in  \eqref{ch1:eq:altmke} and denoted $\hat{\theta}_{n,m}$. 
We describe some of its theoretical properties in Section \ref{ch1:sec:theory:mewe}.

Variations of these estimators have recently been applied by for instance
\citet{arjovsky2017wasserstein} and \citet{genevay2018learning}. In the
settings they consider, the models are likely to be misspecified, and are
supported on low-dimensional manifolds that might not overlap with the
support of the data-generating mechanism. 
While the Wasserstein distance is well-defined in that case, 
the KL divergence or the total variation are not. This motivates
the study of minimum Wasserstein estimators for these settings.

\section{Theoretical results} \label{ch1:sec:theory}

We prove the existence, measurability, and consistency of the
MWE and MEWE under weak assumptions, allowing the model to be 
misspecified and to produce data with certain types of dependencies. Under stronger
assumptions, we study the rate of convergence and the asymptotic distribution of the MWE when $d=1$ and $p=1$.
Throughout, we compare our results to those of
\citet{bassetti2006minimum} and \citet{bassetti_regazzini2006}.

Informally, the consistency of the MWE and MEWE can be understood as follows.
Under some conditions, we expect $\hat\mu_n$ to converge to $\mu_\star$, in
the sense that $\was_p(\hat\mu_n, \mu_\star)\to 0$ as $n \to \infty$.  Consequently, the minimum
of $\theta \mapsto \was_p(\hat\mu_n, \mu_\theta)$ might converge to the minimum of
$\theta \mapsto \was_p(\mu_\star, \mu_\theta)$, denoted by $\theta_\star$, assuming its
existence and unicity. The same can be said for the minimum of $\theta \mapsto \mathbb{E}_m\was_p(\hat\mu_n, \hat{\mu}_{\theta,m})$, provided $m \to\infty$ also.
The parameter $\theta_\star$ is thus the limiting object of interest, also termed the estimand. Beyond its interpretation
as the minimizer of $\theta \mapsto \was_p(\mu_\star, \mu_\theta)$,
this parameter would coincide to the data-generating parameter if we assume that the data are
generated from the model. In the misspecified case, note that $\theta_\star$ is not 
necessarily the parameter that minimizes $\text{KL}(\mu_\star|\mu_\theta)$,
which is the limit of the maximum likelihood estimator under standard regularity conditions.

\subsection{Minimum Wasserstein estimator}\label{ch1:sec:theory:mwe}

\subsubsection{Existence, measurability, and consistency}

We first list assumptions on the data-generating process and on the model that are sufficient 
for the existence, measurability,
and consistency for the MWE.

\begin{assumption}\label{ch1:as:cvgwas}
The data-generating process is such that $\was_p(\hat{\mu}_n, \mu_\star) \to 0$,  $\mathbb{P}$-almost surely as $n\to \infty$.
\end{assumption}

\begin{assumption} \label{ch1:as:weakcvg}
The map $\theta\mapsto \mu_\theta$ is continuous in the sense that $\rho_\mathcal{H}(\theta_n,\theta) \to 0$ implies $\mu_{\theta_n}  \Rightarrow \mu_\theta$ as $n\to \infty$.
\end{assumption}

\begin{assumption}\label{ch1:as:relativelycompact}
For some $\varepsilon>0$, the set $B_\star(\varepsilon) = \{\theta \in \mathcal{H} : \was_p(\mu_{\star},\mu_\theta) \leq \varepsilon_\star + \varepsilon\}$ is bounded, where
$\varepsilon_\star = \inf_{\theta \in \mathcal{H}} \was_p(\mu_\star,\mu_\theta)$.
\end{assumption}

\begin{theorem}[Existence and consistency of the MWE] \label{ch1:theorem:consistent}
Under Assumptions \ref{ch1:as:cvgwas}-\ref{ch1:as:relativelycompact}, there exists a set $E\subset \Omega$
with $\mathbb{P}(E)=1$ such that, for all $\omega \in E$, 
$$\inf_{\theta\in\mathcal{H}} \was_p(\hat{\mu}_n(\omega),\mu_\theta) \to \inf_{\theta\in\mathcal{H}} \was_p(\mu_\star,\mu_\theta),$$
and there exists $n(\omega)$ such that for $n\geq n(\omega)$, the sets $\argmin_{\theta\in\mathcal{H}} \was_p(\hat{\mu}_n(\omega),\mu_\theta)$
are non-empty and form a bounded sequence with 
$$\limsup_{n\to\infty} \argmin_{\theta\in\mathcal{H}} \was_p(\hat{\mu}_n(\omega),\mu_\theta) \subset \argmin_{\theta\in\mathcal{H}} \was_p(\mu_{\star},\mu_\theta).$$
\end{theorem}

For a generic function $f$, let $\varepsilon\mhyphen\argmin_x f = \{x: f(x)\leq \varepsilon + \inf_x f\}$. Theorem \ref{ch1:theorem:consistent} also holds if one replaces  $\argmin_{\theta\in\mathcal{H}} \was_p(\hat{\mu}_n(\omega),\mu_\theta)$ with $\varepsilon_n\mhyphen\argmin_{\theta\in\mathcal{H}} \was_p(\hat{\mu}_n(\omega),\mu_\theta)$, for any sequence $\varepsilon_n$ converging to zero. If  $\theta_\star = \argmin_{\theta\in\mathcal{H}} \was_p(\mu_\star,\mu_\theta)$ is unique, the result can be rephrased as $\hat{\theta}_n\to \theta_\star$ $\mathbb{P}$-almost surely. 

The following theorem derives from a general result by \citet{brown1973} on the
measurability of estimators defined as minimizers.

\begin{theorem}[Measurability of the MWE]  \label{ch1:theorem:measurable}
Suppose that $\mathcal{H}$ is a $\sigma$-compact Borel measurable subset of $\mathbb{R}^{d_\theta}$. Under Assumption \ref{ch1:as:weakcvg}, for any $n \geq 1$ and $\varepsilon > 0$, there exists a Borel measurable function $\hat{\theta}_n : \Omega \to \mathcal{H}$ that satisfies
\begin{align*}
\hat{\theta}_n(\omega) \in
  \begin{cases}
    \argmin_{\theta\in\mathcal{H}} \was_p(\hat{\mu}_n(\omega),\mu_\theta) & \text{if this set is non-empty,}  \\
    \varepsilon \mhyphen \argmin_{\theta\in\mathcal{H}} \was_p(\hat{\mu}_n(\omega),\mu_\theta) & \text{otherwise}.
  \end{cases}
\end{align*}
\end{theorem}

Theorem \ref{ch1:theorem:consistent} generalizes the results of
\citet{bassetti2006minimum}, where the model is assumed to be well-specified in
the sense that $\mu_\star\in\mathcal{M}$. Moreover, Theorem \ref{ch1:theorem:consistent} 
allows for data-generating processes which do not produce independent data points. For instance, if
the data form a stationary and ergodic time series whose marginal distribution
has finite $p$-th moments, then Assumption \ref{ch1:as:cvgwas} still holds. These and other
sufficient conditions for Assumption \ref{ch1:as:cvgwas} to be satisfied are
elaborated upon in the Appendix. Theorem
\ref{ch1:theorem:measurable} is only a minor generalization of the result in
\citet{bassetti2006minimum}, where it is assumed that for each $n \geq 1$, $\argmin_{\theta\in\mathcal{H}}
\was_p(\hat{\mu}_n(\omega),\mu_\theta)$ is non-empty for almost every
$\omega\in\Omega$. In the next section, this small modification also enables the direct
application of results by \citet{pollard1980}.

\subsubsection{Rate of convergence and asymptotic distribution \label{ch1:subsec:asymptoticdistribution}}
Under conditions guaranteeing the consistency of the minimum Wasserstein
estimator, we study its rate of convergence and asymptotic distribution in the case where $p=1$,
$\mathcal{Y} = \mathbb{R}$, $\rho(x,y) = \lvert x-y \rvert$. Under this setup, it can be shown that 
$\was_1(\mu,\nu) = \int_0^1 \lvert F_\mu^{-1}(s) - F_\nu^{-1}(s)\rvert ds = \int_\mathbb{R} \lvert F_\mu(t) - F_\nu(t)\rvert dt,$ where $F_\mu$ and $F_\nu$ denote the cumulative distribution functions (CDFs) of $\mu$ and $\nu$ respectively \citep[see e.g.][Theorem 6.0.2]{ambrosio2005}. Additionally, assume that $\mathcal{H}$ is endowed with a norm: $\rho_\mathcal{H}(\theta,\theta') = \|\theta-\theta'\|_\mathcal{H}$. We also require that $\theta_\star$ is
``well-separated'':
\begin{assumption}\label{ch1:as:wellsep}
For all $\varepsilon > 0$, there exists $\delta > 0$ such that
$$\inf_{\theta \in \mathcal{H}:\|\theta - \theta_\star\|_\mathcal{H} \geq \varepsilon} \was_1(\mu_{\theta_\star},\mu_\theta) >  \delta.$$
\end{assumption}
This assumption is commonly made in the asymptotic study of M-estimators; see e.g. Chapter 5 of \citet{van2000asymptotic} and the Appendix. We focus on the setting in which the model is well-specified, but also discuss some extensions to the misspecified setting in Section \ref{ch1:sec:extensions}.

Our approach to derive asymptotic distributions follows \citet{pollard1980}.
Let $F_\theta$, $F_\star$ and $F_n$ denote the CDFs of $\mu_\theta$,
$\mu_\star$ and $\hat{\mu}_n$ respectively. Informally speaking, we show
that $\sqrt{n}W_1(\hat{\mu}_n,\mu_\theta)$ can be approximated by
$$\int_\mathbb{R}\lvert \sqrt{n}(F_n(t)-F_\star(t)) - \langle
\sqrt{n}(\theta-\theta_\star), D_{\theta_\star}(t) \rangle \rvert dt$$
near $\theta_\star$, for some $D_{\theta_\star} \in (L_1(\mathbb{R}))^{d_\theta}$,  with
$\langle \theta, u \rangle = \sum_{i=1}^{d_\theta}\theta_iu_i$. Results by
\citet{del1999central} and \citet{dede2009} give conditions under which $
\sqrt{n}(F_n-F_\star)$ converges to a zero mean Gaussian process $G_\star$ with
known covariance structure,  for both independent and certain classes of
dependent data. Heuristically, the distribution of $\sqrt{n}(\hat{\theta}_n -\theta_\star)$ is then close to that of
$\argmin_{u\in\mathcal{H}}\int_\mathbb{R}\lvert G_\star(t) - \langle
u,D_{\theta_\star}(t)\rangle\rvert dt$. The required form of $D_{\theta_\star}$ is given in the following assumption:
\begin{assumption}\label{ch1:as:diff}
There exists a non-singular $D_{\theta_\star} \in (L_1(\mathbb{R}))^{d_\theta}$ such that 
$$\int_\mathbb{R}\lvert F_\theta(t)-F_{\theta_\star}(t) - \langle \theta-\theta_\star, D_{\theta_\star}(t)\rangle \rvert dt = o(\|\theta-\theta_\star\|_\mathcal{H}),$$  
as  $\|\theta-\theta_\star\|_\mathcal{H} \to 0.$
\end{assumption}

To provide some intuition into the nature of the  ``derivative" $D_{\theta_\star}$, we consider the following simple example. Let $\mu_{\theta} = \mathcal{N}(\theta,1)$ for $\theta \in \mathbb{R}$, and $\mu_\star = \mu_{\theta_\star}$ for some $\theta_\star$. By Taylor expanding $F_\theta(t) = \Phi(t-\theta)$ around $\theta_\star$ (for fixed  $t$), Assumption \ref{ch1:as:diff} can be shown to hold with $D_{\theta_\star}(t) = - \varphi(t-\theta_\star)$, where $\Phi$ and $\varphi$ denote the CDF and density of a standard Gaussian variable, respectively.
Next, we state a result that holds for a well-specified model producing i.i.d. data, and analogous results for misspecified models and certain types of dependent processes can be found in the Appendix.

\begin{theorem} \label{ch1:theorem:asymptoticdistribution}
Suppose that $Y_i \sim \mu_\star = \mu_{\theta_\star}$ i.i.d.~and that $\theta_\star$ is in the interior of $\mathcal{H}$, and that $\int_0^\infty\sqrt{\mathbb{P}(\lvert Y_0 \rvert > t)} dt < \infty$  Suppose that Assumptions \ref{ch1:as:cvgwas}-\ref{ch1:as:diff} hold and that the minimum of $u\mapsto \int_\mathbb{R}\lvert G_\star(t) - \langle u, D_{\theta_\star}(t)\rangle  \rvert dt$ is almost surely unique. Then, the MWE with $p=1$ satisfies 
$$\sqrt{n}(\hat{\theta}_n -\theta_\star) \Rightarrow \argmin_{u\in\mathcal{H}} \int_\mathbb{R}\lvert G_\star(t) - \langle u, D_{\theta_\star}(t)\rangle  \rvert dt,$$ 
as $n \to \infty$, where $G_\star$ is a zero mean Gaussian process with  $$\mathbb{E}G_\star(s)G_\star(t) = \min\{F_\star(s),F_\star(t)\} - F_\star(s)F_\star(t).$$
\end{theorem}

A similar statement for $p=2$ can potentially be derived by considering the results of
\citet{del2005asymptotics}. The condition $\int_0^\infty\sqrt{\mathbb{P}(\lvert Y_0 \rvert > t)} dt < \infty$ implies the existence of second moments, and is itself implied by the existence of moments of order $2+\varepsilon$ for some $\varepsilon > 0$ \citep[see e.g. Section 2.9 in][]{wellner1996}. 
The uniqueness assumption on the argmin in the limit can be relaxed by considering
convergence to the entire set of minimizing values, as in the Appendix and Section 7 of
\citet{pollard1980}. Still, uniqueness can sometimes be established, using e.g. the results of \citet{cheney1969}. This approach is taken by \citet{bassetti_regazzini2006},
who directly show that Theorem \ref{ch1:theorem:asymptoticdistribution} holds when $\mathcal{M}$ is a
location-scale family supported on a bounded open interval. The existence and form of $D_{\theta_\star}$ can in many cases be derived if the model is differentiable in quadratic mean \citep{lecam1970}, which is elaborated upon in the Appendix. There, one can also find results to verify Assumptions \ref{ch1:as:cvgwas} and \ref{ch1:as:wellsep}. It can in some cases potentially be easier to verify the assumptions for a reparameterization of $\theta$, say $\varphi = r(\theta)$. Provided that the theorem holds for $\hat{\varphi}_n$ and that the inverse map $r^{-1}$ is differentiable, the limiting distribution of $\hat{\theta}_n$ can be derived using a delta method argument.

Computing confidence intervals using the asymptotic distribution provided by Theorem \ref{ch1:theorem:asymptoticdistribution} is hard, due in part to its dependence on unknown quantities. However, the existence of the limiting distribution is in itself sufficient to guarantee the asymptotic validity of appropriately constructed
subsampling confidence intervals \citep[][Theorem 2.2.1]{politis1999subsampling}. This also generalizes to settings with certain kinds of dependent data.
Under slightly stronger assumptions, the closely related $m$ out of $n$ bootstrap
produces asymptotically valid confidence intervals as well; see \citet{bickel2008choice} and references therein. In the numerical experiments of Section \ref{ch1:sec:illustrations}, we 
find that the standard bootstrap \citep{efron1994introduction} works well in practice.

Theorem \ref{ch1:theorem:asymptoticdistribution} also holds for approximations of
the MWE, say $\tilde{\theta}_n$, provided that $\tilde{\theta}_n =
\hat{\theta}_n + o_{\mathbb{P}}(1/\sqrt{n})$, as can be seen from its proof. In
light of the convergence of the MEWE to the MWE as $m\to\infty$ established in
Section \ref{ch1:sec:theory:mewe}, there exists a sequence $m(n)$ (depending on $\omega$) such
that the associated MEWE $\hat{\theta}_{n,m(n)}$ satisfies the conclusion of Theorem \ref{ch1:theorem:asymptoticdistribution}.

\subsubsection{Extensions} \label{ch1:sec:extensions}

Under slightly stronger assumptions, Theorem
\ref{ch1:theorem:asymptoticdistribution} can be extended to the misspecified
setting. In particular, suppose that 
there exists a neighborhood $N$ of $\theta_\star$ and a constant $c>0$ such that for any $\theta \in N$, 
$$\was_1(\mu_{\theta},\mu_\star) \geq \was_1(\mu_{\theta_\star},\mu_\star) + c \|\theta-\theta_\star\|_\mathcal{H}.$$
In the well-specified case, this property is implied by Assumption \ref{ch1:as:diff}.
Then, as elaborated upon in the Appendix, the minimum of $\theta\mapsto \was_1(\hat{\mu}_n,\mu_\theta)$ is attained on the set 
$\mathcal{S}_n = \{\theta: \|\theta-\theta_\star\|_\mathcal{H} \leq 4\was_1(\hat{\mu}_n,\mu_\star)/c\}$ with probability going to one. Since the conditions of Theorem \ref{ch1:theorem:asymptoticdistribution} 
imply that $\was_1(\hat{\mu}_n,\mu_\star) = O_\mathbb{P}(1/\sqrt{n})$, this in turn implies that $\|\hat{\theta}_n - \theta_\star\|_\mathcal{H} = O_\mathbb{P}(1/\sqrt{n})$ also. In other words, the minimum Wasserstein estimator retains its rate of convergence in the misspecified case.

To find its asymptotic distribution, one can observe that with probability going to one, the map $\theta\mapsto \sqrt{n}\was_1(\hat{\mu}_n,\mu_\theta)$ can be approximated uniformly well over $\mathcal{S}_n$ by the map $\theta\mapsto \sqrt{n}\int_\mathbb{R}\lvert F_n(t)-F_{\theta_\star}(t) - \langle \theta-\theta_\star, D_{\theta_\star}(t) \rangle \rvert dt,$ which similarly achieves its minimum on $\mathcal{S}_n$. Therefore, as $n$ gets large, $\sqrt{n}(\hat{\theta}_n-\theta_\star)$ behaves like a minimum of 
$$u\mapsto \int_\mathbb{R}\lvert \sqrt{n}(F_n(t) -F_\star(t)) + \sqrt{n}(F_\star(t)-F_{\theta_\star}(t)) - \langle u, D_{\theta_\star}(t) \rangle \rvert dt.$$
Under the conditions of Theorem \ref{ch1:theorem:asymptoticdistribution}, $\sqrt{n}(F_n -F_\star)$ converges to $G_\star$ in the sense of \citet{del1999central}. In turn, $\sqrt{n}(\hat{\theta}_n-\theta_\star)$ should be distributed as the minimizer(s) of $u\mapsto \int_\mathbb{R}\lvert G_\star(t)+ \sqrt{n}(F_\star(t)-F_{\theta_\star}(t)) - \langle u, D_{\theta_\star}(t) \rangle \rvert dt$ as $n$ grows. A technical complication arises since this function converges pointwise to infinity,  and we therefore leave formal statements for the Appendix.

Extensions to cases with multivariate data are
left for future research. It is unclear whether convergence to
$\theta_\star$ will occur at the same $\sqrt{n}$ rate in higher dimensions.
This is because $\mathbb{E}\was_p(\hat{\mu}_n,\mu_\star)$ is 
on the order of $n^{-1/d}$ whenever $\mu_\star$ is absolutely continuous with respect to the Lebesgue measure and $d > 2p$ \citep[see
e.g.][and references therein]{weed2017sharp}. On the other hand,
\citet{del2017central} show, under some assumptions, that the 2-Wasserstein
distance satisfies the following CLT:
$$\sqrt{n} \left( \was_2^2(\hat{\mu}_n,\mu_\theta) - \mathbb{E}\was_2^2(\hat{\mu}_n,\mu_\theta) \right) \Rightarrow \mathcal{N}\left(0, \sigma^2(\mu_\star,\mu_\theta)\right),$$
where $\sigma^2(\mu_\star,\mu_\theta)$ has a known form and the expectation is
taken with respect to the observations $y_{1:n} \sim \mu_\star^{(n)}.$ Similar
results are expected to hold for other $p$ also. It therefore seems likely that
the distance(s) between the MWE and the minimizer(s) of $\theta \mapsto
\mathbb{E}\was_2^2(\hat{\mu}_n,\mu_\theta)$ converges to zero at the standard
$\sqrt{n}$ rate. If these speculations hold true, one could interpret them in
terms of a bias-variance trade-off: the bias would appear to be on the order of
$n^{-1/d}$, whereas the variance is on the order of $n^{-1/2}$. However, note that the function $\theta \mapsto
\mathbb{E}\was_2^2(\hat{\mu}_n,\mu_\theta)$ depends only on population
properties of $\mu_\star^{(n)}$. As such, it is a reasonable
alternative to the objective function  $\theta \mapsto
\was_2^2(\mu_\star,\mu_\theta)$, and might still yield reasonable
identification of the parameters. For instance, if the model is well-specified and Gaussian with $\theta$ being a location parameter, it seems likely that $\theta \mapsto
\mathbb{E}\was_2^2(\hat{\mu}_n,\mu_\theta)$ is minimized at $\theta_\star$ for any $n$. It is therefore unclear
whether the slow convergence rate of the bias would always be of practical concern.

\subsection{Minimum expected Wasserstein estimator}\label{ch1:sec:theory:mewe}
\subsubsection{Existence, measurability, and consistency}

In order to show similar results for the MEWE as for the MWE, we introduce the following additional assumptions.

\begin{assumption}\label{ch1:as:weakcvg_intermediate}
For any $m\geq 1$, if $\rho_\mathcal{H}(\theta_n,\theta) \to 0$, then $\mu_{\theta_n}^{(m)}  \Rightarrow \mu_\theta^{(m)}$ as $n\to \infty$.
\end{assumption}

\begin{assumption}\label{ch1:as:unicvg}
If $\rho_\mathcal{H}(\theta_n,\theta) \to 0$, then $\mathbb{E}_n\was_p(\mu_{\theta_n},\hat{\mu}_{\theta_n,n}) \to 0$ as $n\to \infty$.
\end{assumption}

Assumption \ref{ch1:as:weakcvg_intermediate} is slightly stronger than Assumption \ref{ch1:as:weakcvg}, stating that we not only need weak convergence of the ``model'' distributions $\mu_\theta$, but also of the sample distributions $\mu_\theta^{(m)}$ for any $m\geq 1$. Assumption \ref{ch1:as:unicvg} is implied by $\sup_{\theta\in\mathcal{H}}\mathbb{E}_n\was_p(\mu_{\theta},\hat{\mu}_{\theta,n}) \to 0$, which in turn might hold when $\mathcal{H}$ is compact and the inequalities in \citet{fournier_guillin2015} hold. 

In the next result, we prove an analogous version of Theorem \ref{ch1:theorem:consistent} for the MEWE as $\min\{n,m\} \to \infty$. For simplicity, we write $m$ as a function of $n$ and require that $m(n) \to \infty$ as $n\to \infty$.

\begin{theorem}[Existence and consistency of the MEWE] \label{ch1:theorem:consistent:mewe}
Under Assumptions \ref{ch1:as:cvgwas}-\ref{ch1:as:relativelycompact} and \ref{ch1:as:weakcvg_intermediate}-\ref{ch1:as:unicvg}, there exists a set $E\subset \Omega$
with $\mathbb{P}(E)=1$ such that, for all $\omega \in E$, 
$$\inf_{\theta\in\mathcal{H}} \mathbb{E}_{m(n)} \was_p(\hat{\mu}_n(\omega),\hat{\mu}_{\theta,m(n)}) \to \inf_{\theta\in\mathcal{H}} \was_p(\mu_\star,\mu_\theta),$$ and there exists
$n(\omega)$ such that, for all $n\geq n(\omega)$, the sets $\argmin_{\theta\in\mathcal{H}} \was_p(\hat{\mu}_n(\omega),\hat{\mu}_{\theta,m(n)})$
are non-empty and form a bounded sequence with 
$$\limsup_{n\to\infty} \argmin_{\theta\in\mathcal{H}} \mathbb{E}_{m(n)} \was_p(\hat{\mu}_n(\omega),\hat{\mu}_{\theta,m(n)}) \subset \argmin_{\theta\in\mathcal{H}} \was_p(\mu_{\star},\mu_\theta).$$
\end{theorem}

\begin{theorem}[Measurability of the MEWE]
Suppose that $\mathcal{H}$ is a $\sigma$-compact Borel measurable subset of $\mathbb{R}^{d_\theta}$. Under Assumption \ref{ch1:as:weakcvg_intermediate}, for any $n \geq 1$ and $m\geq 1$ and $\varepsilon > 0$, there exists a Borel measurable function $\hat{\theta}_{n,m} : \Omega \to \mathcal{H}$ that satisfies
\begin{align*}
\hat{\theta}_{n,m}(\omega) \in
  \begin{cases}
    \argmin_{\theta\in\mathcal{H}} \mathbb{E}_m\was_p(\hat{\mu}_n(\omega),\hat{\mu}_{\theta,m}), & \text{if this set is non-empty,}  \\
    \varepsilon \mhyphen \argmin_{\theta\in\mathcal{H}}\mathbb{E}_m\was_p(\hat{\mu}_n(\omega),\hat{\mu}_{\theta,m}), & \text{otherwise}.
  \end{cases}
\end{align*}
\end{theorem}

The results   above appear to be the first of their kind for the MEWE.

\subsubsection{Convergence to the MWE}

The next result considers the case where the data are fixed, while $m\to \infty$. It shows that the MEWE converges to the MWE, assuming the latter
exists. Using the results of \citet{del2017central} and references therein, one
could potentially derive the rate of this convergence, which we leave for future work.
We formulate the following additional assumption, in which the observed empirical distribution is kept fixed and
$\varepsilon_n = \inf_{\theta \in \mathcal{H}}\was_p(\hat{\mu}_n,\mu_\theta)$.

\begin{assumption}\label{ch1:as:bounded:fixedn}
For some $\varepsilon > 0$, the set $B_n(\varepsilon) = \{\theta\in\mathcal{H} : \was_p(\hat{\mu}_n,\mu_\theta)  \leq \varepsilon_n +\varepsilon\}$ is bounded.
\end{assumption}

\begin{theorem}[MEWE converges to MWE as $m\to \infty$] \label{ch1:theorem:mewe_to_mwe}
Under Assumptions \ref{ch1:as:weakcvg} and \ref{ch1:as:weakcvg_intermediate}-\ref{ch1:as:bounded:fixedn}, 
then 
$$\inf_{\theta\in\mathcal{H}} \mathbb{E}_m \was_p(\hat{\mu}_n,\hat{\mu}_{\theta,m}) \to \inf_{\theta\in\mathcal{H}} \was_p(\hat{\mu}_n,\mu_\theta),$$
and there exists an $\hat{m}$ such that, for all $m\geq \hat{m}$, the sets $\argmin_{\theta\in\mathcal{H}} \mathbb{E}_m\was_p(\hat{\mu}_n,\hat{\mu}_{\theta,m})$ are non-empty and form a bounded sequence with 
$$\limsup_{m\to\infty} \argmin_{\theta\in\mathcal{H}} \mathbb{E}_m \was_p(\hat{\mu}_n,\hat{\mu}_{\theta,m}) \subset \argmin_{\theta\in\mathcal{H}} \was_p(\hat{\mu}_{n},\mu_\theta).$$
\end{theorem}

\section{Computational aspects} \label{ch1:sec:computation}

\subsection{Computing the Wasserstein distance} \label{ch1:sec:distancecalculations}

We recall some strategies to calculate or approximate the Wasserstein distance
between empirical distributions. In the case where $\mathcal{Y} \subset
\mathbb{R}$, the exact computation is cheap, as the main computational task reduces to sorting the samples. However, in
dimensions $d>1$, the cost is in general expensive, which
has motivated a rich literature on fast approximations \citep{peyre2018computational}.
We will write $\was_p(y_{1:n},z_{1:m})$ for $\was_p(\hat\mu_n,\hat\nu_m)$,
where $\hat\mu_n$ and $\hat\nu_m$ stand for the empirical distributions
$n^{-1}\sum_{i=1}^n \delta_{y_i}$ and $m^{-1}\sum_{i=1}^m \delta_{z_i}$.
The Wasserstein distance then takes the form 
\begin{equation} \label{ch1:eq:wass_def_discrete}
    \was_p(y_{1:n}, z_{1:m})^p = \inf_{\gamma\in \Gamma_{n,m}} \sum_{i=1}^n \sum_{j=1}^m \rho(y_i, z_j)^p \gamma_{ij}
\end{equation}
where $\Gamma_{n,m}$ is the set of $n\times m$ matrices with non-negative
entries, columns and rows resp. summing to $m^{-1}$ and $n^{-1}$.

\subsubsection{Exact computation\label{ch1:sec:wassersteincalculation}}

The formulation in \eqref{ch1:eq:wass_def_discrete} is a linear program, and can be solved with generic linear program solvers. However, specialized approaches can be more efficient. 
In the univariate case with $\rho(x,y) = |x-y|$, the optimal transport coupling can be found by sorting the vectors $y_{1:n}$ and $z_{1:m}$
to get the collections of order statistics $\{y_{(i)}\}_{i=1}^n$ and $\{z_{(j)}\}_{j=1}^m$. Suppose that
$m = \ell n$ for some $\ell \geq 1$. Then, the $p$-Wasserstein distance in 
 \eqref{ch1:eq:wass_def_discrete} can be expressed as
\begin{equation}\label{ch1:eq:wass_sorted}
\was_p^p(y_{1:n},z_{1:m}) = \frac{1}{m}\sum_{i=1}^n \sum_{j=1}^\ell |y_{(i)} - z_{(\ell(i-1)+j)}|^p,
\end{equation}
which can be seen from the representation $\was^p_p(\hat{\mu}_n,\hat{\nu}_m) =
\int_0^1 \lvert F_{\mu,n}^{-1}(s) - F_{\nu,m}^{-1}(s)\rvert^p ds$ \citep[see
e.g.][Theorem 6.0.2]{ambrosio2005}. The cost of the Wasserstein distance
computation is thus of order $m\log m$ in the univariate setting. Note that, in some cases, the generation
of $m$ sorted observations can be done directly for a cost of order $m$, for instance by 
generating already-sorted uniforms and applying a quantile function \citep{devroye:1985}.
It should also be noted that the expression $\was^p_p(\mu,\nu) = \int_0^1 \lvert
F_{\mu}^{-1}(s) - F_{\nu}^{-1}(s)\rvert^p ds,$  in combination with a numerical
integrator, could be used whenever the quantile functions of $\mu$ and $\nu$ are
known (as in the g-and-k example of Section \ref{ch1:sec:gandk}). In that case one can directly target
the MWE with a numerical optimizer, as an alternative to computing the MEWE. The same is true if the CDFs are available, 
using the expression $\was_1(\mu,\nu) = \int_\mathbb{R}|F_\mu(t) - F_\nu(t)|dt$
given in Section \ref{ch1:subsec:asymptoticdistribution}.

In multivariate settings, one can solve the problem in
\eqref{ch1:eq:wass_def_discrete} using dual ascent methods \citep[see
e.g.][]{bertsimas1997introduction}.  This includes the Hungarian algorithm,
applicable in the setting where $m=n$, at a cost of order $n^3$. Other
algorithms have a cost of order $n^{2.5}\log(n\, C_{n})$, with
$C_{n}=\max_{1\leq i, j\leq n}\rho(y_i,z_j)$, and can therefore be more
efficient when $C_{n}$ is small \citep[Section 4.1.3]{assignmentproblems}. A
practical alternative is the short-list method, derived from the network
simplex algorithm, presented by \citet{gottschlich2014shortlist} and
implemented in the \texttt{transport} R package \citep{transportpackage}. In
general, simplex algorithms come without guarantees of polynomial running
times, but \citet{gottschlich2014shortlist} show empirically that their method
tends to have sub-cubic cost. When the cost of computing the
Wasserstein distance exactly gets prohibitively large, we can resort to various
approximations.

\subsubsection{Approximations}
In parallel with its increasing popularity as an inferential tool in statistics
and machine learning, there has been fast growth in the number of algorithms
that approximate the Wasserstein distance at reduced computational costs. The
book of \citet{peyre2018computational} provides an overview of many such methods. 
In particular, they provide a thorough discussion of the method introduced by \citet{cuturi2013sinkhorn}, which
regularizes the optimization 
problem in  \eqref{ch1:eq:wass_def_discrete}
using an entropic constraint.
Specifically, the regularized version of  \eqref{ch1:eq:wass_def_discrete} reads:
\begin{equation}
\gamma^\zeta = \argmin_{\gamma \in \Gamma_{n,m}} \sum_{i=1}^n \sum_{j=1}^m \rho(y_i, z_j)^p \gamma_{ij} 
    + \zeta \sum_{i=1}^n \sum_{j=1}^m \gamma_{ij} \log \gamma_{ij},
\end{equation}
which includes a penalty on the entropy of $\gamma$.
The regularized problem can be solved iteratively by Sinkhorn's algorithm \citep{cuturi2013sinkhorn} or iterative Bregman projections \citep{benamou2014iterative} for a total cost of order $nm$. 
Define the dual-Sinkhorn divergence 
$S^\zeta_p(y_{1:n},z_{1:m})^p =  \sum_{i=1}^n \sum_{j=1}^m \rho(y_i, z_j)^p \gamma^\zeta_{ij}$.
If $\zeta$ goes to zero, the dual-Sinkhorn divergence goes to the Wasserstein distance. If $\zeta$ goes to infinity, it converges to the energy distance \citep{ramdas2017wasserstein}. Other fast approximations of the Wasserstein distance include \citet{gibbs_ot2016,altschuler2017near,altschuler2018massively,li2018parallel}.

In the case where $n=m$, computing the Wasserstein distance can be viewed as an assignment problem, which leads to other specialized approaches. 
For instance, \citet{puccetti2017algorithm} proposes a greedy algorithm based on swaps in the assignment, for a cost of $n^2$ per iteration.
When a cost of order $nm$ or $n^2$ is too large, \citet{bernton2019approximate} propose
a new distance generalizing the idea of sorting when $d>1$. It consists in sorting samples according to their
projection via the Hilbert space-filling curve and computing a distance analogous to the one in \eqref{ch1:eq:wass_sorted}, for a computational cost of the order of $m\log m$. A similar idea underlies the sliced Wasserstein distance \citep{rabin2011wasserstein,bonneel2015sliced}, which can be estimated by projecting the data onto $L$ random lines, and by averaging the Wasserstein distances 
computed in the associated one-dimensional spaces, for a total cost on the order of $L m\log m$.

\subsection{Computing the estimators}\label{ch1:sec:optimization}

The exact computation of the MWE and MEWE is in general intractable. This is
also true when $\was_p$ is substituted for any of its approximations mentioned
above.  However, we can envision various schemes to numerically approximate the estimators. 

The calculation of the MEWE can be based on Monte Carlo approximation of the function $\theta \mapsto \mathbb{E}_m\was_p(\hat{\mu}_n, \hat{\mu}_{\theta,m})$ using synthetic samples 
generated given $\theta$.
Assume that a data set $z_{1:m}$ can be sampled from $\mu_{\theta}^{(m)}$ by
setting $z_{1:m} = g_m(u,\theta)$, where $g_m$ is a deterministic function of
the parameter $\theta$ and $u$ a random variable independent of $\theta$. Then,
the  mean $k^{-1} \sum_{i=1}^k \was_p(y_{1:n}, g_m(u^{(i)},\theta))$, where the $u^{(i)}$
are i.i.d., is a natural estimate of $\mathbb{E}_m\was_p(\hat{\mu}_n, \hat{\mu}_{\theta,m})$. By the law of large numbers, we know that
$k^{-1} \sum_{i=1}^k \was_p(y_{1:n}, g_m(u^{(i)},\theta)) \to
\mathbb{E}_m\was_p(\hat{\mu}_n, \hat{\mu}_{\theta,m})$ almost surely as $k\to\infty$. Since this estimator is an average of i.i.d.~random variables, the central limit theorem indicates that the rate of convergence is $\sqrt{k}$.
Moreover, this approximation is a deterministic function of $\theta$, which can be optimized with standard methods.
In turn, this optimization step can be placed within a
Monte Carlo Expectation-Maximization (MCEM) algorithm \citep{wei1990monte},
which would alternate between optimization of $\theta$ and resampling of $u^{(i)}$.
Convergence results for such algorithms, as both the number of iterations and $k$ go to infinity, are reviewed in \citet{neath2013convergence}. 

In practice, we are naturally constrained to finite values of $m$ and $k$. The
incremental cost of increasing $k$ is typically lower than that of increasing
$m$, due in part to the potential for parallelization when calculating the
distances $\was_p(y_{1:n}, g_m(\theta,u^{(i)}))$  for a given $\theta$, and in
part to the algorithmic complexity in $m$, which is super-linear as described 
in the previous section. In the numerical experiments of Section \ref{ch1:sec:illustrations}, we found that $m = 10^4$ and $k=20$ within a single iteration of MCEM
yielded accurate estimators. That is, we draw $u^{(i)}$ for $i=1,\ldots,k$ once and for all, and optimize over $\theta$. We illustrate the effect of choosing different $m$ and $k$ in Section \ref{ch1:sec:mwegamma}.

Several alternatives to the MCEM approach exist. An approach to computing the MEWE was proposed in
\citet{genevay2018learning} based on the Sinkhorn divergence approximation to the
Wasserstein distance. They derive gradients of $S^\zeta_p(y_{1:n},
g_m(u,\theta))$ with respect to $\theta$ while $u$ is fixed, allowing for the
application of stochastic gradient descent. In practice, the gradients can be 
computed with auto-differentiation. A method for computing the MWE was proposed
by \citet{chen2018natural}, in which they pull back the $2$-Wasserstein metric
tensor in $\mathcal{P}_2(\mathcal{Y})$ to $\mathcal{H}$, under which
$\mathcal{H}$ becomes a Riemannian manifold. In turn, this structure allows
them to derive a novel gradient descent algorithm. Alternatively, in the spirit
of Monte Carlo optimization, one can modify  the sampling algorithms used for
the approximate Bayesian computation (ABC) approach described by
\citet{bernton2019approximate} to approximate the MEWE. This has the benefit of
not requiring the synthetic data to be generated via a deterministic function
$g_m$ with fixed-dimensional arguments.  Related discussions can be found in
\citet{wood2010statistical,rubio2013simple}.

\section{Illustrations}\label{ch1:sec:illustrations}

In Sections \ref{ch1:sec:gandk} and \ref{ch1:sec:sumlognormal}, we compute the MEWE in
two well-specified models with intractable likelihoods that produce
i.i.d.~data, taken from the ABC literature.  We empirically estimate the
coverage of bootstrap confidence intervals for the data-generating parameter.
In Section \ref{ch1:sec:gandk}, we also compute the MEWE in a setting where the
data-generating process produces a time series.  In Section \ref{ch1:sec:mwegamma},
we compare the distribution of the MEWE with that of the maximum likelihood
estimator (MLE) in a simple misspecified setting.  We also investigate the
effect of $k$ and $m$ on the distribution of the approximate MEWE.  In Section
\ref{ch1:sec:mwecauchy}, we highlight the robustness of this choice by considering
a heavy-tailed data-generating process for which the MLE is not consistent.
Throughout the numerical experiments, we have chosen $p=1$, as this imposes minimal assumptions on the existence of moments of both the data-generating process and the model.

\subsection{Quantile ``g-and-$\kappa$'' distribution \label{ch1:sec:gandk}}
\subsubsection{Independent data \label{ch1:sec:gandk_independent}}

The g-and-$\kappa$ distribution \citep{tukey1977modern,jorge1984some} is defined in terms of its quantile function:
\begin{equation}
r\in (0,1) \mapsto a + b \left(1 + 0.8\frac{1-\exp(-gz(r)}{1+\exp(-gz(r)}\right) \left(1+z(r)^2\right)^\kappa z(r),
    \label{ch1:eq:gandk}
\end{equation}
where $z(r)$ refers to the $r$-th quantile of the standard Normal distribution.
The model is indexed by the parameter $\theta = (a,b,g,\kappa) \in [0,10]^4$,
and we take $\mu_\star = \mu_{\theta_\star}$ with $\theta_\star = (3,1,2,0.5)$.
The probability density function, and therefore the likelihood of the model, is
analytically intractable; thus the model has become a standard benchmark for ABC methods \citep{sisson2018handbook}. 
Though, the likelihood can be estimated 
by numerically inverting and then differentiating the quantile function, as
described in \citet{rayner2002numerical,bernton2019approximate}.

\begin{figure}[hp]
   \centering
        \begin{subfigure}[b]{0.42\textwidth}
            \centering
            \includegraphics[width=\textwidth]{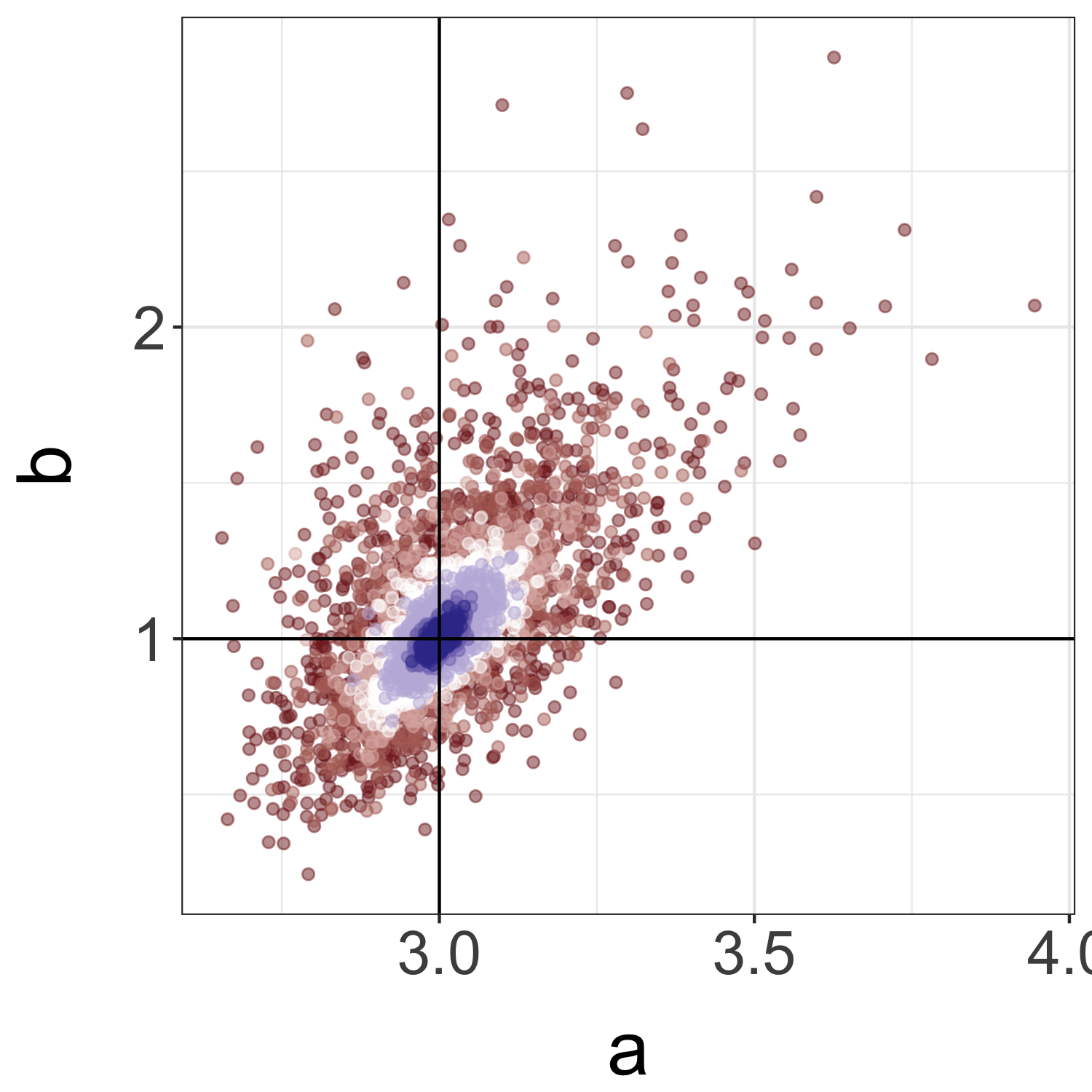}
            \caption{{\small MEWE: $a$ vs $b$.}}   
            \label{ch1:fig:gandk_A_vs_B}
        \end{subfigure}
        \hskip 0.8cm
        \begin{subfigure}[b]{0.42\textwidth}  
            \centering 
            \includegraphics[width=\textwidth]{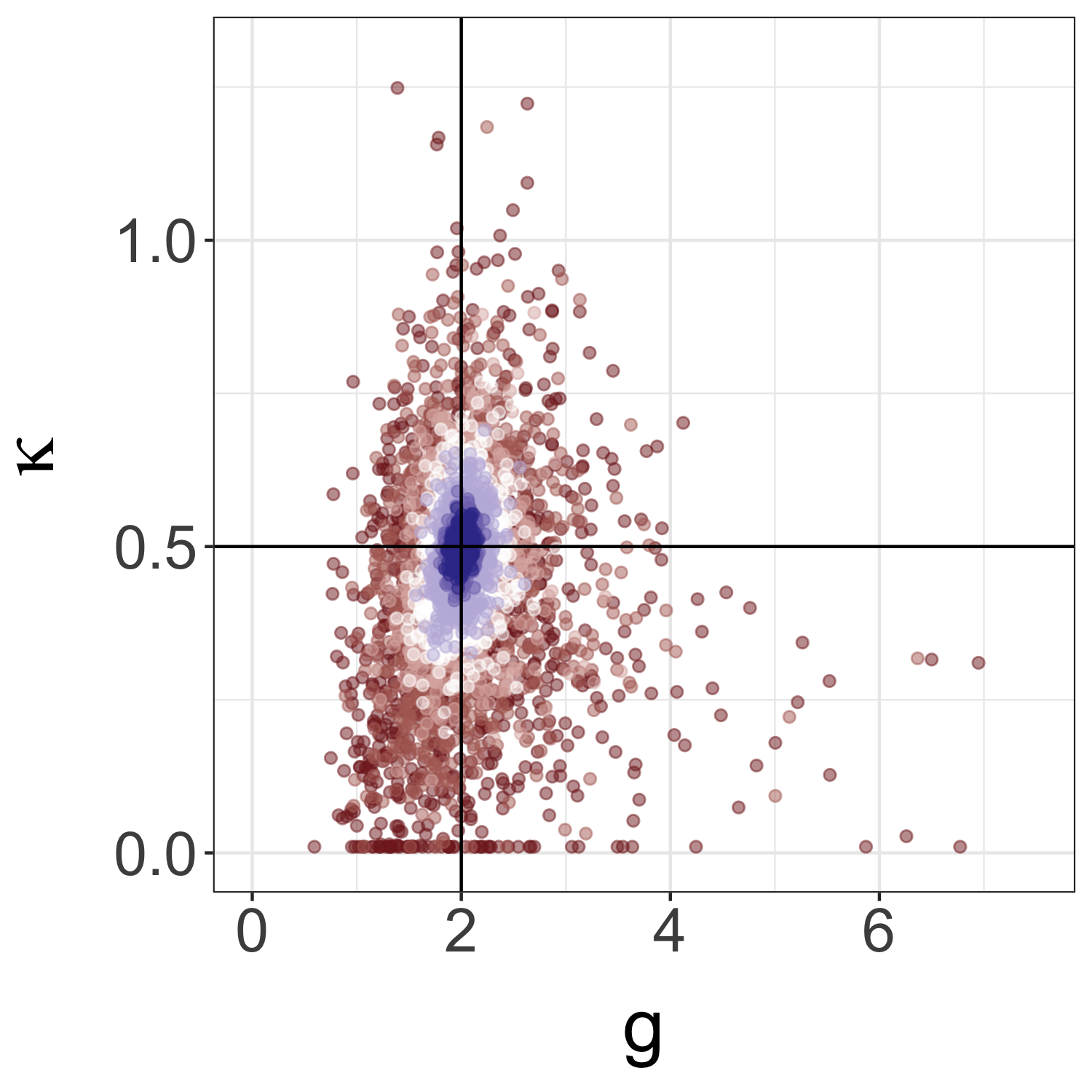}
            \caption{{\small MEWE: $g$ vs $\kappa$.}}      
            \label{ch1:fig:gandk_g_vs_k}
        \end{subfigure}
        \vskip0.5cm
        
        \begin{subfigure}[b]{0.42\textwidth}   
            \centering 
            \includegraphics[width=\textwidth]{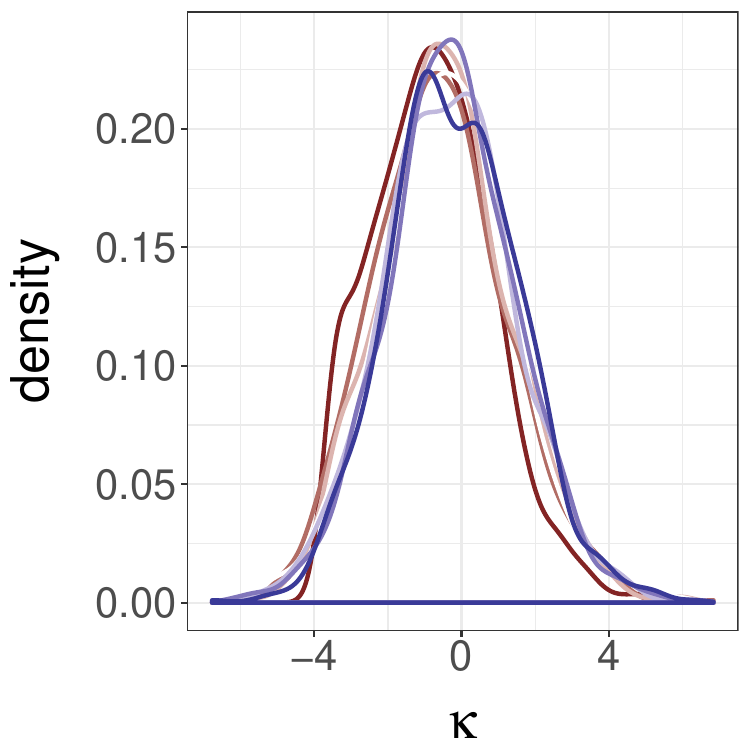}
            \caption{{\small $\sqrt{n}$-scaled estim. of $\kappa$.}}    
            \label{ch1:fig:gandk_rescaled_k}
        \end{subfigure}   
                \hskip 0.8cm
        \begin{subfigure}[b]{0.42\textwidth}   
            \centering 
            \includegraphics[width=\textwidth]{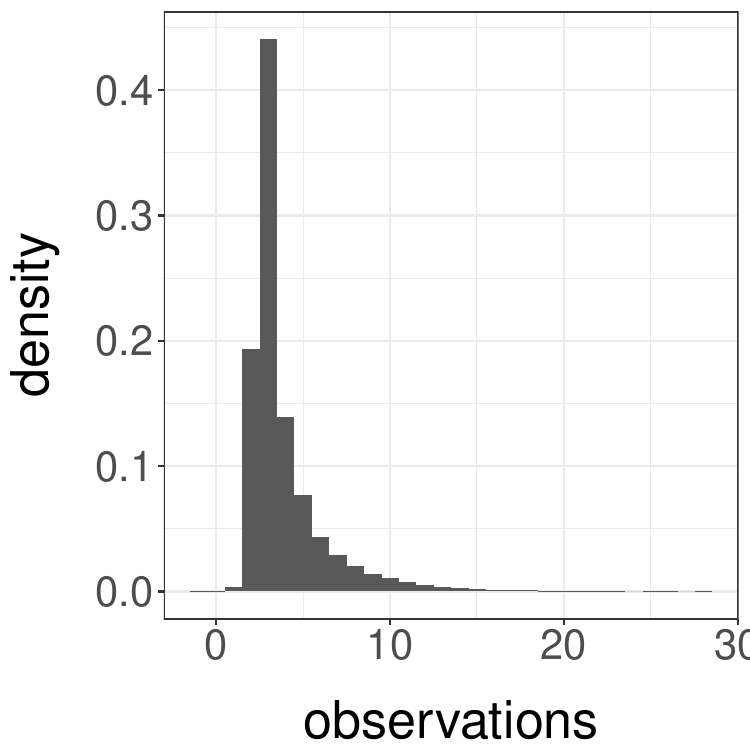}
            \caption{{\small Histogram of data.}}    
            \label{ch1:fig:gandk_hist}
        \end{subfigure}
        \caption{\small Estimators in the well-specified g-and-$\kappa$ model, as described in Section \ref{ch1:sec:gandk_independent}.
        Figures \ref{ch1:fig:gandk_A_vs_B} and
        \ref{ch1:fig:gandk_g_vs_k} show the MEWE's bivariate marginal sampling distributions for $(a,b)$ and $(g,\kappa)$ respectively, as $n$ ranges from $50$ to $10^4$
        (colors from red to white to blue as $n$ increases). For each $n$, we plot $M=1,000$ estimators based on independent data sets. Each estimator was computed with $p=1$, $m = 10^4$, $k=20$, and one iteration of MCEM. Note that for small data sizes ($n = 50$ and $n=100$), the estimator occasionally appears to be on the boundary of the parameter space, which could mean that the optimization procedure failed to converge. The intersections of the black lines indicate data-generating parameters. Figure \ref{ch1:fig:gandk_rescaled_k} shows the MEWE's marginal distribution for $\kappa$ for the different levels of $n$, centered and rescaled by $\sqrt{n}$, illustrating the rate of convergence anticipated by Theorem \ref{ch1:theorem:asymptoticdistribution}. Figure \ref{ch1:fig:gandk_hist} is a histogram of a data set generated with $\theta_\star = (3,1,2,0.5)$ and $n=1,000$.
                \label{ch1:fig:gandk_independent}}
\end{figure}

Sampling i.i.d.~variables from the g-and-$\kappa$ distribution 
can be achieved straightforwardly by plugging independent standard Normals into \eqref{ch1:eq:gandk}
in place of $z(r)$. Therefore, the MEWE with large $m$ can be computed to high
precision. In Figure \ref{ch1:fig:gandk_independent}, we show the behavior of the MEWE with
$p=1$ and $m=10^4$ for different numbers of observed data, and illustrate its
concentration around the data-generating parameter $\theta_\star$. In computing
the MEWE, we used $k=20$ and only one iteration of MCEM. That is, we
approximate the MEWE by sampling $k = 20$ independent $u^{(i)}$ random
variables and minimize $\theta \mapsto k^{-1} \sum_{i=1}^k \was_p(y_{1:n},
g_m(u^{(i)},\theta))$ to form the estimator,
using the \texttt{optim} function in \texttt{R} \citep{Rsoftware}. 

We check the coverage of bootstrap confidence intervals calculated for
$\theta_\star = (3,1,2,0.5)$. We use
the percentile bootstrap \citep{efron1994introduction} for data sets of size
$n=1,000$ and synthetic data sets of size $m=10^4$, and calculate the MEWE with
$k=20$. We draw $400$ data sets from the data-generating process, and $1,000$
bootstrap data sets for each of these. The observed coverage rates of the
resulting $0.95$ confidence intervals were $0.928$ for $a$, $0.945$
for $b$, $0.960$ for $g$, and $0.938$ for $\kappa$. The coverage rates should
approach $0.95$ as $n\to \infty$, $m\to \infty$, and $k\to \infty$ within the
MCEM algorithm.  After a Bonferroni correction, the observed coverage of the confidence sets for
$\theta_\star$ was $0.935$. 

    As mentioned in Section \ref{ch1:sec:wassersteincalculation},
   since the g-and-$\kappa$ distribution has an explicit quantile function 
(insofar as the Normal quantile function can be considered explicit), 
one could instead directly estimate the Wasserstein distance between the g-and-k distribution and
some empirical distribution using a representation of the distance in terms of 
an integral of the difference of quantile functions, combined with a numerical integrator.

\subsubsection{Dependent data \label{ch1:sec:gandk_dependent}}
To illustrate the behavior of the estimator when the data-generating process produces dependent data, we
also generated g-and-$\kappa$ variables using Normals from an AR(1) process. Specifically, we let 
$x_0 \sim \mathcal{N}(0,1)$ and $x_t = \rho x_{t-1} + \eta_t$ for $t \geq 1$, where $\eta_t \sim \mathcal{N}(0,1-\rho^2)$ independently, and $\rho = 0.75$. Hence, these 
variables are marginally distributed as $\mathcal{N}(0,1)$, but are positively correlated. To produce the observation $y_t$ for each $t$, we plugged $x_t$ into \eqref{ch1:eq:gandk} in place of $z(r)$, using the same $\theta_\star$ as in the independent setting. The marginal distribution of the data are therefore the same as before, but
the sequence of observations now forms a stationary and ergodic time series. This setting is covered by the theoretical results of Section \ref{ch1:sec:theory}; Assumption \ref{ch1:as:cvgwas} holds with $\mu_\star = \mu_{\theta_\star}$. The model, as before, is taken to generate i.i.d.~data. 

\begin{figure}[hp]
   \centering
        \begin{subfigure}[b]{0.42\textwidth}
            \centering
            \includegraphics[width=\textwidth]{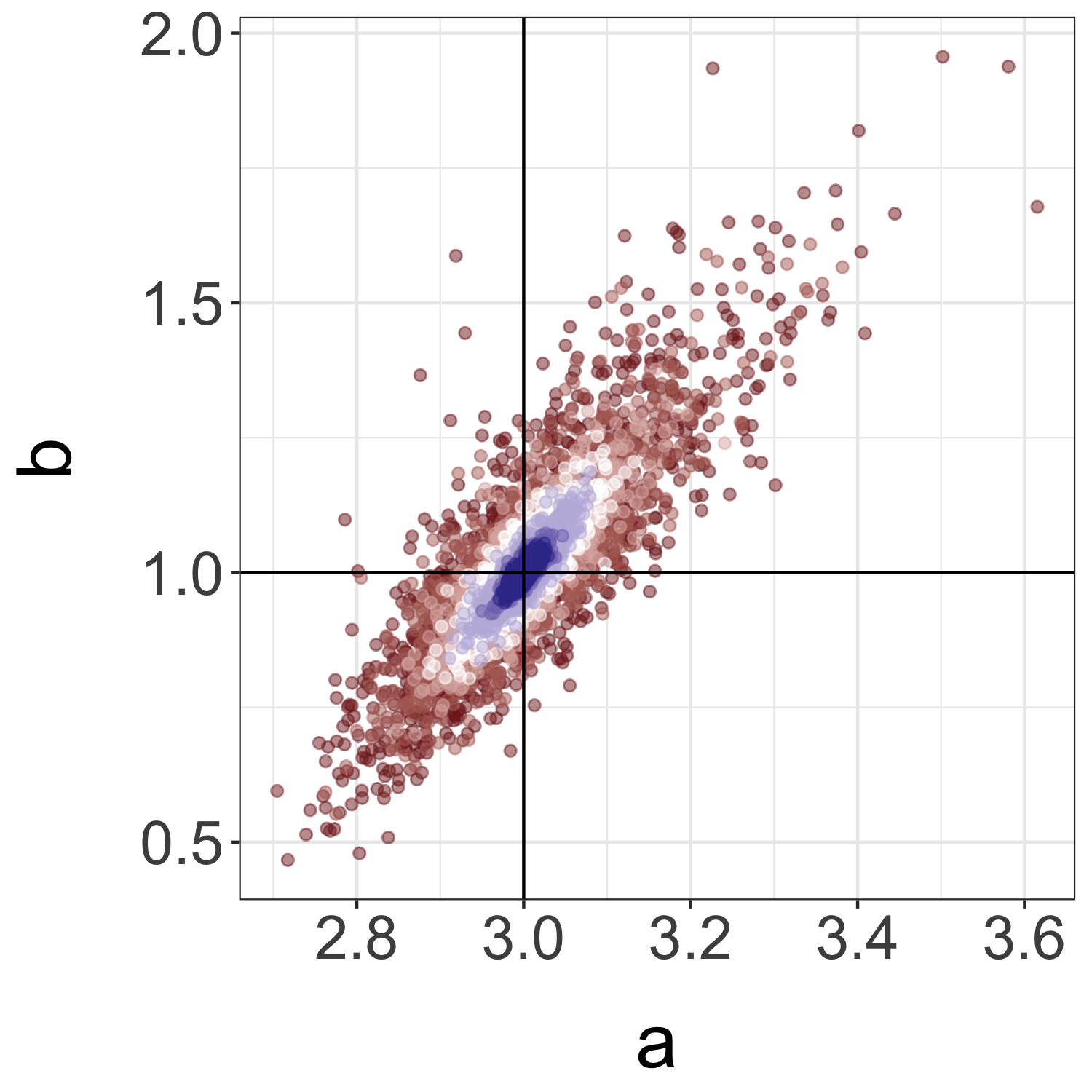}
            \caption{{\small MEWE: $a$ vs $b$.}}   
            \label{ch1:fig:gandk_correlated_A_vs_B}
        \end{subfigure}
	\hskip 0.8cm
        \begin{subfigure}[b]{0.42\textwidth}  
            \centering 
            \includegraphics[width=\textwidth]{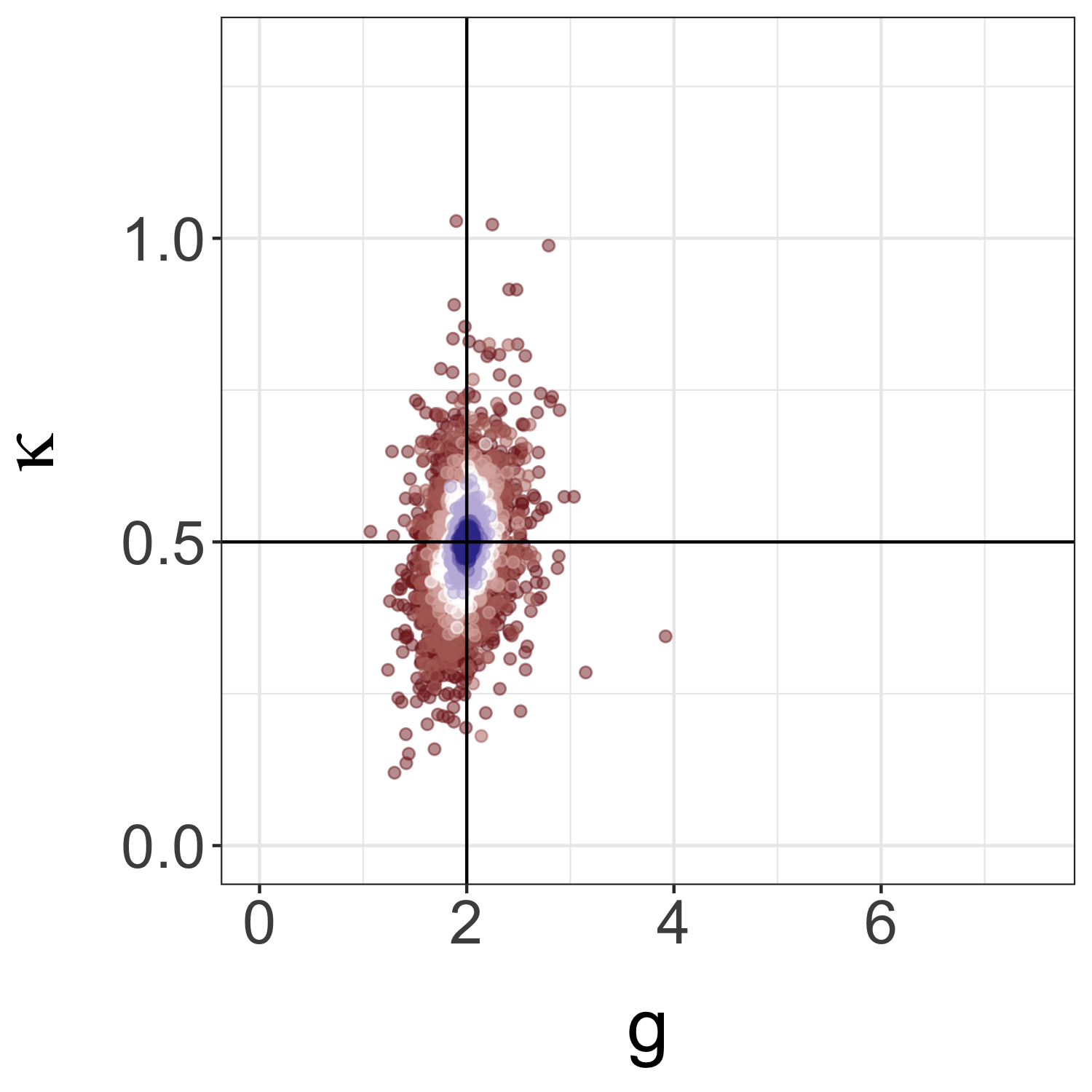}
            \caption{{\small MEWE: $g$ vs $\kappa$.}}      
            \label{ch1:fig:gandk_correlated_g_vs_k}
        \end{subfigure}
        \vskip0.5cm
        
        \begin{subfigure}[b]{0.42\textwidth}   
            \centering 
            \includegraphics[width=\textwidth]{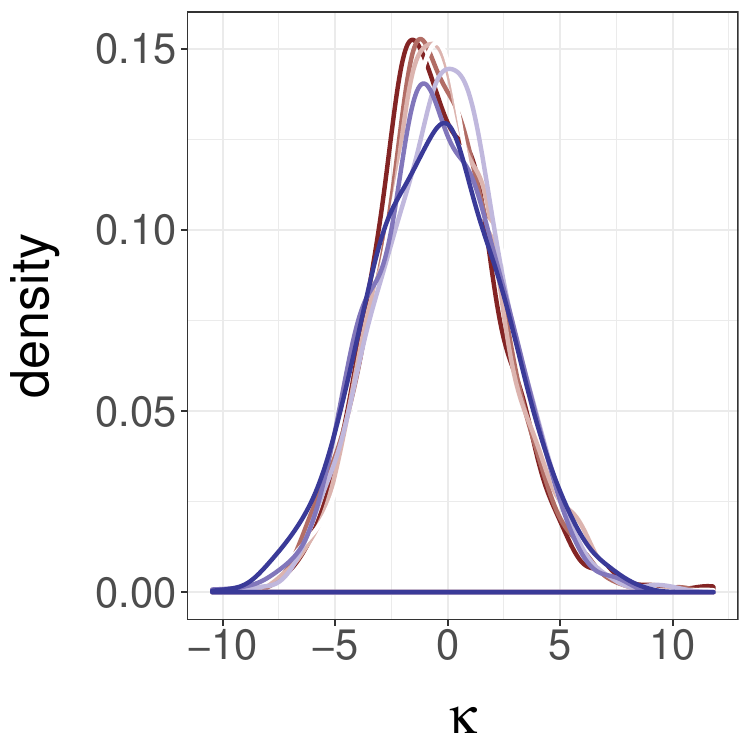}
            \caption{{\small $\sqrt{n}$-scaled estim. of $\kappa$.}}    
            \label{ch1:fig:gandk_correlated_rescaled_k}
        \end{subfigure}       
        \hskip 0.8cm
        \begin{subfigure}[b]{0.42\textwidth}   
            \centering 
            \includegraphics[width=\textwidth]{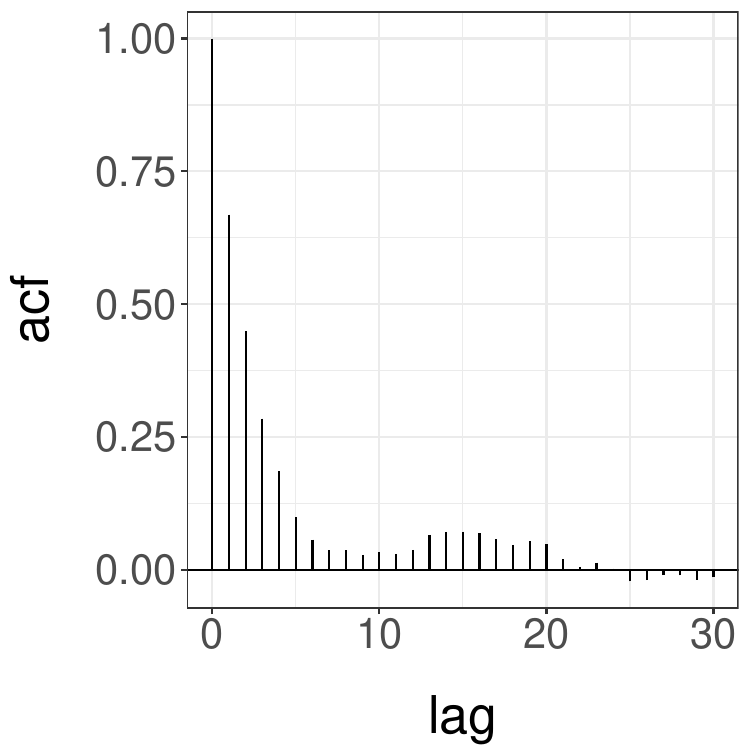}
            \caption{{\small ACF of data.}}    
            \label{ch1:fig:gandk_correlated_acf}
        \end{subfigure}
        \caption{\small Estimators in the g-and-$\kappa$ model with dependent data, as described in Section \ref{ch1:sec:gandk_dependent}.
        Figures \ref{ch1:fig:gandk_correlated_A_vs_B} and
        \ref{ch1:fig:gandk_correlated_g_vs_k} show the MEWE's bivariate marginal sampling distributions for $(a,b)$ and $(g,\kappa)$ respectively, as $n$ ranges from $500$ to $10^5$
        (colors from red to white to blue as $n$ increases). Note that the sample sizes here are $10$ times larger than in the plots for the i.i.d.~setting. For each $n$, we plot $M=1,000$ estimators based on independent data sets. Each estimator was computed with $p=1$, $m = 10^4$, $k=20$, and one iteration of MCEM. The intersections of the black lines indicate data-generating parameters. Figure \ref{ch1:fig:gandk_correlated_rescaled_k} shows the MEWE's marginal distribution for $\kappa$ for the different levels of $n$, centered and rescaled by $\sqrt{n}$, illustrating the rate of convergence anticipated by Theorem \ref{ch1:theorem:asymptoticdistribution}, but that the asymptotic variance is larger than in the i.i.d.~case. Figure \ref{ch1:fig:gandk_correlated_acf} shows the autocorrelation function of a data set generated with $\theta_\star = (3,1,2,0.5)$, $\rho = 0.75$,  and $n=1,000$.
                \label{ch1:fig:gandk_correlated}}
\end{figure}

To approximate the MEWE, we used the same computational approach as in the i.i.d.~setting, with $p=1$, $m = 10^4$, and $k=20$. In Figure \ref{ch1:fig:gandk_correlated}, we show that the MEWE appears to concentrate around $\theta_\star$ at the same rate as in the i.i.d.~setting, but that its asymptotic distribution has higher variance. Note that in Figure \ref{ch1:fig:gandk_correlated}, the data sizes are $10$ times larger than in the plots for the i.i.d.~setting (Figure \ref{ch1:fig:gandk_independent}), as the correlation between the samples effectively reduces the sample size and makes the estimators poorly behaved when $n$ is small.

\subsection{Sum of log-Normal random variables \label{ch1:sec:sumlognormal}}

The distribution of the sum of log-Normal random variables appears in various
settings \citep{Fenton1960,rodrigues2018recalibration}, but no analytical
formula is available for its probability density function, and thus the associated
likelihood function is intractable.  For a given positive integer $L$,
$\gamma\in\mathbb{R}$ and $\sigma > 0$, the model generates an observation $y \in \mathbb{R}$ by
sampling $x_1,\ldots,x_L \sim \mathcal{N}(\gamma,\sigma^2)$ independently, and
defining $y = \sum_{\ell=1}^L \exp(x_\ell)$. Thus, sampling synthetic
observations from the model is simple.  We consider the task of estimating
$\theta = (\gamma,\sigma)$ from data, fixing $L$ to $10$, and
using the MEWE. We generate $n$ observations independently using $\theta_\star = (0,1)$.

\begin{figure}[hp]
        \centering
        \begin{subfigure}[b]{0.42\textwidth}  
            \centering 
            \includegraphics[width=\textwidth]{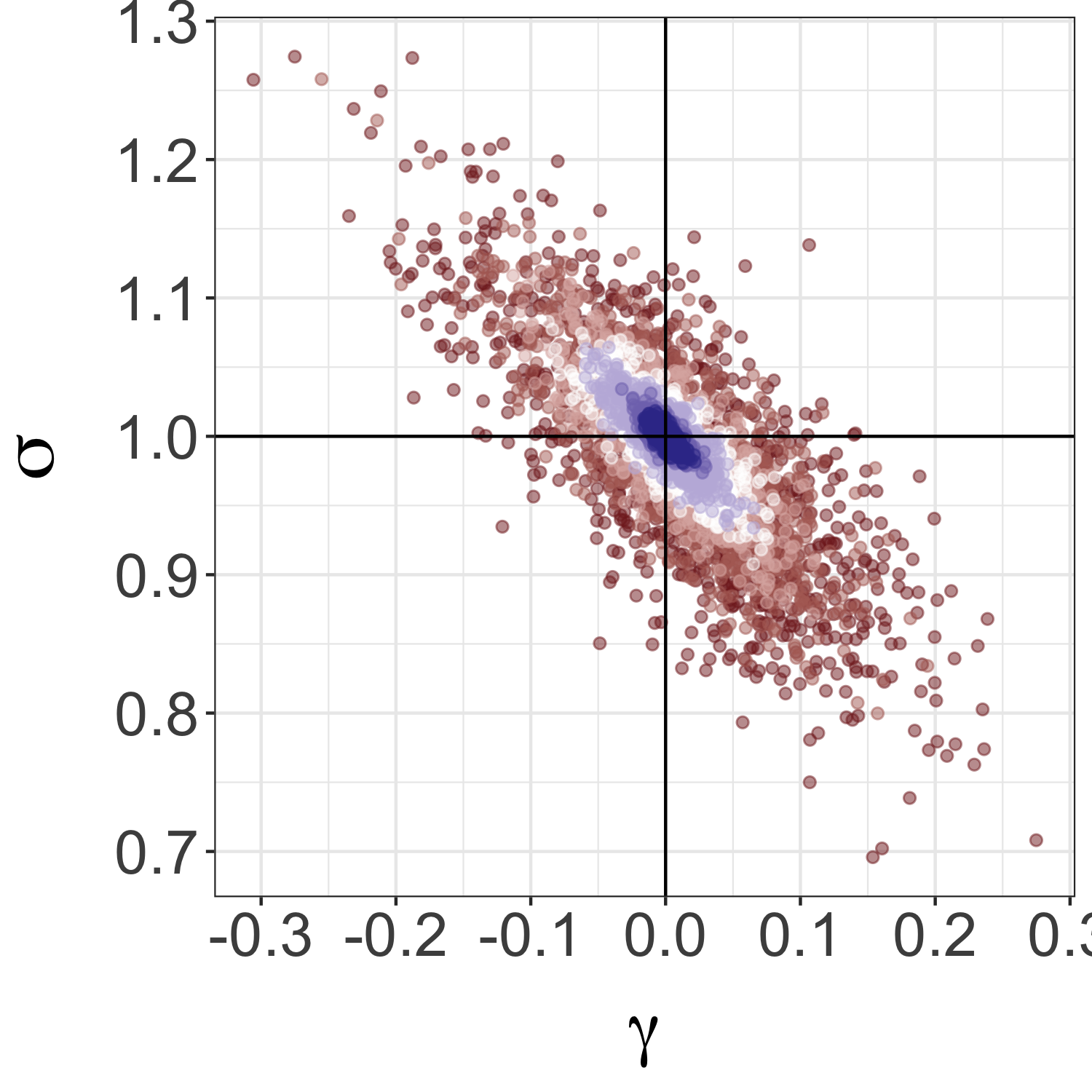}
            \caption{{\small MEWE of $(\gamma,\sigma)$.}}      
            \label{ch1:fig:lognormal_mu_vs_sigma}
        \end{subfigure}
        \hskip 0.8cm
        \begin{subfigure}[b]{0.42\textwidth}   
            \centering 
            \includegraphics[width=\textwidth]{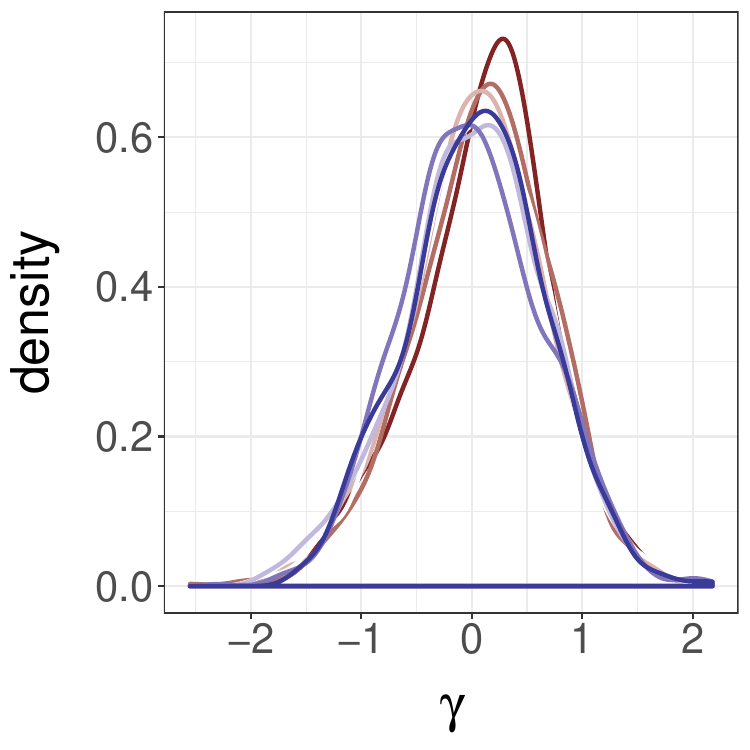}
            \caption{{\small $\sqrt{n}$-scaled estim. of $\gamma$.}}    
            \label{ch1:fig:lognormal_rescaled_mu}
        \end{subfigure}      
        \vskip0.5cm
        
        \begin{subfigure}[b]{0.42\textwidth}   
            \centering 
            \includegraphics[width=\textwidth]{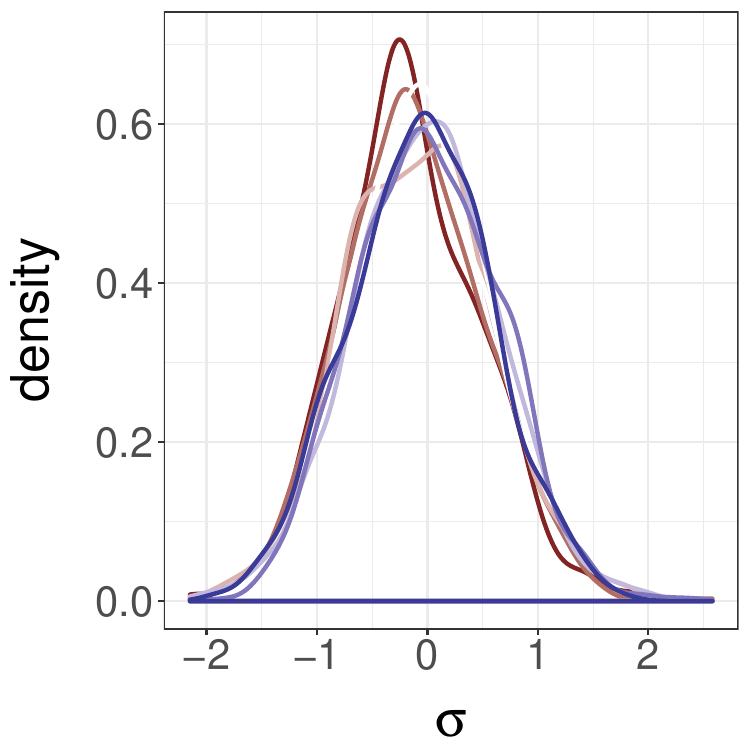}
            \caption{{\small $\sqrt{n}$-scaled estim. of $\sigma$.}}    
            \label{ch1:fig:lognormal_rescaled_sigma}
        \end{subfigure}
        \hskip 0.8cm
        \begin{subfigure}[b]{0.42\textwidth} 
             \centering 
            \includegraphics[width=\textwidth]{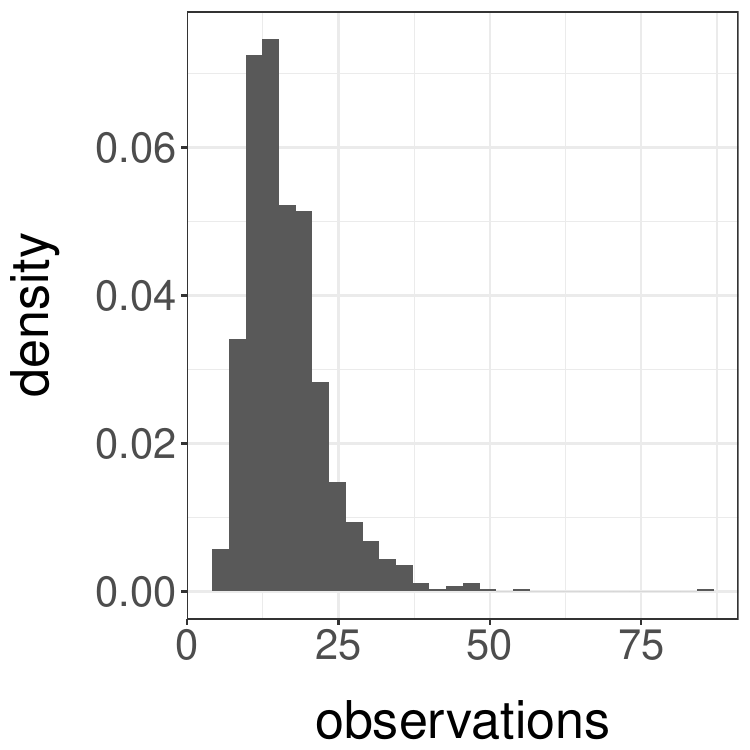}
            \caption{{\small Histogram of data.}}    
            \label{ch1:fig:lognormal_hist}
        \end{subfigure}
        \caption{\small Estimators in the well-specified sum of log-Normals model, as described in Section \ref{ch1:sec:sumlognormal}.
        Figure \ref{ch1:fig:lognormal_mu_vs_sigma} shows the sampling distributions of the MEWE, as $n$ ranges from $50$ to $10^4$
        (colors from red to white to blue as $n$ increases). For each $n$, we plot $M=1,000$ estimators based on independent data sets. 
        Each estimator was computed with $p=1$, $m = 10^4$, $k=20$, and one iteration of MCEM. The intersections of the black lines indicate data-generating parameters. 
        Figures \ref{ch1:fig:lognormal_rescaled_mu} and \ref{ch1:fig:lognormal_rescaled_sigma} show the MEWE's marginal distributions for the different levels of $n$, centered and rescaled by $\sqrt{n}$, illustrating the rate of convergence anticipated by Theorem \ref{ch1:theorem:asymptoticdistribution}. Figure \ref{ch1:fig:lognormal_hist} is a histogram of a data set generated with $\theta_\star = (0,1)$ and $n=1,000$.
        \label{ch1:fig:lognormal}}
\end{figure}

In Figure \ref{ch1:fig:lognormal}, we illustrate the behavior of the MEWE with $p=1$
and $m=10^4$ for different sizes of observed data $n$. The sampling
distribution of the MEWE appears to concentrate around the data-generating
parameter $\theta_\star$ at the $\sqrt{n}$ rate as $n$ increases. In computing
the MEWE, we used $k=20$ and one iteration of MCEM as in the previous section. 

We estimate the coverage of bootstrap confidence intervals calculated for
$\theta_\star = (0,1)$. As before, we use
the percentile bootstrap \citep{efron1994introduction} for data sets of size
$n=1,000$ and synthetic data sets of size $m=10^4$, and calculate the MEWE with
$k=20$. We draw $400$ data sets from the data-generating process, and $1,000$
bootstrap data sets for each. The observed coverage rates were $0.945$ and $0.940$ for $\gamma_\star$ and $\sigma_\star$ respectively, which are close to the limiting $0.95$ coverage rates.
After a Bonferroni correction, the observed coverage of the confidence sets for $\theta_\star$ was $0.960$. 

\subsection{Gamma data fitted with a Normal model}\label{ch1:sec:mwegamma}
\begin{figure}[hp]
        \centering
        \begin{subfigure}[b]{0.42\textwidth}
            \centering
            \includegraphics[width=\textwidth]{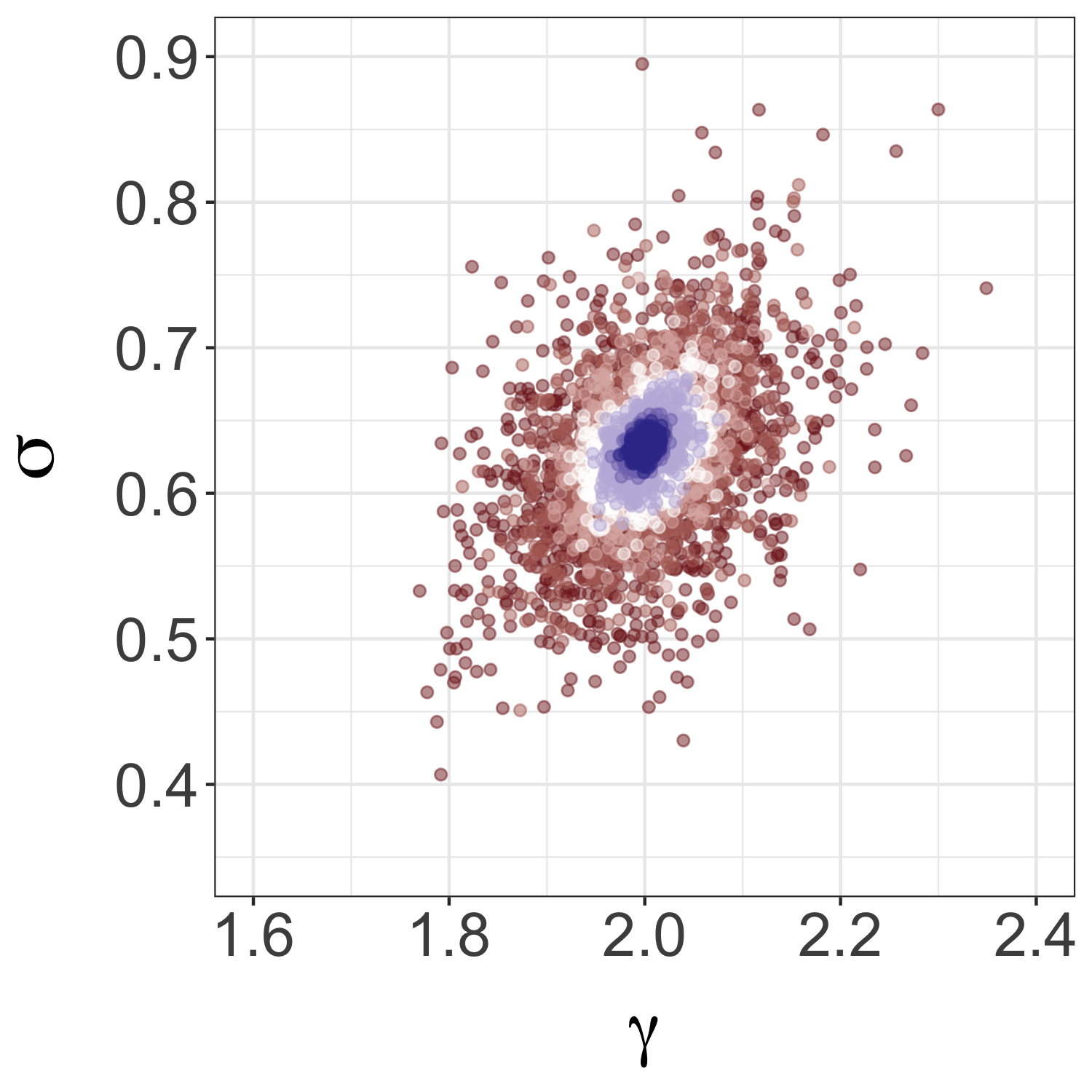}
            \caption{{\small MLE of $(\gamma,\sigma)$.}}    
            \label{ch1:fig:sampling_mle_gamma}
        \end{subfigure}
	\hskip 0.8cm
        \begin{subfigure}[b]{0.42\textwidth}  
            \centering 
            \includegraphics[width=\textwidth]{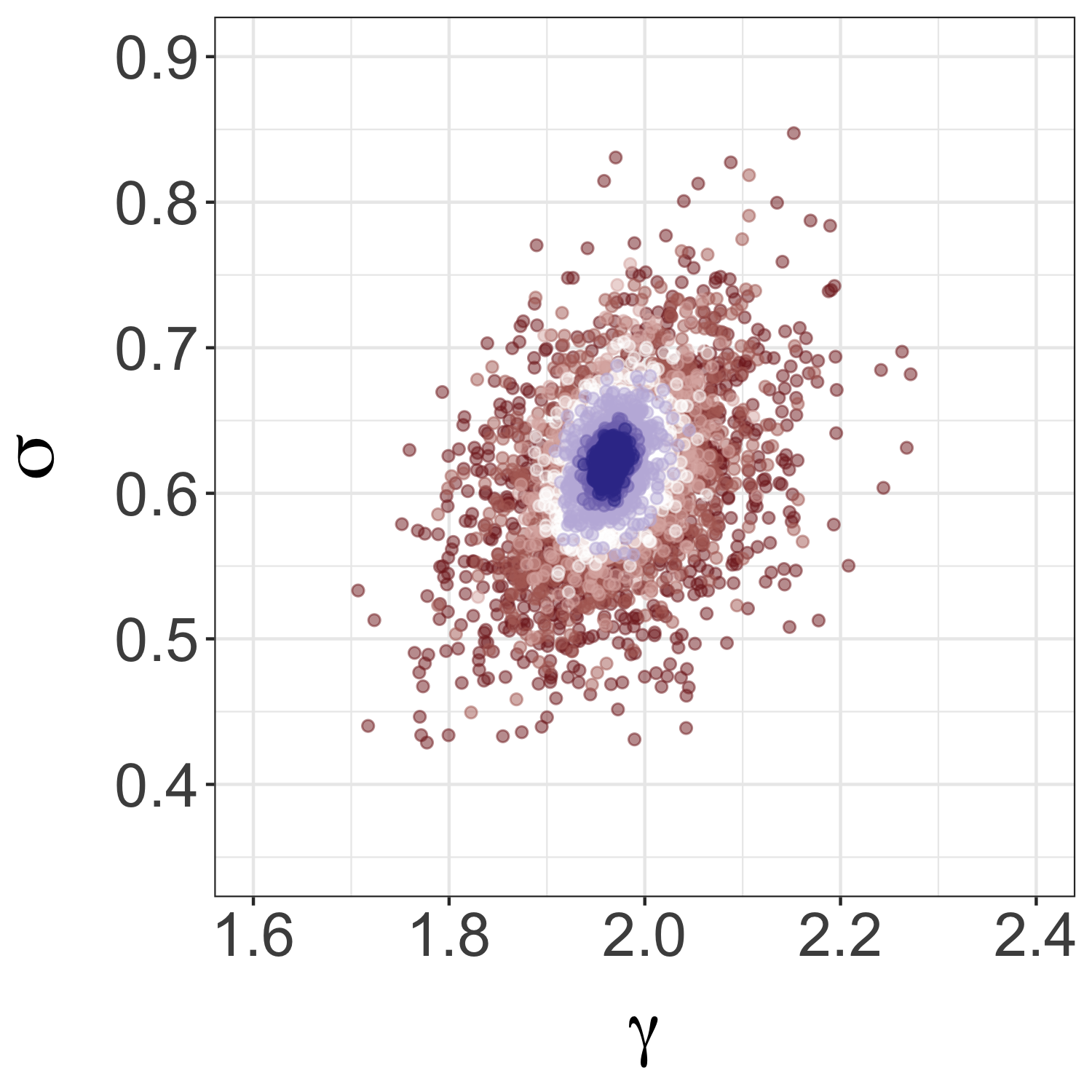}
            \caption{{\small MEWE of $(\gamma,\sigma)$.}}    
            \label{ch1:fig:sampling_mewe_gamma}
        \end{subfigure}
        \vskip0.5cm
        
        \begin{subfigure}[b]{0.42\textwidth}   
            \centering 
            \includegraphics[width=\textwidth]{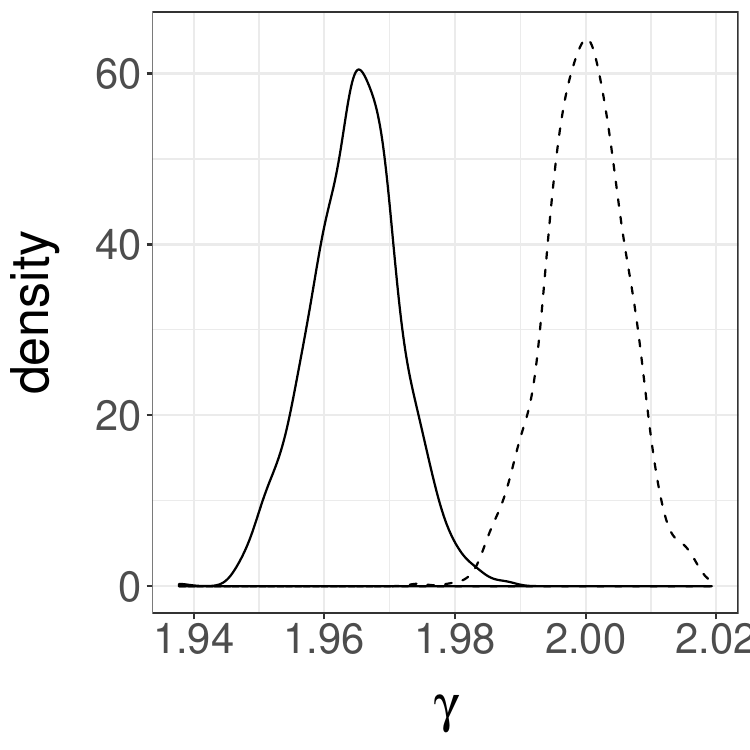}
            \caption{{\small Estimators of $\gamma$.}}    
            \label{ch1:fig:hist_mu_gamma}
        \end{subfigure}       
        \hskip 0.8cm
        \begin{subfigure}[b]{0.42\textwidth}   
            \centering 
            \includegraphics[width=\textwidth]{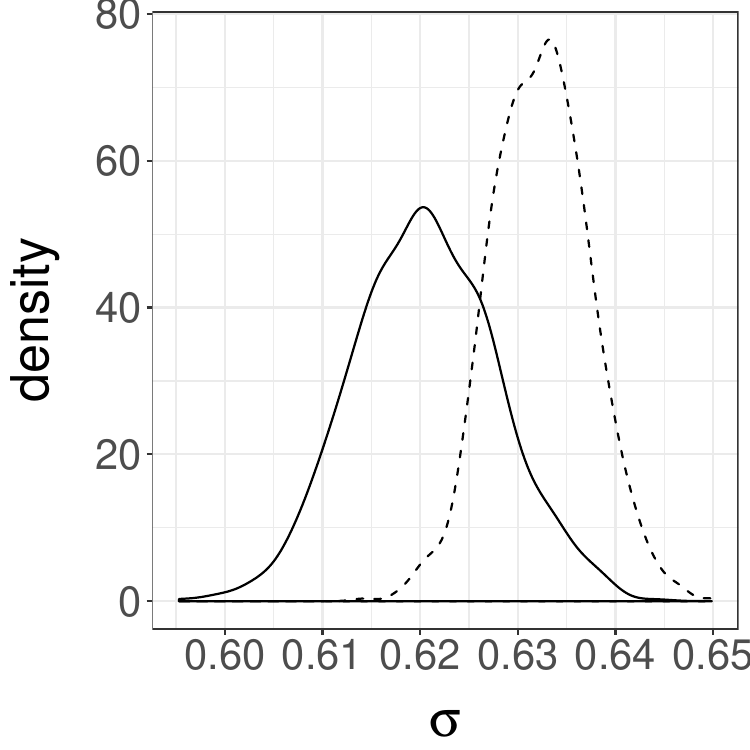}
            \caption{{\small Estimators of $\sigma$.}}    
            \label{ch1:fig:hist_sigma_gamma}
        \end{subfigure}
        \caption{\small Gamma data fitted with a Normal model, as described in Section \ref{ch1:sec:mwegamma}. 
        Figures \ref{ch1:fig:sampling_mle_gamma} and
        \ref{ch1:fig:sampling_mewe_gamma} show the sampling distributions of the MLE
        and MEWE of order 1 respectively, as $n$ ranges from $50$ to $10^4$
        (colors from red to white to blue). Figures \ref{ch1:fig:hist_mu_gamma} and
        \ref{ch1:fig:hist_sigma_gamma} show the marginal densities of the
        estimators of $\gamma$ and $\sigma$ respectively, for $n=10^4$; the MLEs
        are shown in dashed lines and the MEWE in full lines. For the MEWE, we have used $m=10^4$, $k=20$ and one iteration of MCEM. 
                \label{ch1:fig:sampling_gamma}}
\end{figure}

We now consider a misspecified setting. 
Let $\mu_\star$ be a $\text{Gamma}(10,5)$ distribution (parametrized by shape and rate) and $\mathcal{M}=\{\mathcal{N}(\gamma,\sigma^2) : \gamma\in\mathbb{R}, \sigma > 0\}$.
The Normal location-scale model is very simple, yet it is widely used in practice in the form of regression models.
Figure \ref{ch1:fig:sampling_gamma} compares the sampling distributions of the maximum likelihood estimator and approximations of the MEWE of order 1, over $M=1,000$ experiments,
for different values of $n$.
The MEWE converges at the same $\sqrt{n}$ rate
as the MLE, albeit to a distribution that is centered at a different
location. Therefore, despite both estimation techniques leading to similar values for $\gamma$ and $\sigma$, the distributions
of the estimators have very little overlap for large $n$, as observed in Figures \ref{ch1:fig:hist_mu_gamma} and \ref{ch1:fig:hist_sigma_gamma}. For the MEWE, we have again used $m=10^4$, $k=20$, and one iteration of MCEM. 
    
\begin{sidewaysfigure}[hp]
    \centering
        \begin{subfigure}[b]{0.42\textwidth}
            \centering
            \includegraphics[width=\textwidth]{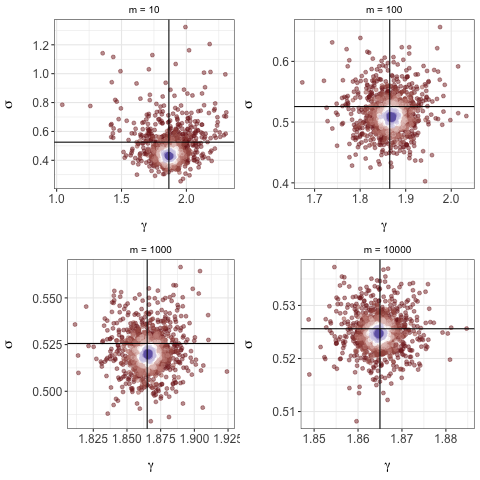}
            \caption{{\small Approximate MEWE for increasing $k$ (colors from red to white to blue as $k$ increases), for different values of $m$.}}   
            \label{ch1:fig:gamma_levelsk}
        \end{subfigure}
       	\hskip1cm
        \begin{subfigure}[b]{0.42\textwidth}
            \centering
            \includegraphics[width=\textwidth]{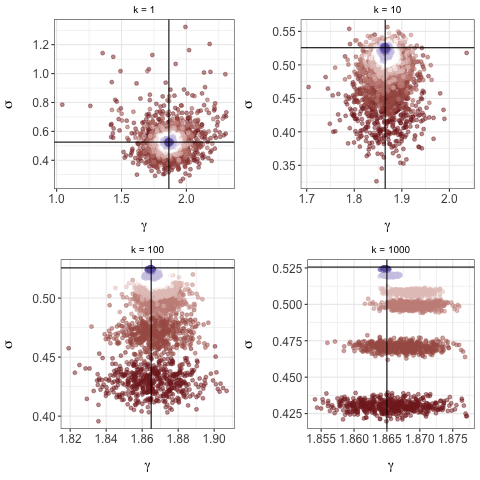}
            \caption{{\small Approximate MEWE for increasing $m$ (colors from red to white to blue as $m$ increases), for different values of $k$.}}   
            \label{ch1:fig:gamma_levelsm}
            \end{subfigure}
         \caption{\label{ch1:fig:gamma_levels}
         \small Gamma data with $n = 100$, fitted with a Normal model, as described in Section \ref{ch1:sec:mwegamma}.
          MEWEs are obtained for different values of $m$ (from $10$ to $10,000$) and $k$ (from $1$ to $1,000$), using one iteration of MCEM,
          $M=500$ times independently. The intersections of the black lines represent the location of the ``exact'' MWE computed with $n=m=10^8$.}
\end{sidewaysfigure}

In Figure \ref{ch1:fig:gamma_levels}, we fix an
observed data set of size $n=100$, and compute $M=500$ instances of the
approximate MEWE for 8 different values of $k$ and $m$, ranging from $1$ to $1,000$ and $10$ to $10,000$ respectively. In Figure
\ref{ch1:fig:gamma_levelsk}, we plot the estimators obtained for all the levels of
$k$, given 4 different values of $m$. In Figure \ref{ch1:fig:gamma_levelsm}, we plot
the estimators obtained for all the levels of $m$, given 4 different values of
$k$. The axis scales are different for each subplot. In both
figures, black points correspond to the ``true'' MWE, calculated using a very large 
value of $m$ ($m=10^8$). 
For low values of $m$, the estimators might be significantly different from the MWE,
as can be seen from the lower-right sub-plots of Figure \ref{ch1:fig:gamma_levelsm}.
When $m$ increases, the estimators converge to the MWE. Increasing $k$ reduces
variation in the estimator.
The changes in $k$ and $m$ had no significant impact on
the number of evaluations of the objective required to locate the maximum
using the \texttt{optim} function in \texttt{R} \citep{Rsoftware}, which uses the Nelder--Mead simplex 
method \citep{nelder1965simplex}.

We check the coverage of bootstrap confidence intervals calculated for
$\theta_\star$ (itself calculated using $n = m = 10^8$ and $k=1$). As before, we use
the percentile bootstrap \citep{efron1994introduction} for data sets of size
$n=1,000$ and synthetic data sets of size $m=10^4$, and calculate the MEWE with
$k=20$. We draw $400$ data sets from the data-generating process, and $1,000$
bootstrap data sets for each of these. The observed coverage rates of the
resulting $0.95$ confidence intervals were $0.960$ and $0.953$ for
$\gamma_\star$ and $\sigma_\star$ respectively. After a Bonferroni correction,
the observed coverage rate of the confidence sets for $\theta_\star =
(\gamma_\star,\sigma_\star)$ was $0.955$.

\subsection{Cauchy data fitted with a Normal model}\label{ch1:sec:mwecauchy}

\begin{figure}[hp]
        \centering
        \begin{subfigure}[b]{0.42\textwidth}  
            \centering 
            \includegraphics[width=\textwidth]{{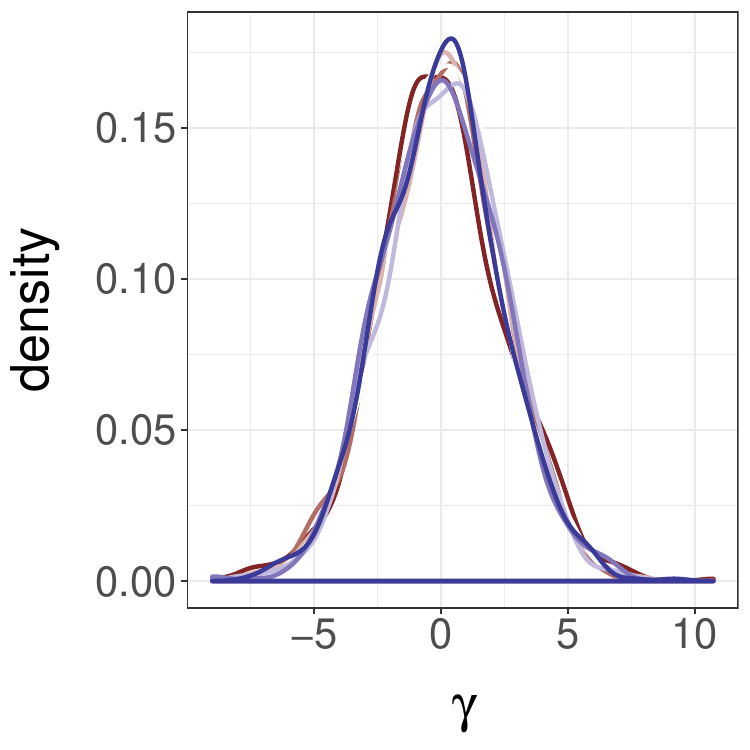}}
            \caption{{\small $\sqrt{n}$-scaled estim. of $\gamma$.}}    
            \label{ch1:fig:cauchy:density}
        \end{subfigure}
        \hskip 0.8cm
         \begin{subfigure}[b]{0.42\textwidth}  
            \centering 
            \includegraphics[width=\textwidth]{{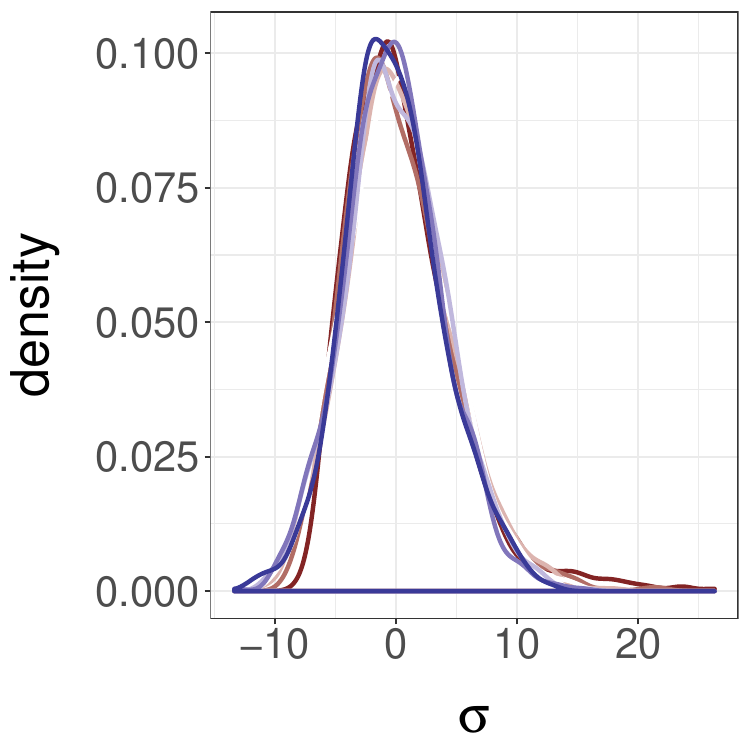}}
            \caption{{\small $\sqrt{n}$-scaled estim. of $\sigma$.}}    
            \label{ch1:fig:cauchy:density_sigma}
        \end{subfigure}

        \begin{subfigure}[b]{0.42\textwidth}  
            \centering 
            \includegraphics[width=\textwidth]{{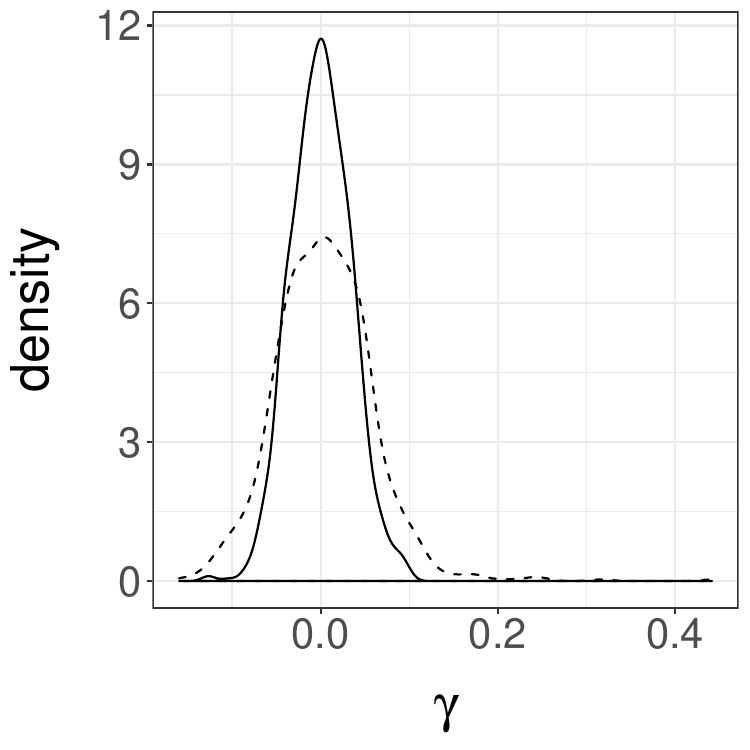}}
            \caption{{\small Estimators of $\gamma$.}}    
            \label{ch1:fig:cauchy:mKLe_mu}
        \end{subfigure}
        \hskip 0.8cm
         \begin{subfigure}[b]{0.42\textwidth}  
            \centering 
            \includegraphics[width=\textwidth]{{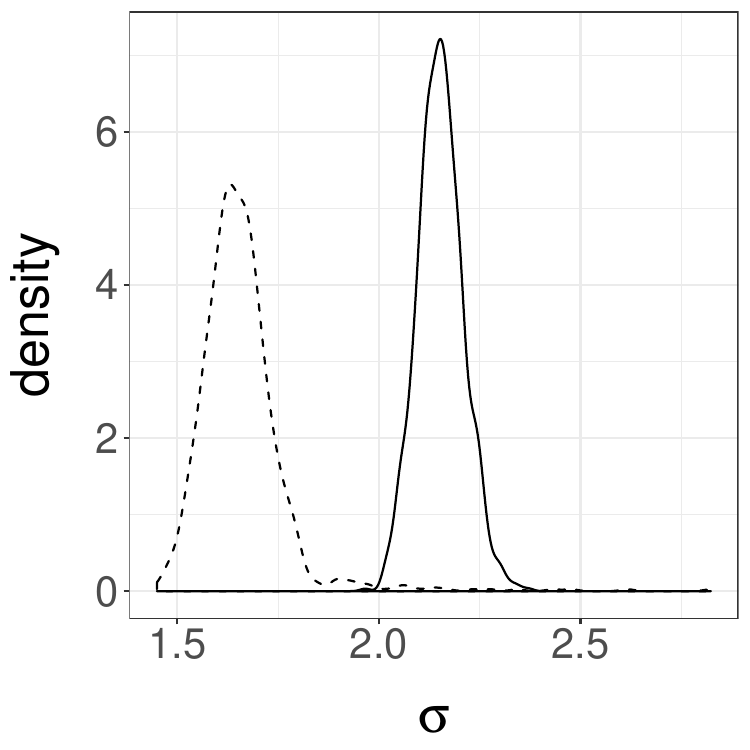}}
            \caption{{\small Estimators of $\sigma$.}}    
            \label{ch1:fig:cauchy:mKLe_sigma}
        \end{subfigure}
        \caption{\small Cauchy data fitted with a Normal model, as described in Section \ref{ch1:sec:mwecauchy}. 
        Marginals distributions of the MEWE of $\gamma$ and $\sigma$, centered by $\theta^\star$ itself computed with $n=m=10^8$, and rescaled by $\sqrt{n}$, are
            shown in Figures \ref{ch1:fig:cauchy:density} and \ref{ch1:fig:cauchy:density_sigma}. 
            Figures \ref{ch1:fig:cauchy:mKLe_mu} and \ref{ch1:fig:cauchy:mKLe_sigma} \label{ch1:fig:cauchy} show the distributions of the MEWE
        for $n=5,000$ (full lines), along with the distribution of an estimator obtained by minimizing an estimate of the Kullback--Leibler divergence (dashed lines).}
\end{figure}

Let $\mu_\star$ be Cauchy with median zero and scale
one, and consider the model $\mathcal{M}=\{\mathcal{N}(\gamma,\sigma^2) :
\gamma\in\mathbb{R}, \sigma > 0\}$. 
We explore the behavior of the MEWE of order 1, over $M=1,000$
repeated experiments.  Figure \ref{ch1:fig:cauchy} shows 
its sampling distributions, for $n$ ranging
from $50$ to $10^4$.  The marginal distribution of the
estimator of $\gamma$ concentrates around $0$, the median of $\mu_\star$. The marginal
distribution of the estimator of $\sigma$ also concentrates to a value close to $2.2$.  The concentration appears to occur at rate
$\sqrt{n}$, as shown by the marginal densities of the rescaled
estimators of $\gamma$ and $\sigma$ in Figures  \ref{ch1:fig:cauchy:density} and \ref{ch1:fig:cauchy:density_sigma}.

In this setting the maximum likelihood estimator would not converge as $n\to \infty$, 
as the maximum likelihood estimator for $\gamma$ is the sample average, and the sample average of independent Cauchy variables
is also Cauchy, with the same location and scale.
As an alternative, we consider an estimator defined by minimizing a sample based estimator of the Kullback-Leibler
divergence between ${\mu}_{\theta}$ and ${\mu}_\star$. For the KL approximation we use the 
function \texttt{KL.divergence} in the \texttt{FNN} package \citep{beygelzimer2013fnn},
which approximates the KL divergence using $\ell$-nearest neighbor estimates described in 
\citet{boltz2009high} (and using the default parameter $\ell=5$). The resulting estimator is termed the minimum KL estimator (MKLE), 
and is a variation of the MDEs discussed by \citet{basu2011statistical}. We compute it using the same approach as for the MEWE, using 
$k = 20$, $m = 10^4$, and one iteration of MCEM. For $n=5,000$ the distributions
of MEWEs and MKLEs are plotted in Figures 
\ref{ch1:fig:cauchy:mKLe_mu} and \ref{ch1:fig:cauchy:mKLe_sigma}. Both estimators appear to be robust in the sense that they converge to well-defined limits, unlike the MLE approach. The estimators of $\gamma$ are concentrated around $0$, but the estimators of $\sigma$ are concentrated around two different values: the MEWEs seem to concentrate around $2.15$ and the MKLEs around $1.65$. The marginal distributions of the MEWE appear to have slightly smaller variance than those of the MKLE.

Note that this example is not covered by the theoretical results of Section \ref{ch1:sec:theory} since the Cauchy distribution does not have a finite first moment. Robustness properties of general minimum distance estimators are discussed in \citet{parr1980minimum}, 
and of the MWE in location models in \citet{bassetti_regazzini2006}. In the location-scale model considered here, if the approximation of the MEWE is computed with $k=1$ and $m = \ell n$ for some $\ell \geq 1$, it can be written
\begin{equation*}
\argmin_{\gamma,\sigma} \sum_{i=1}^n \sum_{j=1}^\ell \lvert y_{(i)} - (\sigma x_{(\ell(i-1)+j)} +\gamma)\rvert.
\end{equation*}
As such, the approximate MEWE can be seen as the coefficients in a median regression \citep{koenker2001quantile} of a vector $\tilde{Y}$ on a vector $\tilde{X}$, where $\tilde{Y}_{\ell(i-1)+1 : \ell i} = y_{(i)}$ for each $i = 1,\dots,n$, and 
$\tilde{X}$ contains the order statistics of an $m$-sample of
$\mathcal{N}(0,1)$ random variables. Quantile regression is often presented as
a robust alternative to linear regression in the presence of outliers, and further 
connections might explain the observed robustness of the MEWE with $p=1$ in this example.

\section{Discussion}\label{ch1:sec:discussion}

The minimum Wasserstein (or Kantorovich) estimation approach
\citep{bassetti2006minimum} has received a renewed attention, due to recent
advances in the field of computational optimal transport
\citep{peyre2018computational}, along with various applications in machine learning.
In the broad context of generative models, these estimators present various
appeals compared to maximum likelihood estimators.  For instance, in Sections
\ref{ch1:sec:gandk} and \ref{ch1:sec:sumlognormal}, we have observed the satisfactory
behavior of minimum expected Wasserstein estimators in models where the
likelihood function is not analytically available. In Sections
\ref{ch1:sec:mwegamma} and \ref{ch1:sec:mwecauchy} we have observed similarities and
differences between MEWE and MLE in misspecified settings, illustrating some
robustness properties of minimum Wasserstein estimation.

Minimum distance estimators were originally developed for obtaining almost
surely convergent estimators  \citep{wolfowitz1957}, and we have showed that
both the MWE and MEWE have this strong consistency property under mild
conditions. We have also proved that the MWE converges to $\theta_\star$ at the
optimal $\sqrt{n}$ convergence rate when the observations are univariate, and
have derived its asymptotic distribution. The generalization of this result to
multivariate data is left for future research. Interestingly, given the known
convergence properties of the Wasserstein distance, it seems reasonable to
conjecture that the rate of the MWE depends (negatively) on the dimension of
the observation space rather than that of the parameter space. Other topics 
for future research include a more general derivation of the limiting
distributions of the estimators, whose existence is needed to justify the asymptotic
coverage of subsampling confidence intervals, as well as the development of a better understanding of
their robustness properties.

\vspace{-0.1cm}
\section*{Acknowledgements}
The bootstrap experiments were in part performed
on the Odyssey cluster supported by the FAS Division of Science, Research
Computing Group at Harvard University. Pierre E. Jacob acknowledges
support from the National Science Foundation through grant DMS-1712872.

\bibliography{biblio}
\bibliographystyle{apalike}

\begin{appendix}
\section{Preliminary results}
Before proving the results stated in the main text, we first provide some preliminary results. 

A sequence of probability measures $(\mu_n)_{n\geq 1}$ is said to converge weakly in $\mathcal{P}_p(\mathcal{Y})$  to
$\mu$ as $n\to \infty$ if $\mu_n\Rightarrow \mu$, i.e. converges weakly in the usual sense, and $\exists\, y_0 \in \mathcal{Y}$ such that
$\int_\mathcal{Y} \rho(y,y_0)^p d{\mu_n}(y) \to \int_\mathcal{Y} \rho(y,y_0)^p
d\mu(y)$. Recall that $\mathcal{Y}$ is a subset of $\mathbb{R}^d$ for $d\geq 1$.
\begin{theorem} \label{theorem:metrize}
The $p$-Wasserstein distance metrizes weak convergence in
$\mathcal{P}_p(\mathcal{Y})$: a sequence $\mu_n$ converges weakly to $\mu$ in
$\mathcal{P}_p(\mathcal{Y})$ if and only if $\was_p(\mu_n,\mu) \to 0$.
\end{theorem}
For a proof, see \citet[Theorem 6.9]{villani2008}. From this result one can deduce the continuity of the $p$-Wasserstein distance.  If $\mu_n$ and
$\nu_n$ converge weakly in $\mathcal{P}_p(\mathcal{Y})$ to $\mu$ and $\nu$,
then $\was_p(\mu_n,\nu_n) \to \was_p(\mu,\nu)$. On the other hand, if $\mu_n$ and
$\nu_n$ converge weakly in the usual sense, the Wasserstein distance is only lower semicontinuous: $$\liminf_{n\to \infty}\was_p(\mu_n,\nu_n) \geq \was_p(\mu,\nu).$$ The following lemma is extended from
\citet{bassetti2006minimum}. Its second condition corresponds to
Assumption \ref{as:sup:weakcvg}, and is implied by the first condition. All limits in the lemma are understood to be as $n\to \infty$.
\begin{lemma} \label{lemma:continuity}
Let $(\theta_n)_{n\geq 1}$ be a sequence in $\mathcal{H}$ and $\theta \in \mathcal{H}$. Suppose that either of the following conditions holds. 
\begin{enumerate}
\item $\rho_\mathcal{H}(\theta_n,\theta) \to 0$ implies that $\was_p(\mu_{\theta_n},\mu_\theta) \to 0$.
\item $\rho_\mathcal{H}(\theta_n,\theta) \to 0$ implies that $\mu_{\theta_n} \Rightarrow \mu_\theta$.
\end{enumerate}
Then, respectively,
\begin{enumerate}
\item $\mathcal{H}\times\mathcal{P}_p(\mathcal{Y}) \ni(\theta,\mu) \mapsto \was_p(\mu_\theta,\mu)$ is continuous.
\item $\mathcal{H}\times\mathcal{P}(\mathcal{Y}) \ni(\theta,\mu) \mapsto \was_p(\mu_\theta,\mu)$ is lower semicontinuous.
\end{enumerate}
\end{lemma}
\begin{proof}
This follows directly from the two assumptions and the continuity/lower semicontinuity derived from Theorem \ref{theorem:metrize}. 
\end{proof}

\begin{lemma}\label{lemma:continuity2}
The function $(\nu,\mu^{(m)}) \mapsto \mathbb{E}\was_p(\nu,\hat{\mu}_m)$ is lower semicontinuous with respect to weak convergence. Furthermore, if $\rho_\mathcal{H}(\theta_n,\theta) \to 0$ implies that $\mu^{(m)}_{\theta_n} \Rightarrow \mu^{(m)}_\theta$, then the map  $(\nu,\theta) \mapsto \mathbb{E}\was_p(\nu,\hat{\mu}_{\theta,m})$ is lower semicontinuous.
\end{lemma}
\begin{proof}
Let $\mu^{(m)}_k \Rightarrow \mu^{(m)}$ and $\nu_k \Rightarrow \nu$. Then there exist versions of the corresponding empirical measures such that $\hat{\mu}_{k,m} \Rightarrow \hat{\mu}_m$ almost surely.

Indeed, 
by Skorokhod's representation theorem, there exists a probability space $(\tilde{\mathbb{P}},\tilde{\Omega},\tilde{\Sigma})$ and random variables $\tilde{X}_k^{1:m}\sim\mu_k^{(m)}$ and $\tilde{X}^{1:m}\sim\mu^{(m)}$ such that $\tilde{X}_k^{1:m} \to \tilde{X}^{1:m}$ $\tilde{\mathbb{P}}$-almost surely. Let $\hat{\mu}_{k,m}$ and $\hat{\mu}_m$ be the empirical distributions of these samples.  By \citet{varadarajan1958weak} and since $\mathcal{Y}$ is separable,  there exists a fixed countable subset $C^{\star}$ of continuous and bounded functions on $\mathcal{Y}$,  such that for any sequence of measures $\mu_n \in \mathcal{P}(\mathcal{Y})$, $\mu_n$ converges weakly to $\mu$ if and only if $\int f d\mu_n \to \int f d\mu$ for all $f \in C^{\star}$. Fix one such $f$. Then,
$$\int f d\hat{\mu}_{k,m} = \frac{1}{m}\sum_{i=1}^m f(\tilde{X}_k^i) \to \frac{1}{m}\sum_{i=1}^m f(\tilde{X}^i) = \int f d\hat{\mu}_{m},$$
on a set of $\tilde{\mathbb{P}}$-probability one, by the continuous mapping theorem. Taking the countable intersection of these sets over $f \in C^\star$, we get that $\hat{\mu}_{k,m} \Rightarrow \hat{\mu}_m$ $\tilde{\mathbb{P}}$-almost surely.

By the lower semicontinuity of the $p$-Wasserstein distance and Fatou's lemma,
$$ \mathbb{E}\was_p(\nu,\hat{\mu}_m) \leq \mathbb{E} \liminf_{k\rightarrow\infty}\was_p(\nu_k,\hat{\mu}_{k,m}) \leq  \liminf_{k\rightarrow\infty}\mathbb{E}\was_p(\nu_k,\hat{\mu}_{k,m}).$$
The lower semicontinuity of $(\nu,\theta) \mapsto \mathbb{E}\was_p(\nu,\hat{\mu}_{\theta,m})$ is proved analogously to Lemma \ref{lemma:continuity}.
\end{proof}

\section{Proofs: MWE}

\subsection{Existence, measurability, and consistency}
For ease of presentation, we recall the assumptions made in the main text.

\begin{assumption}\label{as:sup:cvgwas}
   The data-generating process is such that $\was_p(\hat{\mu}_n, \mu_\star) \to 0$,  $\mathbb{P}$-almost surely as $n\to\infty$.
\end{assumption}

\begin{assumption} \label{as:sup:weakcvg}
The map $\theta\mapsto \mu_\theta$ is continuous in the sense that $\rho_\mathcal{H}(\theta_n,\theta) \to 0$ implies $\mu_{\theta_n}  \Rightarrow \mu_\theta$ as $n\to\infty$.
\end{assumption}

For the next assumption, define $\varepsilon_\star=\inf_{\theta\in\mathcal{H}} \was_p(\mu_\star,\mu_\theta)$; we will use this definition
throughout.

\begin{assumption}\label{as:sup:relativelycompact}
For some $\varepsilon>0$, the set $B_\star(\varepsilon) = \{\theta \in \mathcal{H} : \was_p(\mu_{\star},\mu_\theta) \leq \varepsilon_\star + \varepsilon\}$ is bounded.
\end{assumption}

\begin{theorem}[Existence and consistency of the MWE] \label{theorem:sup:consistent}
Under Assumptions \ref{as:sup:cvgwas}-\ref{as:sup:relativelycompact}, there exists a set $E\subset \Omega$
with $\mathbb{P}(E)=1$ such that, for all $\omega \in E$, 
$$\inf_{\theta\in\mathcal{H}} \was_p(\hat{\mu}_n(\omega),\mu_\theta) \to \inf_{\theta\in\mathcal{H}} \was_p(\mu_\star,\mu_\theta),$$ and there exists
$n(\omega)$ such that, for all $n\geq n(\omega)$, the sets $\argmin_{\theta\in\mathcal{H}} \was_p(\hat{\mu}_n(\omega),\mu_\theta)$ are non-empty and form a bounded sequence with 
$$\limsup_{n\to\infty} \argmin_{\theta\in\mathcal{H}} \was_p(\hat{\mu}_n(\omega),\mu_\theta) \subset \argmin_{\theta\in\mathcal{H}} \was_p(\mu_{\star},\mu_\theta).$$
\end{theorem}

Before giving the proof, we recall a definition and a proposition.

\begin{definition}
A sequence of functions $f_n : \mathcal{H}\to \mathbb{R}$ is said to epi-converge to $f:\mathcal{H}\to \mathbb{R}$ 
if for all $\theta\in\mathcal{H}$,
\[\begin{cases}
\liminf_{n \to \infty}f_n(\theta_n) \geq f(\theta) & \text{for every sequence $\theta_n\to\theta$,} \\
\limsup_{n \to \infty}f_n(\theta_n)\leq f(\theta) & \text{for some sequence $\theta_n\to\theta$.}
\end{cases}\]
\end{definition}

A useful equivalent formulation of epi-convergence can be found in Proposition 7.29 of \citet{rockafellar2009variational}, paraphrased here.

\begin{proposition}[Proposition 7.29 of \citet{rockafellar2009variational}]\label{prop:epiconv_equivalent}
The sequence $f_n : \mathcal{H}\to \mathbb{R}$ epi-converges to $f:\mathcal{H}\to \mathbb{R}$ if and only if
\[\begin{cases}
\liminf_{n \to \infty} \inf_{\theta\in \mathcal{K}}f_n(\theta) \geq \inf_{\theta\in \mathcal{K}} f(\theta) & \text{for every compact set $\mathcal{K}\subset \mathcal{H}$,} \\
\limsup_{n \to \infty}\inf_{\theta\in \mathcal{O}}f_n(\theta)\leq \inf_{\theta\in \mathcal{O}}f(\theta) & \text{for every open set $\mathcal{O}\subset \mathcal{H}$.}
\end{cases}\]
\end{proposition}

In an colloquial sense, epi-convergence is a weak notion of convergence
for which the minimizer of $f_n$ converges to the minimizer of $f$. Showing
that the function $\theta \mapsto \was_p(\hat{\mu}_n,\mu_\theta)$ epi-converges to
$\theta\mapsto \was_p(\mu_\star,\mu_\theta)$ almost surely is the key step in
the proof of Theorem \ref{theorem:sup:consistent}.

\begin{proof}[Proof of Theorem \ref{theorem:sup:consistent}]
First note that, for any $\nu\in\mathcal{P}(\mathcal{Y})$, the lower semicontinuity
of the map $\theta \mapsto \was_p(\nu,\mu_\theta)$ follows from Lemma
\ref{lemma:continuity}, via Assumption \ref{as:sup:weakcvg}. Next, since
$\inf_{\theta \in \mathcal{H}}\was_p(\mu_\star,\mu_\theta) =
\varepsilon_\star$, the set $B_\star(\varepsilon)$ with the $\varepsilon$ of Assumption \ref{as:sup:relativelycompact} is non-empty, by definition of the
infimum. Moreover, since  $\theta \mapsto \was_p(\mu_\star,\mu_\theta)$ is lower
semicontinuous, the set $B_\star(\varepsilon)$ is closed. By Assumption
\ref{as:sup:relativelycompact}, $B_\star(\varepsilon)$ is therefore compact. In
other words, again by lower semicontinuity, the set $\argmin _{\theta \in
    \mathcal{H}}\was_p(\mu_\star,\mu_\theta)$ is non-empty.

We now show that $\theta \mapsto \was_p(\hat{\mu}_n,\mu_\theta)$ epi-converges to $\theta\mapsto \was_p(\mu_\star,\mu_\theta)$ $\mathbb{P}$-almost surely. 
Let $E$ denote the set of probability one from Assumption \ref{as:sup:cvgwas} and let $\omega \in E$. Fix $\mathcal{K} \subset \mathcal{H}$ compact. By lower semicontinuity of $\theta\mapsto \was_p(\hat{\mu}_n(\omega),\mu_\theta)$, we know that 
$$\inf_{\theta\in\mathcal{K}} \was_p(\hat{\mu}_n(\omega),\mu_\theta) = \was_p(\hat{\mu}_n(\omega),\mu_{\theta_n}),$$
for some sequence $\theta_n = \theta_n(\omega) \in \mathcal{K}$. Hence,
\begin{align*}
\liminf_{n\to \infty} &\inf_{\theta\in\mathcal{K}} \was_p(\hat{\mu}_n(\omega),\mu_\theta) = \liminf_{n\to \infty}\was_p(\hat{\mu}_n(\omega),\mu_{\theta_n})\\
& = \lim_{k\to \infty} \was_p(\hat{\mu}_{n_k}(\omega),\mu_{\theta_{n_k}}) \quad \text{$\exists$ subsequence converging to the $\liminf$,} \\
& = \lim_{m\to \infty} \was_p(\hat{\mu}_{n_{k_m}}(\omega),\mu_{\theta_{n_{k_m}}})  \quad \text{$\exists$ subsequence $\theta_{n_{k_m}} \to \bar{\theta}\in\mathcal{K}$ by compactness,} \\ 
& = \liminf_{m\to \infty} \was_p(\hat{\mu}_{n_{k_m}}(\omega),\mu_{\theta_{n_{k_m}}}) \\
& \geq \was_p(\mu_\star,\mu_{\bar{\theta}})  \quad \text{by l.s.c., Assumptions \ref{as:sup:cvgwas} and \ref{as:sup:weakcvg} , $\omega \in E$},\\
& \geq \inf_{\theta\in\mathcal{K}}\was_p(\mu_\star,\mu_{\theta}).
\end{align*}

Fix $\mathcal{O} \subset \mathcal{H}$ open. By definition of the infimum, there exists a sequence $\theta_n \in\mathcal{O}$ such that $\was_p(\mu_\star,\mu_{\theta_n})\to \inf_{\theta\in\mathcal{O}}\was_p(\mu_\star,\mu_\theta)$. Now, $\inf_{\theta\in\mathcal{O}}\was_p(\hat{\mu}_n(\omega),\mu_\theta) \leq \was_p(\hat{\mu}_n(\omega),\mu_{\theta_n})$. Hence,
\begin{align*}
\limsup_{n\to \infty} &\inf_{\theta\in\mathcal{O}} \was_p(\hat{\mu}_n(\omega),\mu_\theta) \leq \limsup_{n\to \infty}\was_p(\hat{\mu}_n(\omega),\mu_{\theta_n})\\
& \leq \limsup_{n\to \infty} \left(\was_p(\hat{\mu}_n(\omega),\mu_\star) + \was_p(\mu_\star,\mu_{\theta_n})\right) \quad \text{by the triangle inequality}, \\
&\leq  \limsup_{n\to \infty} \was_p(\hat{\mu}_n(\omega),\mu_\star) + \limsup_{n\to \infty} \was_p(\mu_\star,\mu_{\theta_n}) \quad \text{by positivity},\\
&=  \limsup_{n\to \infty} \was_p(\mu_\star,\mu_{\theta_n})  \quad \text{by Assumption \ref{as:sup:cvgwas},  $\omega \in E$},\\
& = \inf_{\theta\in\mathcal{O}}\was_p(\mu_\star,\mu_\theta) \quad \text{by definition of $\theta_n$.}
\end{align*}
Using Proposition \ref{prop:epiconv_equivalent}, the sequence of functions $\theta \mapsto \was_p(\hat{\mu}_n(\omega),\mu_\theta)$ epi-converges
to $\theta\mapsto \was_p(\mu_\star,\mu_\theta)$.

Theorem 7.29b) of \citet{rockafellar2009variational} implies that 
\[\limsup_{n\to\infty} \inf_{\theta\in\mathcal{H}} \was_p(\hat{\mu}_n(\omega),\mu_\theta) \leq \inf_{\theta\in\mathcal{H}} \was_p(\mu_\star,\mu_\theta) = \varepsilon_\star.\] So, for all $\alpha > 0$, there exists $n_\alpha(\omega)$, such that for $n\geq n_\alpha(\omega)$, $\inf_{\theta\in\mathcal{H}} \was_p(\hat{\mu}_n(\omega),\mu_\theta) \leq \varepsilon_\star + \alpha$. 
Let $\alpha \in (0,\varepsilon/2)$. The set 
$$\{\theta\in\mathcal{H}: \was_p(\hat{\mu}_n(\omega), \mu_\theta) \leq \varepsilon_\star +\varepsilon/2\}$$
is non-empty for $n\geq n_\alpha(\omega)$, by definition of the infimum. Let $\theta$ belong to this set. Then, by the triangle inequality, $$\was_p(\mu_\star,\mu_\theta) \leq \was_p(\hat{\mu}_n(\omega),\mu_\star) + \was_p(\hat{\mu}_n(\omega),\mu_\theta).$$
By Assumption \ref{as:sup:cvgwas}, there exists an $n_\varepsilon(\omega)$ such that for $n\geq n_\varepsilon(\omega)$, $\was_p(\hat{\mu}_n(\omega),\mu_\star)\leq \varepsilon/2$. So, if $n\geq \max\{n_\alpha(\omega),n_\varepsilon(\omega)\}$, we have that $\was_p(\mu_\star,\mu_\theta) \leq \varepsilon_\star +\varepsilon$. This means that 
$$\{\theta\in\mathcal{H}: \was_p(\hat{\mu}_n(\omega), \mu_\theta) \leq \varepsilon_\star +\varepsilon/2\} \subset  B_\star(\varepsilon).$$
As a consequence, we know that for $n\geq \max\{n_\alpha(\omega),n_\varepsilon(\omega)\}$, $$\inf_{\theta\in\mathcal{H}} \was_p(\hat{\mu}_n(\omega),\mu_\theta)  = \inf_{\theta\in B_\star(\varepsilon)} \was_p(\hat{\mu}_n(\omega),\mu_\theta).$$

By Theorem 7.31a) of \citet{rockafellar2009variational}, this implies
$$\inf_{\theta\in\mathcal{H}} \was_p(\hat{\mu}_n(\omega),\mu_\theta) \to
\inf_{\theta\in\mathcal{H}} \was_p(\mu_\star,\mu_\theta).$$  For $n\geq
\max\{n_\alpha(\omega),n_\varepsilon(\omega)\}$ and by the same reasoning as
for the map $\theta\mapsto \was_p(\mu_\star,\mu_\theta)$, the sets
$\argmin_{\theta\in\mathcal{H}} \was_p(\hat{\mu}_n(\omega),\mu_\theta)$ are
non-empty. By Theorem 7.31b) of \citet{rockafellar2009variational}, the result
follows. The same argument holds for
$\varepsilon_n\mhyphen\argmin_{\theta\in\mathcal{H}}
\was_p(\hat{\mu}_n(\omega),\mu_\theta)$ with $\varepsilon_n\to 0$, since $\inf_{\theta\in\mathcal{H}} \was_p(\hat{\mu}_n(\omega),\mu_\theta)
+ \varepsilon_n \leq \varepsilon_\star + \alpha$ eventually. 

\end{proof}

\begin{theorem}[Measurability of the MWE]  \label{theorem:sup:measurable}
Suppose that $\mathcal{H}$ is a $\sigma$-compact Borel measurable subset of $\mathbb{R}^{d_\theta}$. Under Assumption \ref{as:sup:weakcvg}, for any $n \geq 1$ and $\varepsilon > 0$, there exists a Borel measurable function $\hat{\theta}_n : \Omega \to \mathcal{H}$ that satisfies
\begin{align*}
\hat{\theta}_n(\omega) \in
  \begin{cases}
    \argmin_{\theta\in\mathcal{H}} \was_p(\hat{\mu}_n(\omega),\mu_\theta), & \text{if this set is non-empty,}  \\
    \varepsilon \mhyphen \argmin_{\theta\in\mathcal{H}} \was_p(\hat{\mu}_n(\omega),\mu_\theta), & \text{otherwise}.
  \end{cases}
\end{align*}
\end{theorem}

Before the proof, we first recall a useful result from \citet{brown1973}, also used in \citet{bassetti2006minimum}.

\begin{theorem}[Corollary 1 in \citet{brown1973}]\label{theorem:brown}
Let $X,Y$ be Polish, $D\subset Y\times X$ be Borel, and $f:D\to \mathbb{R}$ be
Borel measurable. Suppose that for all $y\in \text{proj}(D)$, the section $D_y
= \{x : (y,x)\in D\}$ is $\sigma$-compact and that $f_y = f(y,\cdot)$ is lower
semicontinuous with respect to the relative topology on $D_y$. Then
\begin{enumerate}
\item The sets $G=\text{proj}(D)$ and $I=\{y\in G: \text{for some} \; x \in D_y, f(y,x)= \inf f_y\}$ are Borel.
\item For each $\varepsilon >0$, there exists a Borel measurable function $\phi_\varepsilon$ such that for $y\in G$,
$$f(y,\phi_\varepsilon(y))  \begin{cases} 
	= \inf f_y & \text{if} \; y\in I. \\
	\leq \varepsilon + \inf f_y & \text{if} \; y\notin I, \inf f_y\neq -\infty. \\
	\leq -\varepsilon^{-1} & \text{if} \; y\notin I, \inf f_y = -\infty.
\end{cases}$$
\end{enumerate}
\end{theorem}

\begin{proof}[Proof of Theorem \ref{theorem:sup:measurable}]
First note that $\mathcal{Y}^\infty$ endowed with the product topology is
Polish since $(\mathcal{Y},\rho)$ is Polish. Also, $\hat{\mu}_n(\omega)$ depends
on $\omega$ only through $y=Y(\omega)$, where $Y = (Y_t)_{t\in\mathbb{Z}}$. We
can therefore write $\hat{\mu}_n(\omega) =\hat{\mu}_n(y)$, where
$y\in\mathcal{Y}^\infty$, and consider the empirical measure a function on $\mathcal{Y}^{\infty}$. The map $y\mapsto \hat{\mu}_n(y)$ is measurable with respect to the Borel
$\sigma$-algebra on $\mathcal{P}_p(\mathcal{Y})$ with respect to weak convergence. 
Recall also that $(\mathcal{P}_p(\mathcal{Y}),\was_p)$ is Polish since $\mathcal{Y}$ is Polish by Theorem 6.18 of \citet{villani2008}.

Let $\mathcal{D} = \mathcal{Y}^{\infty} \times \mathcal{H}$. 
By Lemma \ref{lemma:continuity} and Assumption \ref{as:sup:weakcvg}, the map $(\mu,\theta)\mapsto \was_p(\mu,\mu_\theta)$ is lower semicontinuous  (and therefore measurable). Hence the map  $\theta \mapsto \was_p(\hat{\mu}_n(y),\mu_\theta)$  is also lower semicontinuous on $\mathcal{H}$ for any $y\in\mathcal{Y}^\infty$. Being the composition of measurable functions, $(y,\theta)\mapsto \was_p(\hat{\mu}_n(y),\mu_\theta)$ is measurable on $\mathcal{D}$. In light of this, the result follows by a direct application of Theorem \ref{theorem:brown}.

\end{proof}

\subsection{Asymptotic distribution}

Let  $p=1$, $\mathcal{Y} = \mathbb{R}$, and $\rho(x,y) = \lvert x - y \rvert$.
In this case we have $$\was_1(\mu,\nu) = \int_0^1 \lvert F_\mu^{-1}(s) -
F_\nu^{-1}(s)\rvert ds = \int_\mathbb{R} \lvert F_\mu(t) - F_\nu(t)\rvert dt,$$
where $F_\mu$ and $F_\nu$ denote the cumulative distribution functions (CDFs)
of $\mu$ and $\nu$ respectively \citep[see e.g.][Theorem 6.0.2]{ambrosio2005}. For this reason, we will occasionally use the notation
$\was_1(\mu,\nu) = \|F_\mu - F_\nu\|_{L_1}$.
Assume also that $\mathcal{H}$ is endowed with a norm:
$\rho_\mathcal{H}(\theta,\theta') = \|\theta-\theta'\|_\mathcal{H}$. We recall results from \citet{del1999central} and \citet{dede2009}, after a few definitions.

\begin{definition}
Suppose that the sequence $\Omega \times \mathbb{R} \ni (\omega,t) \mapsto
X_n(\omega,t)$ for all $n$, and $\Omega \times \mathbb{R} \ni (\omega,t)
\mapsto X(\omega,t)$, are stochastic processes with almost all their sample
paths in $L_1(\mathbb{R})$. Then $X_n$ is said to converge weakly to $X$ in
$L_1(\mathbb{R})$ if $\mathbb{E}f(X_n) \to \mathbb{E}f(X)$ as $n \to \infty$
for all bounded continuous functions $f:L_1(\mathbb{R}) \to \mathbb{R}$.
\end{definition}

\begin{definition}
The stochastic process $\Omega \times \mathbb{R} \ni (\omega,t) \mapsto G_\mu(\omega,t)$ is a $\mu$-Brownian bridge if it is a zero mean Gaussian process with covariance function $\mathbb{E}G_\mu(s)G_\mu(t) = \min\{F_\mu(s),F_\mu(t)\} - F_\mu(s)F_\mu(t).$
\end{definition}

\begin{theorem}[Theorem 2.1a in \citet{del1999central}]\label{theorem:delbarrio}
Suppose that $Y = (Y_t)_{t \in \mathbb{Z}}\sim \mu_\star^\infty$, and define
$F_n(\omega,t) = \hat{\mu}_n(\omega)(-\infty,t]$ and $F_\star(t) =
\mu_\star(-\infty,t]$. The stochastic process $\sqrt{n}(F_n-F_\star)$ converges
weakly in $L_1(\mathbb{R})$ to a
$\mu_\star$-Brownian bridge $G_\star$, if and only if the condition
$\int_0^\infty\sqrt{\mathbb{P}(\lvert Y_0 \rvert > t)} dt < \infty$ is satisfied. 
\end{theorem}

For a stationary sequence, let $\tilde{\alpha}_t = \sup_{u \in \mathbb{R}} \E \lvert \mathbb{P}(Y_t \leq u | \mathcal{F}^0_{-\infty}) - \mathbb{P}(Y_t \leq u)\rvert$. Note that for stationary sequences, $\tilde{\alpha}$-mixing is weaker than $\alpha$-mixing, as defined later in Section \ref{ch1:app:sec:checkassumptions}.

\begin{theorem}[Proposition 3.5 in \citet{dede2009}]\label{theorem:dede}
Suppose that $Y = (Y_t)_{t \in \mathbb{Z}}$ is ergodic and stationary, and that
$$\sum_{k\geq 1} \frac{1}{\sqrt{k}}\int_0^\infty \min\{\sqrt{\tilde{\alpha}_k},\sqrt{\mathbb{P}(\lvert Y_0 \rvert > t)}\}dt < \infty.$$
Then $\sqrt{n}(F_n - F_{\star})$ converges weakly in $L_1(\mathbb{R})$ to a zero mean Gaussian process $G_\star$ with sample paths in $L_1(\mathbb{R})$ and covariance satisfying: for every $f,g \in L_\infty(\mathbb{R})$,
$$ \E f(G_\star)g(G_\star) = \int_{\mathbb{R}^2}f(s)g(u)C(s,u)dsdu,$$
where
$$C(s,u) = \sum_{t\in\mathbb{Z}} \left\{\mathbb{P}(X_0 \leq s, X_t \leq u) - F_{\star}(s)F_{\star}(u)\right\}.$$
\end{theorem}

\citet{dede2009} also provides other conditions, e.g. on $\phi$-mixing coefficients, for which the convergence above holds.
We first consider the well-specified setting, in which our results follow directly from 
\citet{pollard1980}. 

\subsubsection{Well-specified setting}
Suppose that $\mu_\star = \mu_{\theta_\star}$ for some $\theta_\star$ in the interior of
$\mathcal{H}$, and consider the following assumptions:
\begin{assumption}\label{as:sup:wellsep}
For all $\varepsilon > 0$, there exists $\delta > 0$ such that
$$\inf_{\theta \in \mathcal{H}:\|\theta-\theta'\|_\mathcal{H} \geq \varepsilon} \was_1(\mu_{\theta_\star},\mu_\theta) >  \delta.$$
\end{assumption}
\begin{assumption}\label{as:sup:diff}
There exists a non-singular $D_{\theta_\star} \in (L_1(\mathbb{R}))^{d_\theta}$ such that 
$$\int_\mathbb{R}\lvert F_\theta(t)-F_{\theta_\star}(t) - \langle \theta-\theta_\star, D_{\theta_\star}(t)\rangle \rvert dt = o(\|\theta-\theta_\star\|_\mathcal{H}),$$  
as  $\|\theta-\theta_\star\|_\mathcal{H} \to 0.$
\end{assumption}
The following results contain Theorem \ref{ch1:theorem:asymptoticdistribution} of the main text as a special case.

\begin{theorem}\label{theorem:sup:gof}
Suppose that $\mu_\star = \mu_{\theta_\star}$ for some $\theta_\star$ in the interior of
$\mathcal{H}$, and that the conditions of either Theorem \ref{theorem:delbarrio} or Theorem \ref{theorem:dede} are satisfied. 
Under Assumptions \ref{as:sup:cvgwas}-\ref{as:sup:diff}, the goodness-of-fit statistic satisfies 
$$\sqrt{n}\inf_{\theta\in\mathcal{H}}\was_1(\hat{\mu}_n,\mu_\theta) \Rightarrow \inf_{u\in\mathcal{H}} \int_\mathbb{R}\lvert G_\star(t) - \langle u, D_{\theta_\star}(t)\rangle \rvert dt,$$
as $n\to\infty$, where $G_\star$ is given as in Theorem \ref{theorem:delbarrio} or Theorem \ref{theorem:dede} respectively.
\end{theorem}

\begin{theorem} \label{theorem:sup:asymptoticdistribution}
Suppose that the conditions in Theorem \ref{theorem:sup:gof} hold. Suppose also that the random map $\mathcal{H}\ni u \mapsto \int_\mathbb{R}\lvert G_\star(t) - \langle u, D_{\theta_\star}(t)\rangle  \rvert dt$ has an almost surely unique infimum. Then the MWE of order 1 satisfies 
$$\sqrt{n}(\hat{\theta}_n -\theta_\star) \Rightarrow \argmin_{u\in\mathcal{H}} \int_\mathbb{R}\lvert G_\star(t) - \langle u, D_{\theta_\star}(t)\rangle  \rvert dt,$$ 
as $n\to\infty$, where $G_\star$ is given as in Theorem \ref{theorem:delbarrio} or Theorem \ref{theorem:dede}.
\end{theorem}

\begin{proof}
The proofs of these two results follow the steps outlined in
\citet{pollard1980}'s Theorems 4.2 and 7.2 respectively, which also generalize
to the setting where the map $\mathcal{H}\ni u \mapsto \int_\mathbb{R}\lvert
G_\star(t) - \langle u, D_{\theta_\star}(t)\rangle  \rvert dt$ does not necessarily have
a unique minimum (see also Section \ref{sec:asymptoticdistribution:misspecified} below). The delta methods employed therein hold for the
1-Wasserstein distance due to the representation $\was_1(\mu,\nu) = \|F_\mu - F_\nu\|_{L_1}$. Moreover, the
well-separation of $\theta_\star$ provided by Assumption \ref{as:sup:wellsep}, the consistency and measurability of the MWE proved earlier, and Theorems \ref{theorem:delbarrio} and \ref{theorem:dede} proved in \citet{del1999central} and \citet{dede2009}
respectively, guarantee that Pollard's conditions are satisfied. Note that
the measurability concerns outlined in his Section 3 do not apply to
here, as $L_1(\mathbb{R})$ is separable.

\end{proof}

\subsubsection{Misspecified setting}\label{sec:asymptoticdistribution:misspecified}
To study the asymptotic distribution of the MWE in the misspecified setting, we adapt the arguments outlined in Section 7 of \citet{pollard1980}. Define $f(x,u) = \|x-\langle u, D_{\theta_\star}\rangle\|_{L_1}$ and $m(x) = \inf_{u} f(x,u)$. Let $\mathcal{K}$ be the class of all compact, convex, non-empty subsets of a set $L_1(\mathbb{R})$ equipped with its canonical distance. The corresponding Hausdorff metric on $\mathcal{K}$ is defined by $d_H(K_1,K_2) = \inf\{\delta > 0 : K_1 \subset K_2^\delta, K_2 \subset K_1^\delta \}$, where $K^\delta = \cup_{x\in K} \{z\in M: \|z -x\|_{L_1} \leq \delta\}$. Let $K(x,\beta) = \{u: f(x,u) \leq m(x)+\beta\}$. The function $x\mapsto K(x,\beta)$ maps into $\mathcal{K}$ and, by \citet[][Lemma 7.1]{pollard1980}, is measurable.
Let also $$H_n = \sqrt{n}(F_n - F_{\theta_\star}) = \sqrt{n}(F_n - F_{\star}) + \sqrt{n}(F_\star - F_{\theta_\star})$$
and $H_n^\star = G_\star + \sqrt{n}(F_\star - F_{\theta_\star})$.  Let $$M_n = \{\theta \in \mathcal{H}: \was_1(\hat{\mu}_n, \mu_\theta) \leq \inf_\theta \was_1(\hat{\mu}_n,\mu_\theta) + n^{-1/2}\eta_n\},$$ where $\eta_n > 0$ is any sequence such that $\eta_n = o_{\mathbb{P}}(1)$ and $M_n$ is non-empty. That is, $M_n$ is a set of approximate MWEs. 

Consider the following assumption:
\begin{assumption}\label{as:sup:wellsep_additional}
There exists a neighborhood $N$ of $\theta_\star$ and a constant $c_\star>0$ such that for any $\theta \in N$, 
$$\was_1(\mu_{\theta},\mu_\star) \geq \was_1(\mu_{\theta_\star},\mu_\star) + c_\star \|\theta-\theta_\star\|_\mathcal{H}.$$
\end{assumption}
In the well-specified setting, this condition follows from Assumption \ref{as:sup:diff}. The next result concerns the distribution of the set $M_n$ as $n$ becomes large.

\begin{theorem}\label{theorem:asymptotic:misspec}
Suppose Assumptions \ref{as:sup:cvgwas}-\ref{as:sup:wellsep_additional} hold for some $\theta_\star$ in the interior of $\mathcal{H}$, and that the conditions of either Theorem \ref{theorem:delbarrio} or Theorem \ref{theorem:dede} are satisfied.
Then, there exist positive real numbers $\beta_n \to 0$ such that
\begin{enumerate}
\item $\mathbb{P}_\star\left(\{M_n \subset \theta_\star + n^{-1/2}K(H_n,\beta_n)\}\right) \to 1$ as $n \to \infty$, 
where $\mathbb{P}_\star$ denotes inner probability, and
\item if $F_n$ and $G_\star$ are versions of the processes such that $\sqrt{n}(F_n-F_\star)\to G_\star$ in $L_1(\mathbb{R})$ almost surely, then $d_H\left(K(H_n^\star,0),K(H_n,\beta_n)\right) = o_\mathbb{P}(1).$
\end{enumerate}
\end{theorem}
Since $K(H_n^\star,0) = \argmin_{u} \|G_\star + \sqrt{n}(F_\star - F_{\theta_\star}) - \langle u, D_{\theta_\star}\rangle\|_{L_1},$ one can interpret this result as saying that the limit of the set of approximate MWEs $M_n$ behaves distributionally like the limit of the sets $\theta_\star + n^{-1/2}\argmin_{u} \|G_\star + \sqrt{n}(F_\star - F_{\theta_\star}) - \langle u, D_{\theta_\star}\rangle\|_{L_1}$ in the Hausdorff metric sense. Note that the latter sequence does not depend on the data. Since the assumptions guarantee that $\sqrt{n}(F_n-F_\star)\to G_\star$ weakly in $L_1(\mathbb{R})$, there exist versions of these variables that converge almost surely. For simplicity, we assume without loss of generality that these are the variable we work with. As noted by \citet{pollard1980}, establishing the measurability of the sets $\{M_n \subset \theta_\star + n^{-1/2}K(H_n,\beta_n)\} \subset \Omega$ is hard, which is why the result is stated in terms of inner probability. See also \citet[][pp. 67]{pollard1980} for further comments on the sequence $\beta_n$.

\begin{proof}[Proof of Theorem \ref{theorem:asymptotic:misspec}]
Let $\theta \in N$, where $N$ is the set from Assumption \ref{as:sup:wellsep_additional}. By Assumption \ref{as:sup:wellsep} and \citet[][Lemma 4.1]{pollard1980} or the proof of Theorem \ref{theorem:sup:consistent}, we know that the minimizers of $\|F_n-F_\theta\|_{L_1}$ will be attained in $N$ with probability going to one. 
For $\theta \in N$, we have that
\begin{align*}
\|F_n - F_\theta\|_{L_1} &\geq \|F_\star - F_\theta\|_{L_1} - \|F_n - F_\star\|_{L_1} \quad \text{by the triangle inequality,} \\
&\geq \|F_\star - F_{\theta_\star}\|_{L_1} + c_\star\|\theta - \theta_\star\|_{\mathcal{H}} - \|F_n - F_\star\|_{L_1} \quad \text{by Assumption \ref{as:sup:wellsep_additional},}\\
&\geq \|F_n - F_{\theta_\star}\|_{L_1} + c_\star\|\theta - \theta_\star\|_{\mathcal{H}} - 2\|F_n - F_\star\|_{L_1} \quad \text{by the triangle inequality.} 
\end{align*}
Let $\xi_n = \sqrt{n}(4\|F_n - F_\star\|_{L_1} + 2\eta_n)/c_\star$ and $S_n = \{\theta : \sqrt{n}\|\theta - \theta_\star\|_{\mathcal{H}} \leq \xi_n\}$. Then, by the assumptions on $\eta_n$ and $\sqrt{n}(F_n-F_\star)$,  we know that $n^{-1/2}\xi_n = o_\mathbb{P}(1)$. If $\theta \in N \cap S_n^c$, then by the inequality derived above, $\|F_n-F_\theta\|_{L_1} > \|F_n - F_{\theta_\star}\|_{L_1} + 2 (\|F_n - F_\star\|_{L_1} + \eta_n)$. Thus, with inner probability going to one, it has to be that $M_n \subset S_n$.

Next, we approximate $\theta \mapsto \sqrt{n}\|F_n - F_\theta\|_{L_1}$ with the convex map $\theta\mapsto \sqrt{n}\|F_n - F_{\theta_\star} - \langle \theta - \theta_\star, D_{\theta_\star}\rangle \|_{L_1}$ over the set $S_n$. First, note that Assumption \ref{as:sup:diff} implies that the remainder $R_\theta = F_\theta - F_{\theta_\star} - \langle \theta -\theta_\star, D_{\theta_\star}\rangle$ satisfies $$\|R_\theta\|_{L_1} \leq \|\theta - \theta_\star\|_{\mathcal{H}}\cdot \Delta(\|\theta - \theta_\star\|_{\mathcal{H}}),$$ where $\Delta$ is an increasing function such that $\Delta(t) = o(1)$ as $t \to 0$. Define $$\Gamma_n = \sup_{\theta\in S_n} \left| \sqrt{n}\|F_n - F_\theta\|_{L_1} - \sqrt{n}\|F_n - F_{\theta_\star} - \langle \theta - \theta_\star, D_{\theta_\star}\rangle \|_{L_1}\right|.$$ We then have that
\begin{align*}
\Gamma_n &= \sup_{\theta\in S_n} \left| \sqrt{n}\|F_n  - F_{\theta_\star} - \langle \theta - \theta_\star,  D_{\theta_\star}\rangle - R_\theta\|_{L_1} - \sqrt{n}\|F_n - F_{\theta_\star} - \langle \theta - \theta_\star, D_{\theta_\star}\rangle \|_{L_1}\right| \\
&\leq \sup_{\theta\in S_n} \sqrt{n} \|R_\theta\|_{L_1} \quad \text{by the triangle inequality,} \\
&\leq \sup_{\theta\in S_n} \sqrt{n} \|\theta - \theta_\star\|_{\mathcal{H}}\cdot \Delta(\|\theta - \theta_\star\|_{\mathcal{H}}) \quad \text{by Assumption \ref{as:sup:diff},} \\
& \leq \xi_n \Delta \left(\frac{\xi_n}{\sqrt{n}}\right) = o_{\mathbb{P}}(1) \quad \text{by the definitions of $S_n$ and $\xi_n$}.
\end{align*}
Hence, we have uniform control over the difference between $\theta \mapsto \sqrt{n}\|F_n - F_\theta\|_{L_1}$ and its convex approximation over $S_n$. Moreover, the map $\theta\mapsto \sqrt{n}\|F_n - F_{\theta_\star} - \langle \theta - \theta_\star, D_{\theta_\star}\rangle \|_{L_1}$ also attains its minimum on $S_n$ with probability going to one, since for $\theta \in N$ such that $\Delta(\|\theta - \theta_\star\|_{\mathcal{H}}) \leq c_\star/2$,
\begin{align*}
\|F_n - F_{\theta_\star} - \langle \theta &- \theta_\star, D_{\theta_\star}\rangle \|_{L_1} = \|F_n - F_{\theta} + R_\theta\|_{L_1}\\
& \geq \|F_n - F_{\theta}\|_{L_1} - \|R_\theta\|_{L_1} \quad \text{by the triangle inequality,}\\
& \geq \|F_n - F_{\theta_\star}\|_{L_1} + c_\star\|\theta - \theta_\star\|_{\mathcal{H}} - 2\|F_n - F_\star\|_{L_1} \\
&\quad - \|\theta - \theta_\star\|_{\mathcal{H}}\cdot \Delta(\|\theta - \theta_\star\|_{\mathcal{H}}) \quad \text{by Ass. \ref{as:sup:wellsep_additional}, tri. ineq.,}\\
& \geq \|F_n - F_{\theta_\star}\|_{L_1} + \frac{1}{2}c_\star\|\theta - \theta_\star\|_{\mathcal{H}} - 2\|F_n - F_\star\|_{L_1}.
\end{align*}
Hence, if $\theta \in N \cap S_n^c$ and $\Delta(\|\theta - \theta_\star\|_{\mathcal{H}}) \leq c_\star/2$, then 
$$\|F_n-F_\theta\|_{L_1} > \|F_n - F_{\theta_\star}\|_{L_1} +   \eta_n = \|F_n-F_{\theta_\star} -\langle 0, D_{\theta_\star}\rangle \|_{L_1} + \eta_n.$$
In other words, $m(H_n) = \inf_{u \in L_n}f(H_n,u)$ with probability going to one, where we have used the reparameterization $L_n = \{u: u = \sqrt{n}(\theta-\theta_\star), \theta \in S_n\}$, or equivalently $S_n = \theta_\star + n^{-1/2}L_n$. 

Now, since $\Gamma_n = o_\mathbb{P}(1)$, we can find a sequence of positive real numbers $\gamma_n \to 0$ such that $\mathbb{P}(\Gamma_n \leq \gamma_n) \to 1$. Similarly, we can find $\delta_n >0$ and $\alpha_n >0$ such that $\mathbb{P}(\eta_n \leq \delta_n) \to 1$ and $\mathbb{P}(\|H_n - H_n^\star\|_{L_1} \leq \alpha_n) \to 1$. Define $\beta_n = \max\{2\gamma_n +\delta_n, 2\alpha_n\}$. Let $\tau$ be such that $\tau \in L_n$ and $\theta_\star + n^{-1/2}\tau \in M_n$, and suppose that $\Gamma_n\leq \gamma_n$ and $\eta_n\leq \delta_n$. By combining the approximations developed above, we have that
\begin{align*}
m(H_n) &\geq \inf_{u\in L_n}\sqrt{n}\|F_n - F_{\theta_\star + n^{-1/2}u}\|_{L_1} - \gamma_n \\
& \geq \sqrt{n}\|F_n - F_{\theta_\star + n^{-1/2}\tau}\|_{L_1} - \gamma_n - \delta_n \\
& \geq f(H_n, \tau) - 2\gamma_n -\delta_n.
\end{align*}
Since $2\gamma_n + \delta_n \leq \beta_n$, we have that $\tau \in K(H_n,\beta_n)$. This proves the first part of the theorem, as the events considered above all hold with (inner) probability going to one.

By the triangle inequality, $u \in K(H_n^\star,0)$ implies that $u \in K(H_n, 2\|H_n-H_n^\star\|_{L_1})$. Hence, with probability going to one,
$K(H_n^\star,0) \subset K(H_n,\beta_n)$. Similarly, $u \in K(H_n, \beta_n)$ implies that $u \in K(H_n^\star, \beta_n + 2\|H_n-H_n^\star\|_{L_1})$. Recall that $\beta_n + 2\|H_n-H_n^\star\|_{L_1} \to 0$ almost surely. Let $E\subset \Omega$ denote the set on which this occurs. Then, for every  every $\delta >0$, there exists $n(\omega)$ such that for $n \geq n(\omega)$, $K(H_n(\omega),\beta_n) \subset K(H_n^\star(\omega),0)^\delta$. By the definition of the Hausdorff metric, these set inclusions imply that $$d_H\left(K(H_n^\star,0),K(H_n,\beta_n)\right) = o_\mathbb{P}(1).$$
\end{proof}

\subsubsection{Differentiability condition}
The  condition in Assumption \ref{as:sup:diff} can sometimes be established from more familiar concepts of differentiability, such as differentiability in quadratic mean \citep{lecam1970}. The following proposition gives such a result. 
Suppose that the model family is absolutely continuous with respect to the Lebesgue measure $\lambda$ on $\mathbb{R}$, and denote the density $d\mu_\theta/d\lambda$ of $\mu_\theta$ by $ f_\theta$. Let $\xi_\theta(y) = \sqrt{f_\theta(y)}$ for all $y\in\mathbb{R}$. \citet{lecam1970} introduced the concept of differentiability in quadratic mean, which we define below.
\begin{definition}
    The model $\mathcal{M}$ is differentiable in quadratic mean at $\theta_\star$ if there exists 
    $\dot{\xi}_{\theta_\star} \in (L_2(\mathbb{R}))^{d_\theta}$ and $R_{\theta-\theta_\star} \in (L_2(\mathbb{R}))^{d_\theta}$ such that $\xi_\theta = \xi_{\theta_\star}+\langle \theta - \theta_\star, \ \dot{\xi}_{\theta_\star}\rangle + R_{\theta-\theta_\star},$ where $[\int_\mathbb{R} R^2_{\theta-\theta_\star}(y)dy]^{1/2} = o(\|\theta-\theta_\star\|_\mathcal{H})$ as $\|\theta-\theta_\star\|_\mathcal{H}\to 0$.
\end{definition}

Differentiability in quadratic mean holds for many classical models, such as exponential families and many location-scale families \citep[see e.g. Section 12.2 in][]{lehmann2005}.

\begin{proposition}
Suppose that the model family is supported on a set $S\subset \mathbb{R}$ of
bounded Lebesgue measure, and that it is differentiable in quadratic mean at
${\theta_\star}$. Let $$D_{\theta_\star}(t) = \int_{-\infty}^t
2\xi_{\theta_\star}(y)\dot{\xi}_{\theta_\star}(y) dy$$
for $t\in S$ and zero elsewhere. Then, as $\|\theta-\theta_\star\|_\mathcal{H}\to 0$,
$$\int_\mathbb{R}\lvert F_\theta(t)-F_{\theta_\star}(t) - \langle \theta-\theta_\star,
D_{\theta_\star}(t) \rangle \rvert dt = o(\|\theta-\theta_\star\|_\mathcal{H}).$$
\end{proposition}
\begin{proof} Consider
\begin{align*}
&\int_\mathbb{R}\lvert F_\theta(t)-F_{\theta_\star}(t) - \langle \theta-\theta_\star, D_{\theta_\star}(t) \rangle \rvert dt \\ 
&= \int_S \left\lvert \int_{-\infty}^t \xi^2_{\theta}(y) - \xi^2_{\theta_\star}(y) -  2\xi_{\theta_\star}(y)\langle \theta-\theta_\star,\dot{\xi}_{\theta_\star}(y) \rangle dy \right \rvert dt \\
&\leq \int_S \int_\mathbb{R} \lvert \xi^2_{\theta}(y) - \xi^2_{\theta_\star}(y) -  2\xi_{\theta_\star}(y)\langle \theta-\theta_\star,\dot{\xi}_{\theta_\star}(y) \rangle \rvert dy dt \\
&\leq c \int_\mathbb{R} \langle \theta - \theta_\star,  \dot{\xi}_{\theta_\star}(y)\rangle^2 + R^2_{\theta-\theta_\star}(y) + 2\lvert \xi_{\theta_\star}(y) R_{\theta-\theta_\star}(y)\rvert + 2\lvert \langle \theta - \theta_\star,  \dot{\xi}_{\theta_\star}(y)\rangle R_{\theta-\theta_\star}(y)\rvert dy \\
& = o(\| \theta-\theta_\star\|_\mathcal{H}),
\end{align*}
where $c$ is some constant and the last equality follows by applying the Cauchy-Schwarz inequality to the two last terms of the integrand.
\end{proof}

\section{Proofs: MEWE}
\subsection{Existence, measurability, and consistency}
In order to show similar results for the MEWE, we introduce the following assumptions.

\begin{assumption}\label{as:sup:weakcvg_intermediate}
For any $m\geq 1$, if $\rho_\mathcal{H}(\theta_n,\theta) \to 0$, then $\mu_{\theta_n}^{(m)}  \Rightarrow \mu_\theta^{(m)}$ as $n \to \infty$.
\end{assumption}

\begin{assumption}\label{as:sup:unicvg}
If $\rho_\mathcal{H}(\theta_n,\theta) \to 0$, then $\mathbb{E}_n\was_p(\mu_{\theta_n},\hat{\mu}_{\theta_n,n}) \to 0$ as $n \to \infty$.
\end{assumption}

Assumption \ref{as:sup:weakcvg_intermediate} is a slightly stronger version of Assumption \ref{as:sup:weakcvg}, stating that we not only need weak convergence of the ``model'' distributions $\mu_\theta$, but also of the sample distributions $\mu_\theta^{(m)}$ for any $m\geq 1$. Assumption \ref{as:sup:unicvg} is implied by $\sup_{\theta\in\mathcal{H}}\mathbb{E}_n\was_p(\mu_{\theta},\hat{\mu}_{\theta,n}) \to 0$, which in turn might hold when $\mathcal{H}$ is compact and the inequalities in \citet{fournier_guillin2015} hold. In the next result, we prove an analogous version of Theorem \ref{theorem:sup:consistent} for the MEWE as $\min\{n,m\} \to \infty$. For simplicity, we write $m$ as a function of $n$ and require that $m(n) \to \infty$ as $n\to \infty$.

\begin{theorem} \label{theorem:sup:consistent:mewe}
Under Assumptions \ref{as:sup:cvgwas}-\ref{as:sup:relativelycompact} and \ref{as:sup:weakcvg_intermediate}-\ref{as:sup:unicvg}, there exists a set $E\subset \Omega$
with $\mathbb{P}(E)=1$ such that, for all $\omega \in E$,
$$\inf_{\theta\in\mathcal{H}} \mathbb{E}_{m(n)} \was_p(\hat{\mu}_n(\omega),\hat{\mu}_{\theta,m(n)}) \to \inf_{\theta\in\mathcal{H}} \was_p(\mu_\star,\mu_\theta),$$
and there exists $n(\omega)$ such that, for all $n\geq n(\omega)$, the sets $\argmin_{\theta\in\mathcal{H}} \was_p(\hat{\mu}_n(\omega),\hat{\mu}_{\theta,m(n)})$ are non-empty and form a bounded sequence with 
$$\limsup_{n\to\infty} \argmin_{\theta\in\mathcal{H}} \mathbb{E}_{m(n)} \was_p(\hat{\mu}_n(\omega),\hat{\mu}_{\theta,m(n)}) \subset \argmin_{\theta\in\mathcal{H}} \was_p(\mu_{\star},\mu_\theta).$$
\end{theorem}

\begin{proof}[Proof of Theorem \ref{theorem:sup:consistent:mewe}]
For any $\nu\in\mathcal{P}(\mathcal{Y})$, lower semicontinuity of the map $\theta \mapsto \was_p(\nu,\mu_\theta)$ follows from Lemma \ref{lemma:continuity}, via Assumption \ref{as:sup:weakcvg}. Since $\inf_{\theta \in \mathcal{H}}\was_p(\mu_\star,\mu_\theta) = \varepsilon_\star$, $B_\star(\varepsilon)$ with the $\varepsilon$ of Assumption \ref{as:sup:relativelycompact} is non-empty, by definition of the infimum. Moreover, since  $\theta \mapsto \was_p(\mu_\star,\mu_\theta)$ is lower semicontinuous, the set $B_\star(\varepsilon)$ is closed. By Assumption \ref{as:sup:relativelycompact}, $B_\star(\varepsilon)$ is therefore compact. In other words, again by lower semicontinuity, the set $\argmin _{\theta \in \mathcal{H}}\was_p(\mu_\star,\mu_\theta)$ is non-empty.

    We show that $\theta \mapsto \mathbb{E}_{m(n)} \was_p(\hat{\mu}_n,\hat{\mu}_{\theta,m(n)})$ epi-converges to $\theta\mapsto \was_p(\mu_\star,\mu_\theta)$ $\mathbb{P}$-almost surely. 
    Let $E$ denote the set of probability one from Assumption \ref{as:sup:cvgwas} and let $\omega \in E$. Fix $\mathcal{K} \subset \mathcal{H}$ compact. By lower semicontinuity of $\theta\mapsto \mathbb{E}_{m(n)} \was_p(\hat{\mu}_n(\omega),\hat{\mu}_{\theta,m(n)})$, ensured by Lemma \ref{lemma:continuity2} and Assumption \ref{as:sup:weakcvg_intermediate}, we know that $$\inf_{\theta\in\mathcal{K}} \mathbb{E}_{m(n)} \was_p(\hat{\mu}_n(\omega),\hat{\mu}_{\theta,m(n)}) = \mathbb{E}_{m(n)} \was_p(\hat{\mu}_n(\omega),\hat{\mu}_{\theta_n,m(n)}),$$ 
for some sequence $\theta_n = \theta_n(\omega) \in \mathcal{K}$. Hence,
\begin{align*}
&\liminf_{n\to \infty} \inf_{\theta\in\mathcal{K}} \mathbb{E}_{m(n)}\was_p(\hat{\mu}_n(\omega),\hat{\mu}_{\theta,m(n)}) \\ 
&= \liminf_{n\to \infty}\mathbb{E}_{m(n)} \was_p(\hat{\mu}_n(\omega),\hat{\mu}_{\theta_n,m(n)}) \\
& = \lim_{k\to \infty} \mathbb{E}_{m(n_k)} \was_p(\hat{\mu}_{n_k}(\omega),\hat{\mu}_{\theta_{n_k},m(n_k)}) \quad \text{$\exists$ subsequence converging to the $\liminf$,} \\
& = \lim_{\ell \to \infty} \mathbb{E}_{m(n_{k_\ell})} \was_p(\hat{\mu}_{n_{k_\ell}}(\omega),\hat{\mu}_{\theta_{n_{k_\ell}},m({n_{k_\ell}})}) \quad \text{$\exists$ subseq. $\theta_{n_{k_\ell}} \to \bar{\theta}\in\mathcal{K}$ by compactness,} \\ 
& = \liminf_{\ell \to \infty} \mathbb{E}_{m(n_{k_\ell})} \was_p(\hat{\mu}_{n_{k_\ell}}(\omega),\hat{\mu}_{\theta_{n_{k_\ell}},m({n_{k_\ell}})}) \\
& \geq \liminf_{\ell \to \infty} [\was_p(\hat{\mu}_{n_{k_\ell}}(\omega),\mu_{\theta_{n_{k_\ell}}}) - \mathbb{E}_{m(n_{k_\ell})} \was_p(\mu_{\theta_{n_{k_\ell}}},\hat{\mu}_{\theta_{n_{k_\ell}},m({n_{k_\ell}})})] \quad \text{by the triangle ineq.},\\
& \geq \liminf_{\ell \to \infty}  \was_p(\hat{\mu}_{n_{k_\ell}}(\omega),\mu_{\theta_{n_{k_\ell}}}) -  \limsup_{\ell \to\infty}\mathbb{E}_{m(n_{k_\ell})} \was_p(\mu_{\theta_{n_{k_\ell}}},\hat{\mu}_{\theta_{n_{k_\ell}},m({n_{k_\ell}})}) \\
& \geq \was_p(\mu_\star,\mu_{\bar{\theta}})  \quad \text{by l.s.c., Assumptions \ref{as:sup:cvgwas}, \ref{as:sup:weakcvg}, \ref{as:sup:unicvg}, and $\omega \in E$},\\
& \geq \inf_{\theta\in\mathcal{K}}\was_p(\mu_\star,\mu_{\theta}).
\end{align*}

Fix $\mathcal{O} \subset \mathcal{H}$ open. By definition of the infimum, there exists a sequence $\theta_n \in\mathcal{O}$ such that $\was_p(\mu_\star,\mu_{\theta_n})\to \inf_{\theta\in\mathcal{O}}\was_p(\mu_\star,\mu_\theta)$. Now, $$\inf_{\theta\in\mathcal{O}} \mathbb{E}_{m(n)}\was_p(\hat{\mu}_n(\omega),\hat{\mu}_{\theta,m(n)}) \leq \mathbb{E}_{m(n)}\was_p(\hat{\mu}_n(\omega),\hat{\mu}_{\theta_n,m(n)}).$$
Hence,
\begin{align*}
&\limsup_{n\to \infty} \inf_{\theta\in\mathcal{O}}  \mathbb{E}_{m(n)}\was_p(\hat{\mu}_n(\omega),\hat{\mu}_{\theta,m(n)}) \\
&\leq \limsup_{n\to \infty}\mathbb{E}_{m(n)}\was_p(\hat{\mu}_n(\omega),\hat{\mu}_{\theta_n,m(n)})\\
& \leq \limsup_{n\to \infty} [\was_p(\hat{\mu}_n(\omega),\mu_\star) +\was_p(\mu_\star,\mu_{\theta_n}) + \mathbb{E}_{m(n)}\was_p(\mu_{\theta_n},\hat{\mu}_{\theta_n,m(n)})] \quad \text{by  tri.~ineq.}, \\
&=  \limsup_{n\to \infty} \was_p(\mu_\star,\mu_{\theta_n})  \quad \text{by Assumptions \ref{as:sup:cvgwas} and \ref{as:sup:unicvg},  $\omega \in E$},\\
& = \inf_{\theta\in\mathcal{O}}\was_p(\mu_\star,\mu_\theta) \quad \text{by definition of $\theta_n$.}
\end{align*}

Theorem 7.29b) of \citet{rockafellar2009variational} implies that 
\[\limsup_{n\to\infty} (\inf_{\theta\in\mathcal{H}}  \mathbb{E}_{m(n)} \was_p(\hat{\mu}_n(\omega),\hat{\mu}_{\theta,m(n)}) )\leq \inf_{\theta\in\mathcal{H}} \was_p(\mu_\star,\mu_\theta) = \varepsilon_\star.\]
Hence, for all $\alpha > 0$, there exists $n_\alpha(\omega)$, such that $n\geq n_\alpha(\omega)$ implies that $$\inf_{\theta\in\mathcal{H}}  \mathbb{E}_{m(n)} \was_p(\hat{\mu}_n(\omega),\hat{\mu}_{\theta,m(n)})\leq \varepsilon_\star + \alpha.$$ 
Let $\alpha \in (0,\varepsilon/3)$. The set $\{\theta\in\mathcal{H}:  \mathbb{E}_{m(n)} \was_p(\hat{\mu}_n(\omega),\hat{\mu}_{\theta,m(n)}) \leq \varepsilon_\star +\varepsilon/3\}$ is non-empty for $n\geq n_\alpha(\omega)$, by definition of the infimum. Let $\theta$ belong to this set. Then, by the triangle inequality, 
$$\was_p(\mu_\star,\mu_\theta) \leq \was_p(\hat{\mu}_n(\omega),\mu_\star) + \mathbb{E}_{m(n)}\was_p(\hat{\mu}_n(\omega),\hat{\mu}_{\theta,m(n)}) + \mathbb{E}_{m(n)}\was_p(\mu_\theta,\hat{\mu}_{\theta,m(n)}).$$
By Assumption \ref{as:sup:cvgwas}, there exists an $n_\varepsilon(\omega)$ such that for $n\geq n_\varepsilon(\omega)$, $\was_p(\hat{\mu}_n(\omega),\mu_\star)\leq \varepsilon/3$. By Assumption \ref{as:sup:unicvg}, there exists an $\hat{n}(\omega)$ such that for any $n\geq \hat{n}(\omega)$, we have $\mathbb{E}_{m(n)}\was_p(\mu_\theta,\hat{\mu}_{\theta,m(n)}) \leq \varepsilon/3$.  So, if $n\geq \max\{n_\alpha(\omega),n_\varepsilon(\omega),\hat{n}(\omega)\}$, we have $\was_p(\mu_\star,\mu_\theta) \leq \varepsilon_\star +\varepsilon$. This means that $$\{\theta\in\mathcal{H}: \mathbb{E}_{m(n)} \was_p(\hat{\mu}_n(\omega),\hat{\mu}_{\theta,m(n)}) \leq \varepsilon_\star +\varepsilon/3\} \subset  B_\star(\varepsilon).$$
As a consequence, for $n\geq \max\{n_\alpha(\omega),n_\varepsilon(\omega),\hat{n}(\omega)\}$, $$\inf_{\theta\in\mathcal{H}}  \mathbb{E}_{m(n)} \was_p(\hat{\mu}_n(\omega),\hat{\mu}_{\theta,m(n)})  = \inf_{\theta\in B_\star(\varepsilon)} \mathbb{E}_{m(n)} \was_p(\hat{\mu}_n(\omega),\hat{\mu}_{\theta,m(n)}).$$

By Theorem 7.31a) of \citet{rockafellar2009variational}, we have 
$$\inf_{\theta\in\mathcal{H}}  \mathbb{E}_{m(n)}
\was_p(\hat{\mu}_n(\omega),\hat{\mu}_{\theta,m(n)}) \to
\inf_{\theta\in\mathcal{H}} \was_p(\mu_\star,\mu_\theta).$$  Also, for $n\geq
\max\{n_\alpha(\omega),n_\varepsilon(\omega),\hat{n}(\omega)\}$ and by the same
reasoning as for the map $\theta\mapsto \was_p(\mu_\star,\mu_\theta)$, the sets
$\argmin_{\theta\in\mathcal{H}} \mathbb{E}_{m(n)}
\was_p(\hat{\mu}_n(\omega),\hat{\mu}_{\theta,m(n)})$ are non-empty. By Theorem
7.31b) of \citet{rockafellar2009variational}, the result follows. The same
argument holds for the sets $$\varepsilon_n\mhyphen\argmin_{\theta\in\mathcal{H}}
\mathbb{E}_{m(n)} \was_p(\hat{\mu}_n(\omega),\hat{\mu}_{\theta,m(n)})$$ with
$\varepsilon_n\to 0$, since $\inf_{\theta\in\mathcal{H}}
\mathbb{E}_{m(n)} \was_p(\hat{\mu}_n(\omega),\hat{\mu}_{\theta,m(n)}) + \varepsilon_n
\leq \varepsilon_\star + \alpha$ eventually.

\end{proof}

\subsection{Convergence to the MWE}
The next result considers the case where the data and $n$ is fixed, while $m\to \infty$. It shows that the MEWE converges to the MWE, assuming the latter exists. We summarize this condition in the following assumption,  in which the observed empirical distribution is kept fixed and $\varepsilon_n = \inf_{\theta \in \mathcal{H}}\was_p(\hat{\mu}_n,\mu_\theta)$.

\begin{assumption}\label{as:sup:bounded:fixedn}
For some $\varepsilon > 0$, the set $B_n(\varepsilon) = \{\theta\in\mathcal{H} : \was_p(\hat{\mu}_n,\mu_\theta)  \leq \varepsilon_n +\varepsilon\}$ is bounded.
\end{assumption}

\begin{theorem} \label{theorem:sup:mewe_to_mwe}
Under Assumptions \ref{as:sup:weakcvg} and \ref{as:sup:weakcvg_intermediate}-\ref{as:sup:bounded:fixedn}, 
$$\inf_{\theta\in\mathcal{H}} \mathbb{E}_m \was_p(\hat{\mu}_n,\hat{\mu}_{\theta,m}) \to \inf_{\theta\in\mathcal{H}} \was_p(\hat{\mu}_n,\mu_\theta),$$ 
and there exists an $\hat{m}$ such that, for all $m\geq \hat{m}$, the sets $\argmin_{\theta\in\mathcal{H}} \mathbb{E}_m\was_p(\hat{\mu}_n,\hat{\mu}_{\theta,m})$ are non-empty and form a bounded sequence with 
$$\limsup_{m\to\infty} \argmin_{\theta\in\mathcal{H}} \mathbb{E}_m \was_p(\hat{\mu}_n,\hat{\mu}_{\theta,m}) \subset \argmin_{\theta\in\mathcal{H}} \was_p(\hat{\mu}_{n},\mu_\theta).$$
\end{theorem}

\begin{proof}[Proof of Theorem \ref{theorem:sup:mewe_to_mwe}]
Lower semicontinuity of the map $\theta \mapsto \was_p(\hat{\mu}_n,\mu_\theta)$ follows from Lemma \ref{lemma:continuity}, via Assumption \ref{as:sup:weakcvg}. Since $\inf_{\theta \in \mathcal{H}}\was_p(\hat{\mu}_n,\mu_\theta) = \varepsilon_n$, $B_n(\varepsilon)$  with the $\varepsilon$ of Assumption \ref{as:sup:relativelycompact} is non-empty, by definition of the infimum. Moreover, since  $\theta \mapsto\was_p(\hat{\mu}_n,\mu_\theta)$ is lower semicontinuous, the set $B_n(\varepsilon)$ is closed. By Assumption \ref{as:sup:bounded:fixedn}, $B_n(\varepsilon)$ is therefore compact. In other words, by lower semicontinuity, the set $\argmin _{\theta \in \mathcal{H}}\was_p(\hat{\mu}_n,\mu_\theta)$ is non-empty.

We show that $\theta \mapsto \mathbb{E}_m \was_p(\hat{\mu}_n,\hat{\mu}_{\theta,m})$ epi-converges to $\theta\mapsto \was_p(\hat{\mu}_n,\mu_\theta)$ as $m \to \infty$. Fix $\mathcal{K} \subset \mathcal{H}$ compact. By lower semicontinuity of $\theta\mapsto \mathbb{E}_m \was_p(\hat{\mu}_n,\hat{\mu}_{\theta,m})$, ensured by Lemma \ref{lemma:continuity2} and Assumption \ref{as:sup:weakcvg_intermediate}, we know that $$\inf_{\theta\in\mathcal{K}} \mathbb{E}_m \was_p(\hat{\mu}_n,\hat{\mu}_{\theta,m}) = \mathbb{E}_m \was_p(\hat{\mu}_n(\omega),\hat{\mu}_{\theta_m,m}),$$
for some sequence $\theta_m \in \mathcal{K}$. Hence,
\begin{align*}
\liminf_{m\to \infty} &\inf_{\theta\in\mathcal{K}} \mathbb{E}_m\was_p(\hat{\mu}_n,\hat{\mu}_{\theta,m}) \\ 
&= \liminf_{m\to \infty}\mathbb{E}_m \was_p(\hat{\mu}_n,\hat{\mu}_{\theta_m,m}) \\
& = \lim_{k\to \infty} \mathbb{E}_{m_k} \was_p(\hat{\mu}_n,\hat{\mu}_{\theta_{m_k},m_k}) \quad \text{$\exists$ subsequence converging to the $\liminf$,} \\
& = \lim_{\ell \to \infty} \mathbb{E}_{m_{k_\ell}} \was_p(\hat{\mu}_n,\hat{\mu}_{\theta_{m_{k_\ell}},m_{k_\ell}}) \quad \text{$\exists$ subseq. $\theta_{m_{k_\ell}} \to \bar{\theta}\in\mathcal{K}$ by compactness,} \\ 
& = \liminf_{\ell \to \infty} \mathbb{E}_{m_{k_\ell}} \was_p(\hat{\mu}_n,\hat{\mu}_{\theta_{m_{k_\ell}},m_{k_\ell}}) \\
& \geq \liminf_{\ell \to \infty} [\was_p(\hat{\mu}_n,\mu_{\theta_{m_{k_\ell}}}) - \mathbb{E}_{m_{k_\ell}} \was_p(\mu_{\theta_{m_{k_\ell}}},\hat{\mu}_{\theta_{m_{k_\ell}},m_{k_\ell}})] \quad \text{by the triangle ineq.},\\
& \geq \liminf_{\ell \to \infty}  \was_p(\hat{\mu}_n,\mu_{\theta_{m_{k_\ell}}}) -  \limsup_{\ell \to\infty}\mathbb{E}_{m_{k_\ell}} \was_p(\mu_{\theta_{m_{k_\ell}}},\hat{\mu}_{\theta_{m_{k_\ell}},m_{k_\ell}}) \\
& \geq \was_p(\hat{\mu}_n,\mu_{\bar{\theta}})  \quad \text{by l.s.c., Assumptions \ref{as:sup:weakcvg} and  \ref{as:sup:unicvg}},\\
& \geq \inf_{\theta\in\mathcal{K}}\was_p(\hat{\mu}_n,\mu_{\theta}).
\end{align*}

Fix $\mathcal{O} \subset \mathcal{H}$ open. By definition of the inf, there exists a sequence $\theta_m \in\mathcal{O}$ such that $\was_p(\hat{\mu}_n,\mu_{\theta_m})\to \inf_{\theta\in\mathcal{O}}\was_p(\hat{\mu}_n,\mu_\theta)$. Now, $\inf_{\theta\in\mathcal{O}} \mathbb{E}_m\was_p(\hat{\mu}_n,\hat{\mu}_{\theta,m}) \leq \mathbb{E}_m\was_p(\hat{\mu}_n,\hat{\mu}_{\theta_m,m})$. Hence,
\begin{align*}
\limsup_{m\to \infty}& \inf_{\theta\in\mathcal{O}} \mathbb{E}_m\was_p(\hat{\mu}_n,\hat{\mu}_{\theta,m}) \leq \limsup_{m\to \infty}\mathbb{E}_m\was_p(\hat{\mu}_n,\hat{\mu}_{\theta_m,m})\\
& \leq \limsup_{m\to \infty} [\was_p(\hat{\mu}_n,\mu_{\theta_m}) + \mathbb{E}_m\was_p(\mu_{\theta_m},\hat{\mu}_{\theta_m,m})] \quad \text{by the triangle inequality}, \\
&=  \limsup_{m\to \infty} \was_p(\hat{\mu}_n,\mu_{\theta_m})   \quad \text{by Assumption \ref{as:sup:unicvg}},\\
& = \inf_{\theta\in\mathcal{O}}\was_p(\mu_\star,\mu_\theta) \quad \text{by definition of $\theta_m$.}
\end{align*}

Theorem 7.29b) of \citet{rockafellar2009variational} implies that 
\[\limsup_{m\to\infty} (\inf_{\theta\in\mathcal{H}}  \mathbb{E}_m \was_p(\hat{\mu}_n,\hat{\mu}_{\theta,m}) )\leq \inf_{\theta\in\mathcal{H}} \was_p(\hat{\mu}_n,\mu_\theta) = \varepsilon_n.\]
Hence, for all $\alpha > 0$, there exists $m_\alpha$, such that for $m\geq m_\alpha$, $\inf_{\theta\in\mathcal{H}}  \mathbb{E}_m \was_p(\hat{\mu}_n,\hat{\mu}_{\theta,m})\leq \varepsilon_n + \alpha$. 
Let $\alpha \in (0,\varepsilon/2)$. The set 
$$\{\theta\in\mathcal{H}:  \mathbb{E}_m \was_p(\hat{\mu}_n,\hat{\mu}_{\theta,m}) \leq \varepsilon_n +\varepsilon/2\}$$
is non-empty for $m\geq m_\alpha$, by definition of the infimum. Let $\theta$ belong to this set. 
Then, by the triangle inequality, 
$$\was_p(\hat{\mu}_n,\mu_\theta) \leq \mathbb{E}_m \was_p(\hat{\mu}_n,\hat{\mu}_{\theta,m}) + \mathbb{E}_m\was_p(\mu_\theta,\hat{\mu}_{\theta,m}).$$
By Assumption \ref{as:sup:unicvg}, there exists an $\hat{m}$ such that for $m\geq \hat{m}$, $\mathbb{E}_m\was_p(\mu_\theta,\hat{\mu}_{\theta,m})\leq \varepsilon/2$. So, if $m\geq \max\{m_\alpha,\hat{m}\}$, we have that $\was_p(\mu_\star,\mu_\theta) \leq \varepsilon_n +\varepsilon$. This means that 
$$\{\theta\in\mathcal{H}: \mathbb{E}_m \was_p(\hat{\mu}_n,\hat{\mu}_{\theta,m}) \leq \varepsilon_n +\varepsilon/2\} \subset  B_n(\varepsilon).$$
Hence, for $m\geq \max\{m_\alpha,\hat{m}\}$, $\inf_{\theta\in\mathcal{H}}  \mathbb{E}_m \was_p(\hat{\mu}_n,\hat{\mu}_{\theta,m})  = \inf_{\theta\in B_n(\varepsilon)} \mathbb{E}_m \was_p(\hat{\mu}_n,\hat{\mu}_{\theta,m})$.

By Theorem 7.31a) of \citet{rockafellar2009variational}, we know that 
$$\inf_{\theta\in\mathcal{H}}  \mathbb{E}_m \was_p(\hat{\mu}_n,\hat{\mu}_{\theta,m}) \to \inf_{\theta\in\mathcal{H}} \was_p(\hat{\mu}_n,\mu_\theta)$$
as $m\to\infty.$ Also, for any  $m\geq \max\{m_\alpha,\hat{m}\}$, and by the same reasoning as for the map $\theta\mapsto \was_p(\hat{\mu}_n,\mu_\theta)$, the set $\argmin_{\theta\in\mathcal{H}} \mathbb{E}_m \was_p(\hat{\mu}_n,\hat{\mu}_{\theta,m})$ is non-empty. By Theorem 7.31b) of \citet{rockafellar2009variational}, the result follows. 
\end{proof}

\begin{theorem}[Measurability of the MEWE]  \label{theorem:sup:measurable:mewe}
Suppose that $\mathcal{H}$ is a $\sigma$-compact Borel measurable subset of $\mathbb{R}^{d_\theta}$. Under Assumption \ref{as:sup:weakcvg_intermediate}, for any $n \geq 1$ and $m\geq 1$ and $\varepsilon > 0$, there exists a Borel measurable function $\hat{\theta}_{n,m} : \Omega \to \mathcal{H}$ that satisfies
\begin{align*}
\hat{\theta}_{n,m}(\omega) \in
  \begin{cases}
    \argmin_{\theta\in\mathcal{H}} \mathbb{E}_m\was_p(\hat{\mu}_n(\omega),\hat{\mu}_{\theta,m}), & \text{if this set is non-empty,}  \\
    \varepsilon \mhyphen \argmin_{\theta\in\mathcal{H}}\mathbb{E}_m\was_p(\hat{\mu}_n(\omega),\hat{\mu}_{\theta,m}), & \text{otherwise}.
  \end{cases}
\end{align*}
\end{theorem}
\begin{proof}
The proof is identical to that of Theorem \ref{theorem:sup:measurable}, applying Lemma \ref{lemma:continuity2} instead of \ref{lemma:continuity}.
\end{proof}

\section{Checking the assumptions} \label{ch1:app:sec:checkassumptions}
The following proposition gives three data-generating mechanisms for which
$\was_p(\hat{\mu}_n, \mu_\star) \to 0$ $\mathbb{P}$-almost surely, which is Assumption \ref{as:sup:cvgwas}.
The three conditions below are mainly chosen for
illustrative purposes, and are by no means exhaustive.  We first give definitions
that are used in the conditions. We denote by $\mathcal{F}$ the measurable sets of $\Omega$.

\begin{definition}
The stochastic process $Y = (Y_t)_{t \in \mathbb{Z}}$ is stationary if for any $k \in \mathbb{N}$ and $\tau, t_1,\dots t_k \in \mathbb{Z}$ we have that $(Y_{t_1},\dots,Y_{t_k})$ and $(Y_{t_1+\tau},\dots,Y_{t_k+\tau})$
have the same distribution.
\end{definition}

\begin{definition}
The map $T:\Omega \to \Omega$ is $\mathbb{P}$-measure preserving if $\mathbb{P}(T^{-1}(A)) = \mathbb{P}(A)$ for all $A\in \mathcal{F}$.
\end{definition}

\begin{definition}
The map $T:\Omega \to \Omega$ is $\mathbb{P}$-ergodic if it is $\mathbb{P}$-measure preserving, and such that for all $A\in\mathcal{F}$ with $T^{-1}(A)=A$ we have that $\mathbb{P}(A)=0$ or $\mathbb{P}(A)=1$. The stochastic process $Y = (Y_t)_{t \in \mathbb{Z}}$ is ergodic if it can be represented by $Y_t = Y_0 \circ T^t$ for some ergodic $T$ and some random variable $Y_0$.
\end{definition}

\begin{definition}
The stochastic process $Y = (Y_t)_{t \in \mathbb{Z}}$ is $\alpha$-mixing with mixing coefficients 
$$\alpha_t = \sup_{k\in \mathbb{Z}}\sup_{A\in \mathcal{F}_{-\infty}^k, B \in \mathcal{F}_{k+t}^\infty} \lvert\mathbb{P}(A \cap B) - \mathbb{P}(A)\mathbb{P}(B)\rvert,$$
if $\alpha_t \to 0$ as $t \to \infty$, where $\mathcal{F}_{-\infty}^k = \sigma(Y_i : i \leq k)$ and $\mathcal{F}_{k}^\infty = \sigma(Y_i : i \geq k)$.
\end{definition}

\begin{proposition} \label{prop:empirical_as}
Suppose that $Y = (Y_t)_{t \in \mathbb{Z}}$ is a stochastic process such that either
\begin{enumerate}
\item $Y \sim \mu_{\star}^{\infty}$, for some $\mu_\star \in \mathcal{P}_p(\mathcal{Y})$, i.e. the observations are i.i.d, or
\item $(Y_t)_{t \in \mathbb{Z}}$ is ergodic and stationary, represented by $Y_t = Y_0 \circ T^t$, where $Y_0\sim \mu_{\star} \in \mathcal{P}_p(\mathcal{Y})$ and $T$ is an ergodic, measure preserving map, or
\item $(Y_t)_{t \in \mathbb{Z}}$ is $\alpha-$mixing with mixing coefficients $\alpha_t$ such that $\sum_{t=1}^\infty \alpha_t^{1-1/2r} < \infty$, with $Y_t \sim \mu_t$ such that $\mu_t$ converges weakly to $\mu_{\star}$ in $\mathcal{P}_p(\mathcal{Y})$ and satisfies $\sup_t \E\|Y_t\|_\mathcal{Y}^q < \infty$ for some $1\leq \max(r,p) < q < 2r$ (where it is assumed $\rho(x,y) = \|x-y\|_\mathcal{Y}$ for simplicity).
\end{enumerate}
Then there exists a set $E\in\mathcal{F}$ with $\mathbb{P}(E)=1$ such that, for all $\omega\in E$, $\was_p(\hat{\mu}_n(\omega),\mu_\star) \to 0$.
\end{proposition}
\begin{proof}
Under condition 1., Theorem 3 in \citet{varadarajan1958} establishes that there exists a set $E_1$ with $\mathbb{P}(E_1)=1$ such that for all $\omega\in E_1$, $\hat{\mu}_n(\omega)$ converges weakly to $\mu_\star$. By the strong law of large numbers, there exist a set $E_2$ with $\mathbb{P}(E_2)= 1$ and an $x_0 \in \mathcal{X}$ such that $\int_\mathcal{X} \rho(x,x_0)^p d\hat{\mu}_n(\omega)(x) \to \int_\mathcal{X} \rho(x,x_0)^p d\mu_\star(x)$ for all $\omega \in E_2$. Then, in light of Theorem \ref{theorem:metrize}, the claim holds on $E = E_1 \cap E_2$.

\vspace{12pt}
Consider condition 2. By \citet{varadarajan1958weak}, there exists a fixed countable set $C^{\star}$ of continuous and bounded functions on $\mathcal{Y}$, such that for any sequence of measures $\mu_n$ on $\mathcal{Y}$, $\mu_n$ converges weakly to $\mu$ if and only if $\int f d\mu_n \to \int f d\mu$ for all $f \in C^{\star}$.  Fix $f\in C^{\star}$. We know that $f\circ Y_0$ is measurable and that $\E\lvert f\circ Y_0 \rvert < \infty$ since $f$ is bounded, so by Birkhoff's ergodic theorem there exists a set $E_f$ such that $\mathbb{P}(E_f) =1$ and 
$$\int_\mathcal{Y} f d\hat{\mu}_n(\omega) = \frac{1}{n}\sum_{t=1}^nf(Y_t(\omega)) = \frac{1}{n}\sum_{t=1}^nf\circ Y_0 \circ T^t(\omega) \to \int_\mathcal{Y} f d\mu_{\star},$$
for all $\omega \in E_f$. Moreover, since $\mu_{\star} \in \mathcal{P}_p(\mathcal{Y})$ we know $\int_\mathcal{Y} \rho(y,y_0)^p d\mu_{\star}(y) < \infty$ and that there exists a set $E_0$ with $\mathbb{P}(E_0) = 1$ such that 
$$\int \rho(y,y_0)^p d\hat{\mu}_n(y)(\omega)   \to \int \rho(y,y_0)^pd\mu_{\star}(y),$$
for all $\omega\in E_0$. Since $C^\star$ is countable we know that $\mathbb{P}(\cap_{f\in C^\star} E_f \cap E_0) = 1$. In other words, this means that $\was_p(\hat{\mu}_n(\omega), \mu_{\star}) \to 0$ for all $\omega \in E = \cap_{f\in C^\star} E_f \cap E_0$.

\vspace{12pt}
Under condition 3., we first note that since $(Y_t)_{t \in \mathbb{Z}}$ is $\alpha-$mixing, then so is $(f \circ Y_t)_{t \in \mathbb{Z}}$ for any measurable $f$, with mixing coefficients bounded above by $\alpha_t$ since $\sigma(f(Y_i) : i\leq k) \subset \sigma(Y_i : i\leq k)$. Also, since $\mu_t$ converges weakly to $\mu_{\star}$ in $\mathcal{P}_p(\mathcal{Y})$ we have that for all $f \in C^{\star}$,
$$\frac{1}{n}\sum_{t=1}^n \int_\mathcal{Y} f d\mu_t \to \int_\mathcal{Y} fd\mu_{\star},$$
and 
$$\frac{1}{n}\sum_{t=1}^n \int_\mathcal{Y} \|y\|_\mathcal{Y}^p d\mu_t(y) \to \int_\mathcal{Y} \|y\|_\mathcal{Y}^pd\mu_{\star}(y).$$
By \citet{hansen1991} Corollary 4, we know that for all $f\in C^\star$ we have that the zero-mean, $\alpha$-mixing sequence $f(Y_t)-\int_\mathcal{Y} f d\mu_t$ satisfies
$$\frac{1}{n}\sum_{t=1}^n\left\{ f(Y_t)-\int_\mathcal{Y} f d\mu_t\right\} \to 0 \quad \text{$\mathbb{P}$-almost surely.}$$
Similarly, 
$$\frac{1}{n}\sum_{t=1}^n\left\{ \|Y_t\|_\mathcal{Y}^p-\int_\mathcal{Y} \|y\|_\mathcal{Y}^p d\mu_t(y) \right\}\to 0 \quad \text{$\mathbb{P}$-almost surely.}$$
Together this gives us that
$$\int_\mathcal{Y} fd\hat{\mu}_{n} = \frac{1}{n}\sum_{t=1}^n f(Y_t) \to \int_\mathcal{Y} fd\mu_{\star} \quad \text{$\mathbb{P}$-almost surely.}$$
and 
$$\int_\mathcal{Y} \|y\|_\mathcal{Y}^pd\hat{\mu}_{n}(y) =\frac{1}{n}\sum_{t=1}^n \|Y_t\|_\mathcal{Y}^p \to \int_\mathcal{Y} \|y\|_\mathcal{Y}^p d\mu_{\star}(y) \quad \text{$\mathbb{P}$-almost surely.}$$
Then, again by the countability of $C^\star$, we can conclude that $\was_p(\hat{\mu}_n(\omega),\mu_{\star}) \to 0$ for all $\omega$ in a set $E$ defined analogously to the one for the second set of conditions.
\end{proof}

The following proposition can be used to verify Assumption \ref{as:sup:wellsep}.

\begin{proposition}\label{prop:theta_star_exists_compact}
Suppose that either of the conditions of Lemma \ref{lemma:continuity} holds. Suppose that there exists a proper, connected and compact subset $\mathcal{S} \subset \mathcal{H}$ with  positive Lebesgue measure such that $\inf_{\theta\in\mathcal{H} \setminus \mathcal{S}}\was_p(\mu_\star,\mu_\theta)>\inf_{\theta\in\mathcal{H}}\was_p(\mu_\star,\mu_\theta).$ Then there exists a $\theta_\star$ attaining the infimum of $\theta \mapsto \was_p(\mu_\star,\mu_\theta)$. If $\theta_\star$ is unique, then it is well-separated.
\end{proposition}
\begin{proof}
Since $\theta\mapsto \was_p(\mu_\star, \mu_\theta)$ is continuous/lower semicontinuous, it attains a minimum $\theta_\star$ on $\mathcal{S}$. This is also the global minimum by the assumption on $\mathcal{S}$. If $\theta_\star$ is unique, it is well-separated in the sense of Assumption \ref{as:sup:wellsep}, for all $\varepsilon > 0$, there exists $\delta > 0$ such that
$$\inf_{\theta \in \mathcal{H}:\rho_{\mathcal{H}}(\theta,\theta_\star) \geq \varepsilon} \was_p(\mu_\star,\mu_\theta) >  \was_p(\mu_\star,\mu_{\theta_\star}) +\delta.$$
Indeed, let $\varepsilon>0$, and consider $\{\theta \in \mathcal{H}:\rho_{\mathcal{H}}(\theta,\theta_\star) \geq \varepsilon\}$.
Either the set is contained in $\mathcal{H}\setminus\mathcal{S}$, and thus well-separation follows,
or, $\{\theta \in \mathcal{H}:\rho_{\mathcal{H}}(\theta,\theta_\star) \geq \varepsilon\}\cap \mathcal{S}$ is not empty. Then we show that it is compact.
Since $\mathcal{S}$ is compact, there
exists $\bar{\varepsilon} \geq \varepsilon$ such that $\mathcal{S}\subset \{\theta \in \mathcal{H}:\rho_{\mathcal{H}}(\theta,\theta_\star)\leq \bar{\varepsilon}\}$. Therefore,
$$\{\theta \in \mathcal{H}:\rho_{\mathcal{H}}(\theta,\theta_\star) \geq
\varepsilon\}\cap \mathcal{S} = \{\theta \in \mathcal{H}: \bar{\varepsilon}
\geq \rho_{\mathcal{H}}(\theta,\theta_\star) \geq \varepsilon\}\cap \mathcal{S}.$$
Now $\{\theta \in \mathcal{H}: \bar{\varepsilon} \geq \rho_{\mathcal{H}}(\theta,\theta_\star) \geq \varepsilon\}$ is compact. 
An intersection of compact sets is compact. Therefore, $\theta \mapsto \was_p(\mu_\star,\mu_\theta)$ being continuous/lower semicontinuous,
an infimum is attained on $\{\theta \in \mathcal{H}:\rho_{\mathcal{H}}(\theta,\theta_\star) \geq \varepsilon\}\cap \mathcal{S}$, and by uniqueness of $\theta_\star$, well-separation follows.
\end{proof}

\end{appendix}

\end{document}